\documentclass[aps,pra,10pt,a4paper,superscriptaddress,onecolumn,preprintnumbers,floatfix,nofootinbib, showkeys]{revtex4-2}

\usepackage{graphicx}
\usepackage[utf8]{inputenc}
\usepackage{tocloft}
\usepackage{epsf}
\usepackage{bm}
\usepackage{amsmath}
\usepackage{amsfonts}
\usepackage{amssymb}
\usepackage{color}
\usepackage{caption}
\usepackage{subcaption}
\usepackage{booktabs} 
\usepackage{float}
\usepackage{empheq}

\usepackage[citecolor=blue,urlcolor=blue,unicode=true,pdfusetitle, bookmarks=true,bookmarksnumbered=true,bookmarksopen=true,bookmarksopenlevel=3, breaklinks=false,pdfborder={0 0 1},backref=false,colorlinks=true, linkcolor=blue]{hyperref}

\numberwithin{equation}{section}

\makeatletter

\renewcommand*{\p@subsection}{}

\renewcommand*{\p@subsubsection}{}

\renewcommand*{\p@paragraph}{}
\makeatother

\providecommand{\U}[1]{\protect\rule{.1in}{.1in}}

\newcommand{\be}{\begin{equation}}
\newcommand{\ee}{\end{equation}}

\newcommand{\mincir}{\raise
-3.truept\hbox{\rlap{\hbox{$\sim$}}\raise4.truept\hbox{$<$}\ }}
\newcommand{\magcir}{\raise
-3.truept\hbox{\rlap{\hbox{$\sim$}}\raise4.truept\hbox{$>$}\ }}

\begin{document}
\title{III. Interacting Dark Energy: Summary of Models, Pathologies, and Constraints
}

\author{Marcel van der Westhuizen}
\email{marcelvdw007@gmail.com}
\affiliation{Centre for Space Research, North-West University, Potchefstroom 2520, South
Africa}
\author{Amare Abebe}
\email{amare.abebe@nithecs.ac.za}
\affiliation{Centre for Space Research, North-West University, Potchefstroom 2520, South
Africa}
\affiliation{National Institute for Theoretical and Computational Sciences (NITheCS),
South Africa}

\author{Eleonora Di Valentino}
\email{e.divalentino@sheffield.ac.uk}
\affiliation{School of Mathematical and Physical Sciences, University of Sheffield, Hounsfield Road, Sheffield S3 7RH, United Kingdom}

\begin{abstract}

We present an overview of the main results from our two companion papers that are relevant for observational constraints on interacting dark energy (IDE) models. We provide analytical solutions for the dark matter and dark energy densities, $\rho_{\rm dm}$ and $\rho_{\rm de}$, as well as the normalized Hubble function $h(z)$, for eight IDE models. These include five linear IDE models, namely $Q=3H(\delta_{\rm dm} \rho_{\rm dm} + \delta_{\rm de} \rho_{\rm de})$ and four special cases: $Q=3H\delta(\rho_{\rm dm}+\rho_{\rm de})$, $Q=3H\delta(\rho_{\rm dm}-\rho_{\rm de})$, $Q=3H\delta \rho_{\rm dm}$, and $Q=3H\delta \rho_{\rm de}$, together with three non-linear IDE models: $Q=3H\delta \left( \tfrac{\rho_{\rm dm} \rho_{\rm de}}{\rho_{\rm dm}+\rho_{\rm de}} \right)$, $Q=3H\delta \left( \tfrac{\rho_{\rm dm}^2}{\rho_{\rm dm}+\rho_{\rm de}} \right)$, and $Q=3H\delta \left( \tfrac{\rho_{\rm de}^2}{\rho_{\rm dm}+\rho_{\rm de}} \right)$. For these eight models, we present conditions to avoid imaginary, undefined, and negative energy densities. In seven of the eight cases, negative densities arise if energy flows from DM to DE, implying a strong theoretical preference for energy transfer from DE to DM. We also provide conditions to avoid future big rip singularities and evaluate how each model addresses the coincidence problem in both the past and the future. Finally, we propose a set of approaches and simplifying assumptions that can be used when constraining IDE models, by defining regimes that restrict the parameter space according to the behavior researchers are willing to tolerate.
\end{abstract}
\date{\today}
\maketitle
\date{\today }

\section{Introduction}
In our two companion papers titled ``I. Linear Interacting Dark Energy: Analytical Solutions and Theoretical Pathologies''\cite{vanderWesthuizen:2025I} and ``II. Non-Linear Interacting Dark Energy: Analytical Solutions and Theoretical Pathologies'' \cite{vanderWesthuizen:2025II}, we provided an in-depth study of the background dynamics of eight different interacting dark energy (IDE) models. The eight interaction kernels studied are phenomenological and were chosen due to their simplicity, which allowed new analytical solutions for the evolution of dark matter (DM) and dark energy (DE) to be obtained from the modified conservation equations. These specific interaction kernels are also some of the most widely studied (see our companion papers and the references therein), which prompted us to fill in gaps within the literature, especially with regard to the often-overlooked presence of both negative energy densities and future big rip singularities. We also discussed how these models address the coincidence problem, and the possibility of phantom crossing appearing in either the effective DE equation of state or the reconstructed dynamical DE equation of state, such that at some point we have either $w^{\rm eff}_{\rm de}(z) = -1$ or $\tilde{w}(z)=-1$.

The relevancy of IDE models to modern cosmology is discussed in detail in our companion papers, but a brief summary of five of the greatest motivations to study these models is given below for the reader who is only interested in an overview of the topic:
\begin{enumerate}
    \item \textbf{The cosmological constant problem:} The predicted energy density of a cosmological constant $\Lambda$ is approximately 120 orders of magnitude smaller than the predicted value. \cite{Weinberg:1988cp}. This does not directly motivate research into IDE models, but provides a reason to consider DE models beyond the $\Lambda$CDM model.
    \item \textbf{The coincidence problem:} The densities of DM and DE are observed to have the same order of magnitude today, even though they are predicted to differ by many orders of magnitude in both the past and the future \cite{Amendola_2000, Zimdahl_2001, Chimento_2003, Farrar_2004, Wang_2004, Olivares_2006, Sadjadi_2006, Quartin_2008, Campo_2009, Caldera_Cabral_2009_DSA, He_2011, delcampo2015interactiondarksector, von_Marttens_2019}. This provided the initial motivation to specifically study IDE models. 
    \item \textbf{The Hubble tension:} The $4\sigma-6\sigma$ discrepancy in the estimation of the present expansion rate $H_0$ from late-time probes such as Type Ia Supernova and early-time probes such as CMB. The potential of IDE models to address this tension has caused a resurgence in their popularity in recent years \cite{Kumar:2016zpg, Murgia:2016ccp, Kumar:2017dnp, DiValentino:2017iww, Kumar:2021eev, Pan:2023mie, Benisty:2024lmj, Yang:2020uga, Forconi:2023hsj, Pourtsidou:2016ico, DiValentino:2020vnx, DiValentino:2020leo, Nunes:2021zzi, Yang:2018uae, vonMarttens:2019ixw, Lucca:2020zjb, Zhai:2023yny, Bernui:2023byc, Hoerning:2023hks, Giare:2024ytc, Escamilla:2023shf, vanderWesthuizen:2023hcl, Silva:2024ift, DiValentino:2019ffd, Li:2024qso, Pooya:2024wsq, Halder:2024uao, Castello:2023zjr, Yao:2023jau, Mishra:2023ueo, Nunes:2016dlj, Silva:2025hxw, Zheng_2017, Kumar_2019, Anchordoqui_2021, Pan_2019, Guo_2021, Gao:2021xnk, Di_Valentino_2021_H0_review, Gariazzo_2022, Wang_2022, Califano_2023, Pan_2024, Liu_2023, Liu_2024, Sabogal_2025,Yang:2025uyv, Vagnozzi:2023nrq}.   
    \item \textbf{The $S_8$ discrepancy:} The $2\sigma-5\sigma$ discrepancy between early-time and late-time measurements of the parameter $S_8$, which is related to the clumping of matter on cosmological scales. IDE models have been investigated to possibly alleviate this tension along with the $H_0$ tension \cite{Kumar:2017dnp, Kumar_2019, Di_Valentino_2020_rhode, Anchordoqui_2021, Kumar:2021eev, Gariazzo_2022, lucca2021darkenergydarkmatterinteractions, Sabogal_2024, Liu_2023, Liu_2024, Sabogal_2025, yang2025probingcoldnaturedark, Liu_2025}. 
    \item \textbf{Hints of dynamical dark energy}: Recent measurements from DESI collaboration of baryonic acoustic oscillations (BAO) provide a $2.8 \sigma-4.2\sigma$ preference for dynamical DE over the $\Lambda$CDM model \cite{DESI:2025zgx,DESI:2025fii,DESI:2025qqy,DESI:2025wyn} (see also~\cite{DESI:2024mwx,Cortes:2024lgw,Shlivko:2024llw,Luongo:2024fww,Yin:2024hba,Gialamas:2024lyw,Dinda:2024kjf,Najafi:2024qzm,Wang:2024dka,Ye:2024ywg,Tada:2024znt,Carloni:2024zpl,Chan-GyungPark:2024mlx,DESI:2024kob,Ramadan:2024kmn,Notari:2024rti,Orchard:2024bve,Hernandez-Almada:2024ost,Pourojaghi:2024tmw,Giare:2024gpk,Reboucas:2024smm,Giare:2024ocw,Chan-GyungPark:2024brx,Menci:2024hop,Li:2024qus,Li:2024hrv,Notari:2024zmi,Gao:2024ily,Fikri:2024klc,Jiang:2024xnu,Zheng:2024qzi,Gomez-Valent:2024ejh,RoyChoudhury:2024wri,Lewis:2024cqj,Wolf:2024eph,Wolf:2024stt,Wolf:2025jed, Wolf:2025jlc,Shajib:2025tpd,Giare:2025pzu,Chaussidon:2025npr,Kessler:2025kju,Pang:2025lvh,RoyChoudhury:2025dhe,Scherer:2025esj,Teixeira:2025czm,Specogna:2025guo,Cheng:2025lod,Cheng:2025hug,Ozulker:2025ehg,Gialamas:2025pwv,Lee:2025pzo}). IDE models provide a natural mechanism for the dynamical behavior of DE, thus reinvigorating the interest in these models. 
\end{enumerate}
In this paper, we summarize the main results obtained from our two companion papers, as well as provide additional insight into approaches that can be used to constrain these models with the latest cosmological data. To this end, we provide new expressions for the normalized Hubble function $h(z)$ for each interaction in Section~\ref{h(z)}. Furthermore, understanding the parameter space is paramount when constraining IDE models, as non-physical behavior may arise in sections of the parameter space. We therefore provide conditions to avoid common pitfalls for these models in Section~\ref{pitfalls}, as summarized in Table~\ref{tab:Com_real} (imaginary/undefined densities), Table~\ref{tab:Com_PEC} (conditions to avoid negative energy densities) and Table~\ref{tab:Com_AE_BR} (accelerated expansion and big rip avoidance). The potential of each model to either alleviate or solve the coincidence problem in both the past and future is summarized in Table \ref{tab:Com_CP}. In Section \ref{regimes}, we illustrate how these conditions can be applied in practice by defining regimes that restrict the allowed parameter space of these models when constraining them with observational data. These regimes allow researchers to choose for themselves which behavior of the models they are willing to tolerate. Lastly, in Section \ref{conclusions}, we provide our main conclusions from this study and discuss future work.  

\section{Analytical solutions for the energy densities and Hubble function for 8 IDE models.} \label{h(z)}

In order to obtain observational constraints for IDE models, it is useful to have an analytical expression for the Hubble parameter $H$, which describes the background expansion of the universe and is given by:
\begin{gather} \label{DSA.H}
\begin{split}
H^2 = \left(\frac{\dot{a}}{a}\right)^2 = \frac{8\pi G}{3} \left(\rho_{\text{r}} + \rho_{\text{bm}} + \rho_{\text{dm}} + \rho_{\text{de}}\right), \quad 
\rho_{\text{r}} = \rho_{\text{(r,0)}} a^{-4}, \quad 
\rho_{\text{bm}} = \rho_{\text{(bm,0)}} a^{-3}.
\end{split}
\end{gather}
Equation \eqref{DSA.H} is the Hubble function for a flat universe containing DM, DE, radiation (r), and baryons (bm). Only the expressions for $\rho_{\text{dm}}$ and $\rho_{\text{de}}$ will differ from non-interacting models, depending on the interaction kernel $Q$. It should be noted that baryonic matter $\rho_{\text{bm}}$ and $\rho_{\text{dm}}$ are separately conserved and evolve independently as in the $\Lambda$CDM model. 

In most cases, due to the complexity of the interaction kernel $Q$, the conservation equations cannot be solved analytically, and numerical methods need to be used to find the evolution of the energy densities of the dark components. However, exact solutions can be found for simple phenomenological IDE models. The expressions for $\rho_{\text{dm}}$ and $\rho_{\text{de}}$, and the normalized Hubble function $h(z) = H(z)/H_0$, for five linear IDE models (derived in companion paper I) and three non-linear IDE models (derived in companion paper II) follow below. 

\subsection{Linear IDE models}
\subsubsection{Linear IDE model 1: $Q=3 H (\delta_{\text{dm}} \rho_{\text{dm}} + \delta_{\text{de}}  \rho_{\text{de}})$}

This is the most general IDE model we studied, with the four following linear models being special cases of this interaction kernel, depending on the sign and magnitude of coupling constants $\delta_{\text{dm}}$ and $\delta_{\text{de}}$. Due to the presence of a determinant $\Delta$ in both $\rho_{\text{dm}}$ and $\rho_{\text{de}}$, care should be taken to avoid imaginary values, using the conditions provided in Table~\ref{tab:Com_real}. Negative energies may also be avoided using the set of three conditions given in Table~\ref{tab:Com_PEC}. The densities of DM and DE evolve with the scale factor according to the expressions \eqref{eq:rho_dm_Q_linear} and \eqref{eq:rho_de_Q_linear}, where $w$ is the DE equation of state. This interaction has been studied in  
\cite{Chimento_2012, Pan_2015, Pan_2017, von_Marttens_2019, Quartin_2008, He_2008, Caldera_Cabral_2009_DSA, Caldera_Cabral_2009_structure, Pan_2020, Nojiri_2005_IDE, de_Haro_2023, Valiviita:2008iv, He_2009, He_2011, Costa_2014, Eingorn_2015, Costa_2017, Sadjadi_2006, carrasco2023discriminatinginteractingdarkenergy, Arevalo_2017, Cid_2019, Lepe_2016, Costa_2019, An_2019, Mahata_2015}.

\begin{gather}
\begin{split} \label{eq:rho_dm_Q_linear}
\rho_{\text{dm}} = -& \frac{\delta_{\text{dm}}-\delta_{\text{de}}+w+\Delta}{4w \Delta} 
\Bigl[\rho_{\text{(de,0)}}\Bigl(\delta_{\text{dm}}-\delta_{\text{de}}+w-\Delta\Bigr)
+\rho_{\text{(dm,0)}}\Bigl(\delta_{\text{dm}}-\delta_{\text{de}}-w-\Delta\Bigr)\Bigr] 
a^{\frac{3}{2}\Bigl(\delta_{\text{dm}}-\delta_{\text{de}}-w-2+\Delta\Bigr)} \\[1mm]
+& \frac{\delta_{\text{dm}}-\delta_{\text{de}}+w-\Delta}{4w \Delta} 
\Bigl[\rho_{\text{(de,0)}}\Bigl(\delta_{\text{dm}}-\delta_{\text{de}}+w+\Delta\Bigr)
+\rho_{\text{(dm,0)}}\Bigl(\delta_{\text{dm}}-\delta_{\text{de}}-w+\Delta\Bigr)\Bigr] 
a^{\frac{3}{2}\Bigl(\delta_{\text{dm}}-\delta_{\text{de}}-w-2-\Delta\Bigr)} ,
\end{split}   
\end{gather}
\begin{gather}
\begin{split} \label{eq:rho_de_Q_linear}
\rho_{\text{de}} = +& \frac{\delta_{\text{dm}}-\delta_{\text{de}}-w+\Delta}{4w \Delta} 
\Bigl[\rho_{\text{(de,0)}}\Bigl(\delta_{\text{dm}}-\delta_{\text{de}}+w-\Delta\Bigr)
+\rho_{\text{(dm,0)}}\Bigl(\delta_{\text{dm}}-\delta_{\text{de}}-w-\Delta\Bigr)\Bigr] 
a^{\frac{3}{2}\Bigl(\delta_{\text{dm}}-\delta_{\text{de}}-w-2+\Delta\Bigr)} \\[1mm]
-& \frac{\delta_{\text{dm}}-\delta_{\text{de}}-w-\Delta}{4w \Delta} 
\Bigl[\rho_{\text{(de,0)}}\Bigl(\delta_{\text{dm}}-\delta_{\text{de}}+w+\Delta\Bigr)
+\rho_{\text{(dm,0)}}\Bigl(\delta_{\text{dm}}-\delta_{\text{de}}-w+\Delta\Bigr)\Bigr] 
a^{\frac{3}{2}\Bigl(\delta_{\text{dm}}-\delta_{\text{de}}-w-2-\Delta\Bigr)} ,
\end{split}   
\end{gather}
where $\Delta$ is the determinant:
\begin{gather}
\begin{split} \label{eq:determinant_Q_linear}
\Delta = \sqrt{(\delta_{\text{dm}}+\delta_{\text{de}}+w)^2 - 4\delta_{\text{de}}\delta_{\text{dm}}}\,. 
\end{split}   
\end{gather}
The normalized Hubble function $h(z)$ for this model is obtained by substituting \eqref{eq:rho_dm_Q_linear} and \eqref{eq:rho_de_Q_linear} into \eqref{DSA.H}, while applying the transformations $a=(1+z)^{-1}$ and $\tfrac{8 \pi G}{3H_0^2} \rho_{\rm{(i,0)}} = \Omega_{\rm{(i,0)}}$, which result in the following:

\begin{gather}
\begin{split} \label{hz_Q_linear}
h(z) = \Bigg\{ -& \frac{1}{2 \Delta} 
\Bigl[\Omega_{\text{(de,0)}}\Bigl(\delta_{\text{dm}}-\delta_{\text{de}}+w-\Delta\Bigr)
+\Omega_{\text{(dm,0)}}\Bigl(\delta_{\text{dm}}-\delta_{\text{de}}-w-\Delta\Bigr)\Bigr] 
(1+z)^{-\tfrac{3}{2}\Bigl(\delta_{\text{dm}}-\delta_{\text{de}}-w-2+\Delta\Bigr)} \\[1mm]
+& \frac{1}{2 \Delta} 
\Bigl[\Omega_{\text{(de,0)}}\Bigl(\delta_{\text{dm}}-\delta_{\text{de}}+w+\Delta\Bigr)
+\Omega_{\text{(dm,0)}}\Bigl(\delta_{\text{dm}}-\delta_{\text{de}}-w+\Delta\Bigr)\Bigr] 
(1+z)^{-\tfrac{3}{2}\Bigl(\delta_{\text{dm}}-\delta_{\text{de}}-w-2-\Delta\Bigr)} \\[1mm]
+& \Omega_{\text{(bm,0)}}(1+z)^3 + \Omega_{\text{(r,0)}}(1+z)^4 
\Bigg\}^{\tfrac{1}{2}}.
\end{split}   
\end{gather}

\subsubsection{Linear IDE model 2: $Q=3H\delta( \rho_{\text{dm}}+\rho_{\text{de}})$}

This interaction changes the dynamics in both the distant past and future, during DM and DE domination, respectively.  
Since $Q \neq 0$ when either $\rho_{\text{dm}}$ or $\rho_{\text{de}}$ becomes zero, there is no mechanism to avoid negative energies for this interaction. This model has both negative DM and DE densities in the iDMDE regime, but this can be avoided with a sufficiently small interaction in the iDEDM regime, as given by the conditions in Table~\ref{tab:Com_PEC}. Due to the presence of a determinant $\Delta$ in both $\rho_{\text{dm}}$ and $\rho_{\text{de}}$, imaginary values may appear, but these can be avoided using the conditions provided in Table~\ref{tab:Com_real}. The densities of DM and DE evolve with the scale factor according to the expressions \eqref{eq:rho_dm_Q_dm+de} and \eqref{eq:rho_de_Q_dm+de}. This interaction was previously studied in \cite{Olivares_2005, Califano_2024, Pan_2020, Ar_valo_2022, Olivares_2006, Valiviita:2008iv,  He_2009, He_2011, Campo_2009, Wang:2016lxa, Arevalo_2017, Feng_2016, Huang_2019, AbdollahiZadeh:2019lsx, Pan_2020_Field, Costa_2014, Costa_2017, An_2017, An_2018, Bachega_2020, Li_2020, Halder_2021, Mukhopadhyay_2021, Xiao_2021, Aljaf_2021, Califano_2024, Zhang_2018, Zhang_2019, Liu_2022, Bachega_2020, Califano_2024, Sharma_2021, Yang_2019}.

\begin{gather}
\begin{split} \label{eq:rho_dm_Q_dm+de}
\rho_{\text{dm}} = & -\frac{w+\Delta}{4w\Delta}\Bigl[
\rho_{\text{(de,0)}}(w-\Delta) + \rho_{\text{(dm,0)}}(-w-\Delta)
\Bigr]a^{\tfrac{3}{2}(-w-2+\Delta)} \\[1mm]
& + \frac{w-\Delta}{4w\Delta}\Bigl[
\rho_{\text{(de,0)}}(w+\Delta) + \rho_{\text{(dm,0)}}(-w+\Delta)
\Bigr]a^{\tfrac{3}{2}(-w-2-\Delta)}, 
\end{split}   
\end{gather}
\begin{gather}
\begin{split} \label{eq:rho_de_Q_dm+de}
\rho_{\text{de}} = & +\frac{-w+\Delta}{4w\Delta}\Bigl[
\rho_{\text{(de,0)}}(w-\Delta) + \rho_{\text{(dm,0)}}(-w-\Delta)
\Bigr]a^{\tfrac{3}{2}(-w-2+\Delta)} \\[1mm]
& -\frac{-w-\Delta}{4w\Delta}\Bigl[
\rho_{\text{(de,0)}}(w+\Delta) + \rho_{\text{(dm,0)}}(-w+\Delta)
\Bigr]a^{\tfrac{3}{2}(-w-2-\Delta)},
\end{split}   
\end{gather}
where $\Delta$ is the determinant:
\begin{gather}
\begin{split} \label{eq:determinant_Q_dm+de}
\Delta = \sqrt{w(4\delta + w)}\,.
\end{split}   
\end{gather}

The normalized Hubble function $h(z)$ for this model is:

\begin{gather}
\begin{split} \label{hz_Q_dm+de}
h(z) = \Bigg\{ -& \frac{1}{2 \Delta} 
\Bigl[\Omega_{\text{(de,0)}}(w-\Delta)
+ \Omega_{\text{(dm,0)}}(-w-\Delta)\Bigr] 
(1+z)^{-\tfrac{3}{2}(-w-2+\Delta)} \\[1mm]
+& \frac{1}{2 \Delta} 
\Bigl[\Omega_{\text{(de,0)}}(w+\Delta)
+ \Omega_{\text{(dm,0)}}(-w+\Delta)\Bigr] 
(1+z)^{-\tfrac{3}{2}(-w-2-\Delta)} \\[1mm]
+&  \Omega_{\text{(bm,0)}}(1+z)^3 + \Omega_{\text{(r,0)}}(1+z)^4
\Bigg\}^{\tfrac{1}{2}}.
\end{split}   
\end{gather}

\subsubsection{Linear IDE model 3: $Q=3H\delta( \rho_{\text{dm}}-\rho_{\text{de}})$}

This interaction is a sign-changing interaction; therefore, the sign of $\delta$ will only determine the initial direction of energy transfer, which will switch at some point in the cosmological evolution. Regardless of the choice of $\delta$, this model will always have either negative DM or DE densities, as seen in Table~\ref{tab:Com_PEC}. This model has a determinant $\Delta$, but due to the square in each term, imaginary values will always be avoided. The densities of DM and DE evolve with the scale factor according to the expressions \eqref{eq:rho_dm_Q_dm-de} and \eqref{eq:rho_de_Q_dm-de}. This interaction was previously studied in \cite{Sun_2012,Zhang_2014, halder2024phasespaceanalysissignshifting, Pan_2019, Pan_2024, Aljaf_2021, Khurshudyan_2015}.

\begin{gather}
\begin{split} \label{eq:rho_dm_Q_dm-de}
\rho_{\text{dm}} = -& \frac{2\delta+w+\Delta}{4w \Delta} 
\Bigl[\rho_{\text{(de,0)}}(2\delta+w-\Delta)
+ \rho_{\text{(dm,0)}}(2\delta-w-\Delta)\Bigr] 
a^{\tfrac{3}{2}(2\delta-w-2+\Delta)} \\[1mm]
+& \frac{2\delta+w-\Delta}{4w \Delta} 
\Bigl[\rho_{\text{(de,0)}}(2\delta+w+\Delta)
+ \rho_{\text{(dm,0)}}(2\delta-w+\Delta)\Bigr] 
a^{\tfrac{3}{2}(2\delta-w-2-\Delta)} , 
\end{split}   
\end{gather}

\begin{gather}
\begin{split} \label{eq:rho_de_Q_dm-de}
\rho_{\text{de}} = & +\frac{2\delta-w+\Delta}{4w \Delta} 
\Bigl[\rho_{\text{(de,0)}}(2\delta+w-\Delta)
+ \rho_{\text{(dm,0)}}(2\delta-w-\Delta)\Bigr] 
a^{\tfrac{3}{2}(2\delta-w-2+\Delta)} \\[1mm]
&- \frac{2\delta-w-\Delta}{4w \Delta} 
\Bigl[\rho_{\text{(de,0)}}(2\delta+w+\Delta)
+ \rho_{\text{(dm,0)}}(2\delta-w+\Delta)\Bigr] 
a^{\tfrac{3}{2}(2\delta-w-2-\Delta)} ,
\end{split}   
\end{gather}
where $\Delta$ is the determinant:
\begin{gather}
\begin{split} \label{eq:determinant_Q_dm-de}
\Delta = \sqrt{4\delta^2 + w^2}\,.
\end{split}   
\end{gather}
The normalized Hubble function $h(z)$ for this model is:

\begin{gather}
\begin{split} \label{hz_Q_dm-de}
h(z) = \Bigg\{ -& \frac{1}{2 \Delta} 
\Bigl[\Omega_{\text{(de,0)}}(2\delta+w-\Delta)
+ \Omega_{\text{(dm,0)}}(2\delta-w-\Delta)\Bigr] 
(1+z)^{-\tfrac{3}{2}(2\delta-w-2+\Delta)} \\[1mm]
+& \frac{1}{2 \Delta} 
\Bigl[\Omega_{\text{(de,0)}}(2\delta+w+\Delta)
+ \Omega_{\text{(dm,0)}}(2\delta-w+\Delta)\Bigr] 
(1+z)^{-\tfrac{3}{2}(2\delta-w-2-\Delta)} \\[1mm]
+&   \Omega_{\text{(bm,0)}}(1+z)^3 + \Omega_{\text{(r,0)}}(1+z)^4
\Bigg\}^{\tfrac{1}{2}}.
\end{split}   
\end{gather}

\subsubsection{Linear IDE model 4: $Q=3H\delta \rho_{\text{dm}}$}

This interaction mostly changes the dynamics in the distant past during DM domination. For this interaction, $Q=0$ if $\rho_{\text{dm}}=0$; therefore, we can guarantee that $\rho_{\text{dm}} \geq 0$ at all times. In contrast, this interaction will have past negative DE densities in the iDMDE regime, but this can be avoided with a sufficiently small interaction in the iDEDM regime, as given by the conditions in Table~\ref{tab:Com_PEC}. The densities of DM and DE evolve with the scale factor according to the expressions \eqref{eq:rho_dm_Q_dm} and \eqref{eq:rho_de_Q_dm}. This interaction was previously studied in \cite{Wang_2004, von_Marttens_2019, Mishra:2023ueo, Valiviita:2008iv, He_2008, V_liviita_2010, Izquierdo_2017, Pan_2020, Ar_valo_2022, vanderWesthuizen:2023hcl, Rodriguez_Benites_2024, Nojiri_2005_IDE, vanderWesthuizen:2023hcl, M.B.Gavela_2009, He_2009, Honorez_2010, He_2011, Khyllep_2022, pooya2024growthmatterperturbationsinteracting, carrasco2023discriminatinginteractingdarkenergy, Kumar:2016zpg, Li_2020_2, Guo_2021, Di_Valentino_2021_H0_review, Wang_2022, Costa_2014, Costa_2017, Santos_2017, An_2017, An_2018, Grand_n_2019, von_Marttens_2019, Bachega_2020, Aljaf_2021, Califano_2024, Li_2024, yan2025investigatinginteractingdarkenergy, Zhang_2018, Zhang_2019, Liu_2022, Caprini_2016, Yang_2019, Bachega_2020, Califano_2024, wang2025prospectsconstraininginteractingdark, Costa_2018, Li_2020, Halder_2021, Mukhopadhyay_2021, Xiao_2021, Guo_2017, Feng_2019_2, Feng_2020, Zhao_2020, Li_2020_2, Guo_2018, li2025probingsignchangeableinteractiondark, Feng_2016, Feng_2018, Sadri_2019, Guo_2018, B_gu__2019, Pan_2020_Field, Banerjee_2024, li2025probingsignchangeableinteractiondark, Guin:2025xki, zhang2009crossingphantomdivide, Rezaei_2020, Zhao_2023, nagpal2025darksectorinteractionsprobing}.

\begin{gather}
\begin{split} \label{eq:rho_dm_Q_dm}
\rho_{\text{dm}} = \rho_{\text{(dm,0)}}\, a^{-3(1-\delta)},
\end{split}   
\end{gather}
\begin{gather}
\begin{split} \label{eq:rho_de_Q_dm}
\rho_{\text{de}} = \Biggl(\rho_{\text{(de,0)}} 
+ \rho_{\text{(dm,0)}} \left(\frac{\delta}{\delta + w}\right) 
\Bigl[1 - a^{3(\delta+w)}\Bigr]\Biggr) a^{-3(w+1)}.
\end{split}   
\end{gather} 
The normalized Hubble function $h(z)$ for this model is:

\begin{gather}
\begin{split} \label{hz_Q_dm}
h(z) = \Bigg\{ & \Omega_{\text{(dm,0)}} (1+z)^{3(1-\delta)} \\[1mm]
& + \Biggl(\Omega_{\text{(de,0)}} 
+ \Omega_{\text{(dm,0)}} \left(\frac{\delta}{\delta+w}\right) 
\Bigl[1 - (1+z)^{-3(\delta+w)}\Bigr]\Biggr)(1+z)^{3(w+1)} \\[1mm]
& +  \Omega_{\text{(bm,0)}}(1+z)^3 + \Omega_{\text{(r,0)}}(1+z)^4 \Bigg\}^{\tfrac{1}{2}}.
\end{split}   
\end{gather}

\subsubsection{Linear IDE model 5: $Q=3H\delta \rho_{\text{de}}$}

This interaction mostly changes the dynamics for the late-time expansion and the distant future during DE domination. For this interaction, $Q=0$ if $\rho_{\text{de}}=0$; therefore, we can guarantee that $\rho_{\text{de}} \geq 0$ at all times. In contrast, this interaction will have future negative DM densities in the iDMDE regime, but this can be avoided with a sufficiently small interaction in the iDEDM regime, as given by the conditions in Table~\ref{tab:Com_PEC}. The densities of DM and DE evolve with the scale factor according to the expressions \eqref{eq:rho_dm_Q_de} and \eqref{eq:rho_de_Q_de}. This interaction was previously studied in \cite{Valiviita:2008iv, He_2009, von_Marttens_2019, vanderWesthuizen:2023hcl, He_2008, He_2009, Bahamonde_2018, Izquierdo_2018, Pan_2020, Panotopoulos_2020, Deogharia_2021, von_Marttens_2020, Ar_valo_2022, vanderWesthuizen:2023hcl,Nojiri_2005_IDE, de_Haro_2023, vanderWesthuizen:2023hcl,M.B.Gavela_2009, He_2009, Honorez_2010, He_2011, Marcondes_2016, Yang_2016, Khyllep_2022, pooya2024growthmatterperturbationsinteracting, Silva_2024,Panotopoulos_2020,  carrasco2023discriminatinginteractingdarkenergy,Salvatelli_2013, Di_Valentino_2017, Kumar:2017dnp, Kumar_2019, Li_2020_2, Yang_2020_2, Kumar:2021eev, Lucca_2020,  Di_Valentino_2020,  Di_Valentino_2020_rhode, Di_Valentino_2021_H0_review, lucca2021darkenergydarkmatterinteractions, Anchordoqui_2021, Wang_2022, Yang_2023, Bernui_2023, Califano_2023, Pan_2024, Sabogal_2024, Giare:2024smz, Sabogal_2025,Clemson_2012, Costa_2014, Costa_2017, An_2017, Santos_2017, An_2018, Grand_n_2019, von_Marttens_2019, Aljaf_2021, Nunes_2022, Zhai_2023, Li_2024, Benisty:2024lmj, yan2025investigatinginteractingdarkenergy,Zhang_2018, Zhang_2019, Liu_2022, Zhao_2023, Giar__2024, Silva_2024,Caprini_2016, Bachega_2020, Yang_2020_3, Califano_2023, Califano_2024, wang2025prospectsconstraininginteractingdark,Costa_2018, Li_2020, Halder_2021, Mukhopadhyay_2021, Xiao_2021,Guo_2017, Kumar:2017dnp, Feng_2019, Feng_2019_2, Zhao_2020, Yang_2020, Yang_2020_2, Li_2020_2,Guo_2018, Giar__2024, Sabogal_2025, silva2025newconstraintsinteractingdark, li2025probingsignchangeableinteractiondark,Feng_2016, Nayak_2020, Sinha_2020,Pan_2020_Field, Ghodsi_Yengejeh_2023, Guin:2025xki,Zheng_2017, Di_Valentino_2021_closed, Yang_2021, Joseph_2022, Zhao_2023, Forconi_2024, mbewe2024viscouscosmologicalfluidslargescale, yang2025probingcoldnaturedark}.
\begin{gather}
\begin{split} \label{eq:rho_dm_Q_de}
\rho_{\text{dm}} = \Biggl(\rho_{\text{(dm,0)}}  
+ \rho_{\text{(de,0)}} \left(\frac{\delta}{\delta+w}\right) 
\Bigl[1 - a^{-3(\delta+w)}\Bigr]\Biggr)a^{-3},
\end{split}   
\end{gather}
\begin{gather}
\begin{split} \label{eq:rho_de_Q_de}
\rho_{\text{de}} = \rho_{\text{(de,0)}}\, a^{-3(\delta+w+1)}.
\end{split}   
\end{gather}
The normalized Hubble function $h(z)$ for this model is:
\begin{gather}
\begin{split} \label{hz_Q_de}
h(z) = \Bigg\{ & \Biggl(\Omega_{\text{(dm,0)}}  
+ \Omega_{\text{(de,0)}} \left(\frac{\delta}{\delta+w}\right) 
\Bigl[1 - (1+z)^{3(\delta+w)}\Bigr]\Biggr)(1+z)^{3} \\[1mm]
& + \Omega_{\text{(de,0)}} (1+z)^{3(\delta+w+1)} 
+  \Omega_{\text{(bm,0)}}(1+z)^3 + \Omega_{\text{(r,0)}}(1+z)^4\Bigg\}^{\tfrac{1}{2}}.
\end{split}   
\end{gather}

\subsection{Non-linear IDE models}

\subsubsection{Non-linear IDE model 1: $Q=3H\delta \left( \frac{\rho_{\text{dm}} \rho_{\text{de}} }{\rho_{\text{dm}}+\rho_{\text{de}}} \right)$}

This interaction changes the dynamics in both the distant past and future during DM and DE domination, respectively. Furthermore, since $Q=0$ when either $\rho_{\text{dm}}=0$ or $\rho_{\text{de}}=0$, both the DM and DE densities will remain positive for any choice of parameters. The densities of DM and DE evolve with the scale factor according to the expressions \eqref{NLID1_dm_BG} and \eqref{NLID1_de_BG}. This interaction was previously studied in \cite{Arevalo:2011hh, Chimento_2012, Li_2014, Bolotin_2015,Arevalo:2011hh, von_Marttens_2019,Li_2014, Li_2023,Lip:2010dr, paliathanasis2024compartmentalizationcoexistencedarksector,He_2008, von_Marttens_2019, Arevalo:2011hh, Aljaf_2021, vanderwesthuizen2025compartmentalizationdarksectoruniverse,Yang_2023,Yang_2023,Zhang_2006, carrasco2023discriminatinginteractingdarkenergy,Arevalo_2017, Cid_2019,Pan_2020_Field,Sebastianutti_2024,Ma_2010, mazumder2011interactingholographicdarkenergy, Feng_2016,Zhang_2006, Li_2014, Wang_2013}.
\begin{equation}
\begin{split} 
\rho_{\text{dm}} &= \rho_{\text{(dm,0)}}\, a^{-3(1-\delta)} 
\left[\frac{1+r_0 a^{3(w+\delta)}}{1+r_0}\right]^{-\tfrac{\delta}{w+\delta}}, 
\label{NLID1_dm_BG}
\end{split} 
\end{equation}
\begin{equation}
\begin{split} 
\rho_{\text{de}} &= \rho_{\text{(de,0)}}\, a^{-3(1+w)} 
\left[\frac{1+r_0 a^{3(w+\delta)}}{1+r_0}\right]^{-\tfrac{\delta}{w+\delta}},
\label{NLID1_de_BG}
\end{split} 
\end{equation}
where $r=\frac{\rho_{\text{(dm,0)}}}{\rho_{\text{(de,0)}}}=\frac{\Omega_{\text{(dm,0)}}}{\Omega_{\text{(de,0)}}}$. The normalized Hubble function $h(z)$ for this model is: 
\begin{gather}
\begin{split} \label{hz_Q_dmde}
h(z) = \Bigg\{ & \Bigl[\Omega_{\text{(dm,0)}}(1+z)^{3(1-\delta)} 
+ \Omega_{\text{(de,0)}}(1+z)^{3(1+w)}\Bigr] 
\left[\frac{1+\left(\tfrac{\Omega_{\text{(dm,0)}}}{\Omega_{\text{(de,0)}}}\right)(1+z)^{-3(w+\delta)}}
{1+\left(\tfrac{\Omega_{\text{(dm,0)}}}{\Omega_{\text{(de,0)}}}\right)}\right]^{-\tfrac{\delta}{w+\delta}} \\[1mm]
& +\Omega_{\text{(bm,0)}}(1+z)^3 + \Omega_{\text{(r,0)}}(1+z)^4 \Bigg\}^{\tfrac{1}{2}}.
\end{split}   
\end{gather}

\subsubsection{Non-linear IDE model 2: $Q=3H\delta \left( \frac{\rho_{\text{dm}}^2}{\rho_{\text{dm}}+\rho_{\text{de}}} \right)$}

Similar to the kernel $Q=3H\delta \rho_{\text{dm}}$, this interaction primarily affects the dynamics in the distant past during DM domination but has a smaller impact on the late-time dynamics during DE domination. For this interaction, $Q=0$ if $\rho_{\text{dm}}=0$; therefore, we can guarantee that $\rho_{\text{dm}} \geq 0$ at all times. In contrast, this interaction leads to past negative DE densities in the iDMDE regime, which may be avoided with a small interaction in the iDEDM regime, as specified by the conditions in Table~\ref{tab:Com_PEC}. The densities of DM and DE evolve with the scale factor according to the expressions \eqref{NLID2_dm_BG} and \eqref{NLID2_de_BG}. This interaction was previously studied in \cite{Arevalo:2011hh, Bolotin_2015, von_Marttens_2019, carrasco2023discriminatinginteractingdarkenergy, Arevalo_2017, Cid_2019, Rodriguez_Benites_2024, von_Marttens_2019, Aljaf_2021}. 
\begin{equation}
\begin{split} 
\rho_{\text{dm}} &= \rho_{\text{(dm,0)}}\, a^{-3\left(1-\tfrac{w\delta}{w-\delta}\right)} 
\left[\frac{(w+\delta r_0)a^{-3w}+r_0(w-\delta)}{w(1+r_0)}\right]^{\tfrac{\delta}{w-\delta}}, 
\label{NLID2_dm_BG}
\end{split} 
\end{equation}
\begin{equation}
\begin{split} 
\rho_{\text{de}} &= \rho_{\text{(de,0)}}\, a^{-3\left(1-\tfrac{w\delta}{w-\delta}\right)} 
\left(\frac{(w+\delta r_0)a^{-3w}-\delta r_0}{w}\right)  
\left[\frac{(w+\delta r_0)a^{-3w}+r_0(w-\delta)}{w(1+r_0)}\right]^{\tfrac{\delta}{w-\delta}} .
\label{NLID2_de_BG}
\end{split} 
\end{equation}
The normalized Hubble function $h(z)$ for this model is:
\begin{gather}
\begin{split} \label{hz_Q_dmdm}
h(z) = \Bigg\{ & \Biggl[\Omega_{\text{(dm,0)}} 
+ \Omega_{\text{(de,0)}} \left(
\frac{\Bigl[w+\delta \left(\tfrac{\Omega_{\text{(dm,0)}}}{\Omega_{\text{(de,0)}}}\right)\Bigr](1+z)^{3w} 
- \delta \left(\tfrac{\Omega_{\text{(dm,0)}}}{\Omega_{\text{(de,0)}}}\right)}{w}
\right)\Biggr] 
(1+z)^{3\left(1-\tfrac{w\delta}{w-\delta}\right)} \\[1mm]
& \times \left[
\frac{\Bigl[w+\delta \left(\tfrac{\Omega_{\text{(dm,0)}}}{\Omega_{\text{(de,0)}}}\right)\Bigr](1+z)^{3w} 
+ \left(\tfrac{\Omega_{\text{(dm,0)}}}{\Omega_{\text{(de,0)}}}\right)(w-\delta)}
{w\Bigl[1+\tfrac{\Omega_{\text{(dm,0)}}}{\Omega_{\text{(de,0)}}}\Bigr]}
\right]^{\tfrac{\delta}{w-\delta}}  +  \Omega_{\text{(bm,0)}}(1+z)^3 + \Omega_{\text{(r,0)}}(1+z)^4 \Bigg\}^{\tfrac{1}{2}}.
\end{split}   
\end{gather}

\subsubsection{Non-linear IDE model 3: $Q=3H\delta \left( \frac{\rho_{\text{de}}^2}{\rho_{\text{dm}}+\rho_{\text{de}}} \right)$}

Similar to the kernel $Q=3H\delta \rho_{\text{de}}$, this interaction primarily affects the dynamics of the late-time expansion and the distant future during DE domination, but has a smaller impact on the past dynamics during DM domination. For this interaction, $Q=0$ if $\rho_{\text{de}}=0$; therefore, we can guarantee that $\rho_{\text{de}} \geq 0$ at all times. In contrast, this interaction leads to future negative DM densities in the iDMDE regime, but this can be avoided with a sufficiently small interaction in the iDEDM regime, as specified by the conditions in Table~\ref{tab:Com_PEC}. The densities of DM and DE evolve with the scale factor according to the expressions \eqref{NLID3_dm_BG} and \eqref{NLID3_de_BG}. This interaction was previously studied in \cite{Arevalo:2011hh, Bolotin_2015, von_Marttens_2019, carrasco2023discriminatinginteractingdarkenergy, Arevalo_2017, Cid_2019, von_Marttens_2019, Aljaf_2021}.

\begin{equation}
\begin{split} 
\rho_{\text{dm}} &=  \rho_{\text{(dm,0)}}\, a^{-3\left(1+ \tfrac{w^2}{w-\delta}\right)}  
\left(\frac{(w r_0+\delta )a^{3w}-\delta}{w r_0}\right) 
\left[\frac{(w r_0+\delta )a^{3w}+w-\delta}{w \left(1+ r_0 \right)}\right]^{\tfrac{\delta}{w-\delta}}, 
\label{NLID3_dm_BG}
\end{split} 
\end{equation}

\begin{equation}
\begin{split} 
\rho_{\text{de}} &=  \rho_{\text{(de,0)}}\, a^{-3\left(1+ \tfrac{w^2}{w-\delta}\right)}  
\left[\frac{(w r_0+\delta )a^{3w}+w-\delta}{w \left(1+ r_0\right)}\right]^{\tfrac{\delta}{w-\delta}} .
\label{NLID3_de_BG}
\end{split} 
\end{equation}
The normalized Hubble function $h(z)$ for this model is:
\begin{gather}
\begin{split} \label{hz_Q_dede}
h(z) = \Bigg\{ & \Biggl[\Omega_{\text{(dm,0)}} 
\left(\frac{\Bigl[w \left( \tfrac{\Omega_{\text{(dm,0)}}}{\Omega_{\text{(de,0)}}}\right)+\delta \Bigr](1+z)^{-3w}-\delta}{w \left( \tfrac{\Omega_{\text{(dm,0)}}}{\Omega_{\text{(de,0)}}}\right)}\right) 
+ \Omega_{\text{(de,0)}}\Biggr] (1+z)^{3\left(1+\tfrac{w^2}{w-\delta}\right)} \\[1mm]
& \times \left[\frac{\Bigl[w \left( \tfrac{\Omega_{\text{(dm,0)}}}{\Omega_{\text{(de,0)}}}\right)+\delta \Bigr](1+z)^{-3w}+w-\delta}{w \left(1+ \tfrac{\Omega_{\text{(dm,0)}}}{\Omega_{\text{(de,0)}}}\right)}\right]^{\tfrac{\delta}{w-\delta}}  +  \Omega_{\text{(bm,0)}}(1+z)^3 + \Omega_{\text{(r,0)}}(1+z)^4 \Bigg\}^{\tfrac{1}{2}}.
\end{split}   
\end{gather}

\section{Conditions to avoid pitfalls in the parameter space for each IDE model} \label{pitfalls}

In this section, we summarize the main constraints obtained from our study. First, in order to observationally constrain these models, it is important to identify when the DM and DE densities become imaginary or undefined, so that these regions of parameter space can be avoided, as summarized in Table~\ref{tab:Com_real}. Negative DM and DE densities may also be regarded as non-physical by most researchers. The derived positive energy conditions for each interaction kernel are summarized in Table~\ref{tab:Com_PEC}. The conditions required to ensure accelerated expansion in the distant future, while avoiding a future big rip singularity, are given in Table~\ref{tab:Com_AE_BR}. 
To provide intuition regarding the range of allowed parameters, we also include examples where we have substituted $\Omega_{\rm{(dm,0)}}=0.266$, $\Omega_{\rm{(de,0)}}=0.685$ (implying $r_0=\tfrac{\Omega_{\rm{(dm,0)}}}{\Omega_{\rm{(de,0)}}}=0.388$), and $w=-1$. In Table~\ref{tab:Com_AE_BR}, the example value is given with $w=-1.1$, to illustrate that IDE models can avoid a big rip even when DE lies in the phantom regime ($w<-1$). For completeness, we have also included Table~\ref{tab:Com_CP}, which compares how the eight interactions address the coincidence problem in both the past and the future. The magnitude of the problem is indicated by the deviation from $\zeta=0$, with $\zeta=-3w$ corresponding to the non-interacting $w$CDM model. This table is only relevant for the iDEDM regime, as the iDMDE regime exacerbates the problem~\cite{vanderWesthuizen:2023hcl}. 

\begin{figure}
    \centering
    \includegraphics[width=0.9 \linewidth]{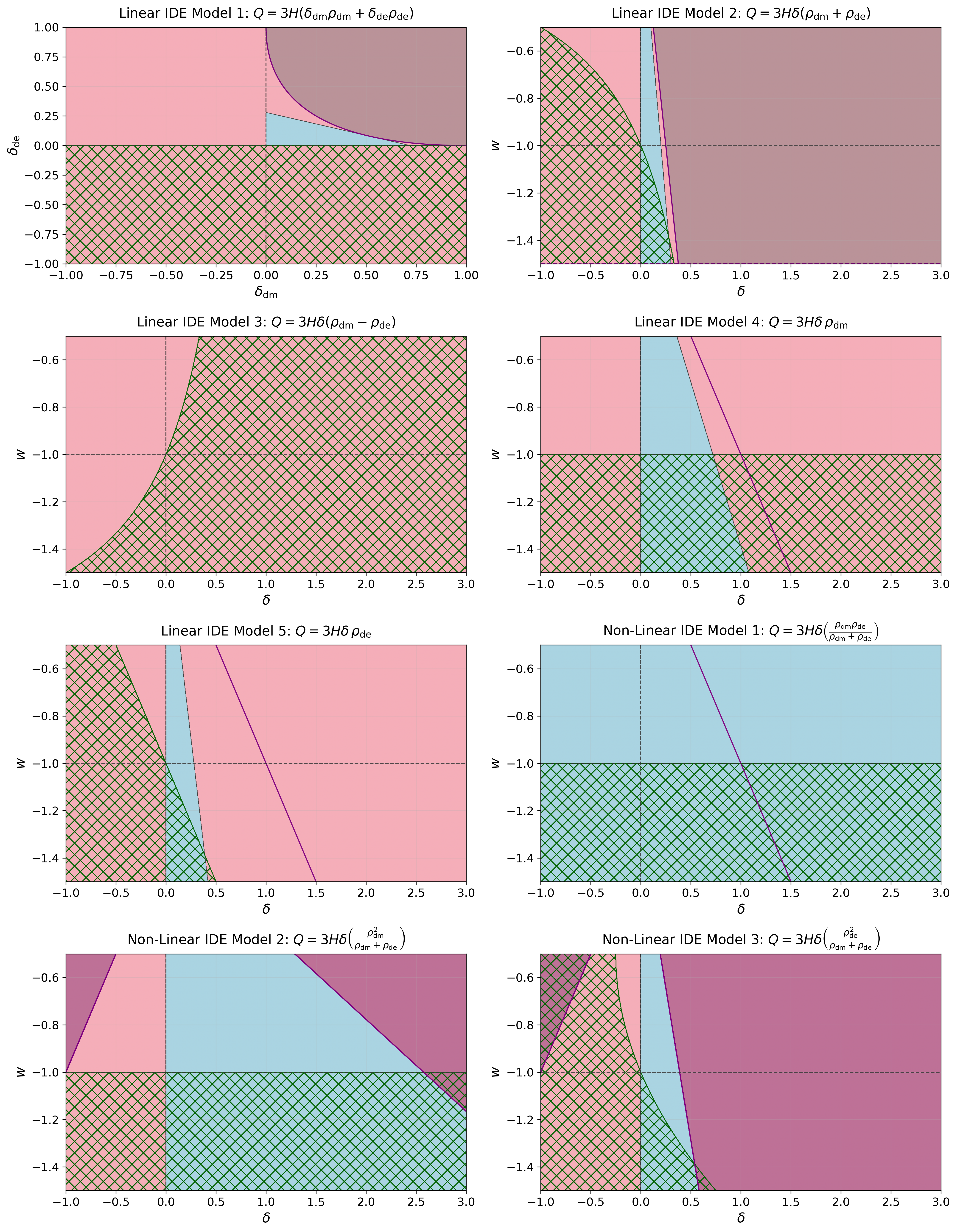}
    \caption{2D Portraits of the parameter space for each of the 8 IDE models, using Table \ref{tab:Com_real}, \ref{tab:Com_PEC} and \ref{tab:Com_AE_BR}. Blue areas indicate where the model has positive energy densities throughout all of cosmic evolution. Pink areas indicate that negative energies will occur in either the past or future expansion. The gray overlay indicates the presence of imaginary energy densities, while the purple areas show undefined energy densities. Lastly, the green mesh indicates the presence of future big rip singularities. The purple areas in the bottom two panels may have undefined values at some scale factor, but it is not guaranteed, see equations \eqref{NLID2_dm_BG}, \eqref{NLID2_de_BG}, \eqref{NLID3_dm_BG} and \eqref{NLID3_de_BG}. For all cases $\Omega_{\rm{(dm,0)}}=0.266$, $\Omega_{\rm{(de,0)}}=0.685$ (implying $r_0=0.388$). For linear IDE model 1, we set $w=-1$. }
    \label{fig:Q_parameter}
\end{figure}

The conditions in Table \ref{tab:Com_real}, \ref{tab:Com_PEC} and \ref{tab:Com_AE_BR} are illustrated in Figure \ref{fig:Q_parameter}, easing comparison  between models and providing a more intuitive understanding of the parameter space of each model. For seven of the eight models, energy transfer from DM to DE ($\delta<0$) results in negative densities. This creates a strong theoretical preference for scenarios where energy flows in the opposite direction, from DE to DM ($\delta>0$). Energy flow from DE to DM also alleviates the coincidence problem and makes big rip future singularities less likely.

\begin{table}[H]
\centering
\renewcommand{\arraystretch}{1.2} 
\setlength{\tabcolsep}{10pt}     
\begin{tabular}{|c|c|c|}
\hline
\textbf{Interaction $Q$} 
 & Conditions to avoid imaginary $\rho_{\text{dm/de}}$
 & Conditions to avoid undefined $\rho_{\text{dm/de}}$ \\ \hline \hline

$3 H (\delta_{\text{dm}} \rho_{\text{dm}} + \delta_{\text{de}}  \rho_{\text{de}})$
 & $(\delta_{\text{dm}}+\delta_{\text{de}}+w)^2>4\delta_{\text{de}}\delta_{\text{dm}}$
 & $w\ne0$ ; $(\delta_{\text{dm}}+\delta_{\text{de}}+w)^2-4\delta_{\text{de}}\delta_{\text{dm}} \neq 0$
\\ \hline

$3H\delta( \rho_{\text{dm}}+\rho_{\text{de}})$
 & $\delta\le-\frac{w}{4}$
 & $w\ne0$ ; $\delta\ne-\frac{w}{4}$
\\ \hline

$3H\delta( \rho_{\text{dm}}-\rho_{\text{de}})$
 & $\rho_{\text{dm/de}}$ \text{ always real}
 & $w\ne0$
\\ \hline

$3H \delta \rho_{\text{dm}}$
 & $\rho_{\text{dm/de}}$ \text{ always real}
 & $\delta\ne -w$
\\ \hline

$3H\delta \rho_{\text{de}}$
 & $\rho_{\text{dm/de}}$ \text{ always real}
 & $\delta\ne -w$
\\ \hline

$3H\delta \left( \frac{\rho_{\text{dm}} \rho_{\text{de}} }{\rho_{\text{dm}}+\rho_{\text{de}}} \right)$
 & $\rho_{\text{dm/de}}$ \text{ always real}
 & $\delta\ne -w$
\\ \hline

$3H\delta \left( \frac{\rho^2_{\text{dm}} }{\rho_{\text{dm}}+\rho_{\text{de}}} \right)$
 & $\rho_{\text{dm/de}}$ \text{ always real}
 & $w<0$ ; $w<\delta\le-\frac{w}{r_0}$
\\ \hline

$3H\delta \left( \frac{\rho^2_{\text{de}} }{\rho_{\text{dm}}+\rho_{\text{de}}} \right)$
 & $\rho_{\text{dm/de}}$ \text{ always real}
 &  $w<0$ ; $w<\delta\le-w r_0$
\\ \hline

\end{tabular}
\caption{Conditions required to avoid imaginary or undefined energy densities for the different interaction kernels.}
\label{tab:Com_real}
\end{table}

\begin{table}[H]
\centering
\renewcommand{\arraystretch}{1.2} 
\setlength{\tabcolsep}{10pt}     
\begin{tabular}{|c|c|c|c|}
\hline
\textbf{Interaction $Q$} 
 & \text{$\rho_{\text{dm/de}}>0$ domain} 
 & \text{$\rho_{\text{dm/de}}>0$ conditions} 
 & \text{Example values} 
\\ \hline \hline

$3 H (\delta_{\text{dm}} \rho_{\text{dm}} + \delta_{\text{de}}  \rho_{\text{de}})$
 & \text{DE $\rightarrow$ DM}
 & \multicolumn{2}{|c|}{$\; \delta_{\text{dm}}\ge0 ; \; \delta_{\text{de}}\ge0 ; \;\delta_{\text{dm}}r_0+ \delta_{\text{de}}\le-\frac{w r_0}{(1+r_0)}$}
\\ \hline

$3H\delta( \rho_{\text{dm}}+\rho_{\text{de}})$
 & \text{DE $\rightarrow$ DM}
 & $0 \le \delta \le -\frac{w r_0}{(1+r_0)^2}$
 & $0 \le \delta \le 0.201$
\\ \hline

$3H\delta( \rho_{\text{dm}}-\rho_{\text{de}})$
 & \text{No viable domain}
 & \text{No viable domain}
 & \text{No viable domain}
\\ \hline

$3H \delta \rho_{\text{dm}}$
 & \text{DE $\rightarrow$ DM}
 & $0 \leq \delta \leq -\frac{w}{(1 + r_0)}$
 & $0 \leq \delta \leq 0.720$
\\ \hline

$3H\delta \rho_{\text{de}}$
 & \text{DE $\rightarrow$ DM}
 & $0 \leq \delta \leq -\frac{w}{\left(1+\frac{1}{r_0}\right)}$
 & $0 \leq \delta \leq 0.280$
\\ \hline

$3H\delta \left( \frac{\rho_{\text{dm}} \rho_{\text{de}} }{\rho_{\text{dm}}+\rho_{\text{de}}} \right)$
 & \text{DE $\leftrightarrow$ DM}
 & $\forall \delta$
 & $\forall \delta$
\\ \hline

$3H\delta \left( \frac{\rho^2_{\text{dm}} }{\rho_{\text{dm}}+\rho_{\text{de}}} \right)$
 & \text{DE $\rightarrow$ DM}
 & $0 \leq \delta \leq -\frac{w}{r_0}$
& $0 \leq \delta \leq 2.575$
\\ \hline

$3H\delta \left( \frac{\rho^2_{\text{de}} }{\rho_{\text{dm}}+\rho_{\text{de}}} \right)$
 & \text{DE $\rightarrow$ DM}
 & $0 \leq \delta \leq -w r_0$
& $0 \leq \delta \leq 0.388$
\\ \hline

\end{tabular}
\caption{Conditions required to ensure positive energy densities for the different interaction kernels.}
\label{tab:Com_PEC}
\end{table}

\begin{table}[H]
\centering
\renewcommand{\arraystretch}{1.2} 
\setlength{\tabcolsep}{10pt}     
\begin{tabular}{|c|c|c|}
\hline
\textbf{Model} 
 & \text{Accelerated expansion $\left[w=-1 \right]$} 
 & \text{No big rip if $w<-1$ $\left[w=-1.1 \right]$ } 
\\ \hline \hline

$Q=3 H (\delta_{\text{dm}} \rho_{\text{dm}} + \delta_{\text{de}}  \rho_{\text{de}})$
 &   $ \delta_{\text{dm}} \left(3w+1\right)-\delta_{\text{de}} \ge w + \frac{1}{3}$
 & $ \delta_{\text{dm}} \left(w+1\right)-\delta_{\text{de}} \le w + 1$
\\ \hline

$Q=3H\delta( \rho_{\text{dm}}+\rho_{\text{de}})$
 & $\delta\le \frac{1}{3}+\frac{1}{9 w}$ ; $\left[\delta \leq 0.222 \right]$
 & $\delta\ge 1+\frac{1}{ w}$ ; $\left[\delta \geq 0.091 \right]$
\\ \hline

$Q=3H\delta( \rho_{\text{dm}}-\rho_{\text{de}})$
 & $\delta \ge \frac{\frac{1}{3} + w}{2+3w}$ ; $\left[\delta \geq 0.666 \right]$
 & $\delta \le \frac{1 + w}{2+w}$ ; $\left[\delta \le -0.111 \right]$
\\ \hline

$Q=3H \delta \rho_{\text{dm}}$
 & $\forall \delta$ if $w\le-\frac{1}{3}$ 
 & Big rip Inevitable
\\ \hline

$Q=3H\delta \rho_{\text{de}}$
 & $\delta \leq -w-\frac{1}{3} $ ; $\left[\delta \leq 0.666\right]$
 & $\delta \geq -w-1 $ ; $\left[\delta \geq 0.1 \right]$

\\ \hline

$Q=3H\delta \left( \frac{\rho_{\text{dm}} \rho_{\text{de}} }{\rho_{\text{dm}}+\rho_{\text{de}}} \right)$
 &  $\forall \delta$ if $w\le-\frac{1}{3}$ 
 & Big rip Inevitable
\\ \hline

$Q=3H\delta \left( \frac{\rho^2_{\text{dm}} }{\rho_{\text{dm}}+\rho_{\text{de}}} \right)$
 & $\forall \delta$ if $w\le-\frac{1}{3}$ 
 & Big rip Inevitable
\\ \hline

$Q=3H\delta \left( \frac{\rho^2_{\text{de}} }{\rho_{\text{dm}}+\rho_{\text{de}}} \right)$
 & $\delta \leq w(3w+1) $ ; $\left[\delta \leq 2 \right]$
 & $\delta \geq  w(w+1)  $ ; $\left[\delta \geq 0.11 \right]$
\\ \hline

$w$CDM 
 & $w\le-\frac{1}{3}$ 
 & Big rip Inevitable
\\ \hline

\end{tabular}
\caption{Conditions required to ensure accelerated expansion and to avoid a future big rip singularity for the different interaction kernels. Values in square brackets are example values, where we have set $r_0=0.388$ and $w=-1$.}
\label{tab:Com_AE_BR}
\end{table}
  
\begin{table}[H]
\centering
\renewcommand{\arraystretch}{1.2} 
\setlength{\tabcolsep}{8pt}      
\begin{tabular}{|c|c|c|}
\hline 
\textbf{Model with $\delta>0$ (iDEDM)} & \text{Coincidence problem (Past)} & \text{Coindence problem (Future)}  \\ \hline\hline

$Q=3 H (\delta_{\text{dm}} \rho_{\text{dm}} + \delta_{\text{de}}  \rho_{\text{de}})$  
  & \text{Solved } [$\zeta=0$]  
  & \text{Solved } [$\zeta=0$] \\ \hline

$Q=3H\delta( \rho_{\text{dm}}+\rho_{\text{de}})$  
  & \text{Solved } [$\zeta=0$]  
  & \text{Solved } [$\zeta=0$] \\ \hline

$Q=3H\delta( \rho_{\text{dm}}-\rho_{\text{de}})$  
  & \text{Solved } [$\zeta=0$]  
  & \text{Solved } [$\zeta=0$, $\rho_{\rm{dm}}<0$] \\ \hline

$Q=3H \delta \rho_{\text{dm}}$  
  & \text{Solved } [$\zeta=0$]  
  & \text{Alleviated } [$\zeta=-3(w+\delta)$]  \\ \hline

$Q=3H\delta \rho_{\text{de}}$  
  & \text{Alleviated } [$\zeta=-3(w+\delta)$]  
  & \text{Solved } [$\zeta=0$] \\ \hline

$Q=3H\delta \left( \frac{\rho_{\text{dm}} \rho_{\text{de}}}{\rho_{\text{dm}}+\rho_{\text{de}}}\right)$  
  & \text{Alleviated } [$\zeta=-3(w+\delta)$]  
  & \text{Alleviated } [$\zeta=-3(w+\delta)$]   \\ \hline

$Q=3H\delta \left( \frac{\rho^2_{\text{dm}} }{\rho_{\text{dm}}+\rho_{\text{de}}} \right)$ 
  & \text{Solved } [$\zeta=0$]  
  & \text{No change } [$\zeta=-3w$]  \\ \hline

$Q=3H\delta \left(  \frac{\rho^2_{\text{de}} }{\rho_{\text{dm}}+\rho_{\text{de}}}\right)$  
  & \text{No change } [$\zeta=-3w$]  
  & \text{Solved } [$\zeta=0$]  \\ \hline

$w\text{CDM}$  
  & $\zeta=-3w$  
  & $\zeta=-3w$ \\ \hline

$\Lambda\text{CDM}$  
  & $\zeta=-3$  
  & $\zeta=-3$ \\ \hline

\end{tabular}
\caption{Potential of the different interaction kernels to address the coincidence problem.}
\label{tab:Com_CP}
\end{table}

\section{Some approaches and simplifying assumptions when constraining IDE models} \label{regimes}

From the expressions for $h(z)$ in Section~\ref{h(z)}, we can see that the phenomenological IDE models we have studied introduce at least three additional parameters compared to $\Lambda$CDM that need to be constrained. First, we have the interaction constant $\delta$ (or both $\delta_{\rm{dm}}$ and $\delta_{\rm{de}}$). Second, DE may not be a vacuum and is described by the DE equation of state $w$. Lastly, there is an additional parameter that arises because $\Omega_{\rm{(bm,0)}}$ and $\Omega_{\rm{(dm,0)}}$ are separately conserved, instead of simply having $\Omega_{\rm{(m,0)}}$ as the only parameter. We refer to this base model as i$w$CDM, to differentiate it from the other regimes detailed below. Thus, our first regime is:
\begin{enumerate}
    \item \textit{Interacting dark energy} (i$w$CDM): The base IDE regime where $w$ is a free parameter, $\Omega_{\rm{(bm,0)}}$ and $\Omega_{\rm{(dm,0)}}$ are separately conserved, and no \textit{a priori} bounds are imposed on the possible values of the coupling constant, with $\delta \in \lbrack -\infty,+\infty]$. This approach makes no additional assumptions and allows the posteriors to best reflect the data. 
\end{enumerate}
More regimes can be introduced by adopting simplifying assumptions that are often used in the literature to circumvent the presence of additional parameters. These approaches can be classified into the following regimes:
\begin{enumerate}
    \setcounter{enumi}{1}
    \item \textit{Interacting baryonic and dark matter} (i$w$CM): One commonly used approach is to group baryonic and dark matter together under cold matter, $\rho_{\rm{m}}=\rho_{\rm{bm}}+\rho_{\rm{dm}}$. This assumption implies that baryonic matter also participates in the interaction, which is highly unlikely given local solar system tests and fifth-force constraints~\cite{Wang:2016lxa, carroll2021quantumfieldtheoryeveryday}. An alternative approach is to fix the baryonic matter density from other measurements, such as big bang nucleosynthesis, as done in~\cite{vanderwesthuizen2025compartmentalizationdarksectoruniverse}.
    
    \item \textit{Interacting vacuum energy} (i$\Lambda$CDM): To remove the free parameter $w$, it can be fixed to $w=-1$, which assumes that the energy exchange occurs between the vacuum $\Lambda$ and CDM, as in~\cite{Sebastianutti_2024}. Caution should be taken with this assumption, as the vacuum scenario with $w=-1$ has also been shown to cause gravitational instabilities~\cite{Lucca_2020, lucca2021darkenergydarkmatterinteractions}.
    
    \item \textit{Interacting quintessence or phantom dark energy} (i$q$CDM and i$p$CDM): Another common approach is to use the stability criterion defined by the doom factor $\mathbf{d}$~\cite{M.B.Gavela_2009} to identify two valid regimes free from instabilities: the iDMDE regime with $w>-1$ and the iDEDM regime with $w<-1$. In practice, researchers often assume small deviations from vacuum behavior, and thus take either $w=-0.999$ (the i$q$CDM regime with $\delta<0$) or $w=-1.001$ (the i$p$CDM regime with $\delta>0$)~\cite{Valentino_2020_DE, Di_Valentino_2020, Gariazzo_2022, Nunes_2022, Yang_2021, Zhai_2023, Giar__2024}. Our study highlights a shortcoming of this approach, as the i$q$CDM regime with $\delta<0$ inevitably leads to negative energies in most cases. 
\end{enumerate}
Besides these assumptions, the conditions given in Section~\ref{pitfalls} may also be used to set parameter priors that \textit{a priori} prevent negative energies or big rip scenarios. This allows us to define the following additional regimes:
\begin{enumerate}
    \setcounter{enumi}{4}
    \item \textit{Positive IDE} (+i$w$CDM): The MCMC exploration is restricted to the region of parameter space where both DM and DE have positive energy densities, as specified by the conditions in Table~\ref{tab:Com_PEC}. This avoids non-physical scenarios but may artificially constrain the posterior distributions to values that would not be obtained from using the data alone without additional assumptions.
    
    \item \textit{Finite IDE} (fi$w$CDM): The MCMC exploration is restricted to the region of parameter space where the scale factor remains finite and a future big rip singularity is avoided, as specified by the conditions in Table~\ref{tab:Com_AE_BR}. This approach has the same advantages and disadvantages as the regime above.
\end{enumerate}

These regimes can also be combined to form additional cases such as +fi$w$CDM, +i$\Lambda$CDM, +i$\Lambda$CM, etc. The main point is that researchers should clearly state their assumptions when constraining IDE models. Alternatively, one may keep the parameter priors open, as in the base i$w$CDM case, and then use the posteriors to predict the behavior of the model, e.g., whether the data hint at negative DE densities in the past or a big rip singularity in the future. The choice of regime therefore depends on which features of the models researchers are willing to tolerate.

\section{Conclusions} \label{conclusions}

In this work, we have summarized the main conclusions from our two companion papers that are relevant for future observational constraints. Specifically, in Section~\ref{h(z)} we have provided the normalized Hubble function $h(z)$ for each interaction:  
$Q=3 H (\delta_{\text{dm}} \rho_{\text{dm}} + \delta_{\text{de}} \rho_{\text{de}})$ in \eqref{hz_Q_linear},  
$Q=3H\delta(\rho_{\text{dm}}+\rho_{\text{de}})$ in \eqref{hz_Q_dm+de},  
$Q=3H\delta(\rho_{\text{dm}}-\rho_{\text{de}})$ in \eqref{hz_Q_dm-de},  
$Q=3H\delta \rho_{\text{dm}}$ in \eqref{hz_Q_dm},  
$Q=3H\delta \rho_{\text{de}}$ in \eqref{hz_Q_de},  
$Q=3H\delta \left(\tfrac{\rho_{\text{dm}} \rho_{\text{de}}}{\rho_{\text{dm}}+\rho_{\text{de}}}\right)$ in \eqref{hz_Q_dmde},  
$Q=3H\delta \left(\tfrac{\rho_{\text{dm}}^2}{\rho_{\text{dm}}+\rho_{\text{de}}}\right)$ in \eqref{hz_Q_dmdm}, and  
$Q=3H\delta \left(\tfrac{\rho_{\text{de}}^2}{\rho_{\text{dm}}+\rho_{\text{de}}}\right)$ in \eqref{hz_Q_dede}.  

In Section~\ref{pitfalls}, for all eight interactions, we provide: conditions to avoid imaginary or undefined energy densities (Table~\ref{tab:Com_real}); conditions to avoid negative energy densities (Table~\ref{tab:Com_PEC}); conditions to ensure accelerated expansion and avoid a future big rip (Table~\ref{tab:Com_AE_BR}); and a summary of how these interactions address the coincidence problem (Table~\ref{tab:Com_CP}). These results are visualised in Figure \ref{fig:Q_parameter}. 
From these results, we find that scenarios with energy flow from DE to DM are often theoretically more stable, as they tend to avoid negative energies and future singularities, alleviate the coincidence problem, and are broadly consistent with thermodynamic considerations~\cite{Pav_n_2008}. Nevertheless, whether such scenarios are realized in nature must ultimately be decided by data, and future observational tests will be decisive.  For now, some discussions on the possibility of negative DE can be found in \cite{Poulin:2018zxs, Wang:2018fng, Visinelli:2019qqu, Calderon:2020hoc, Escamilla:2023shf, guedezounme2025phantomcrossingdarkinteraction}.

In Section~\ref{regimes}, we provide a brief discussion of simplifying assumptions that are often implicitly or explicitly used when constraining IDE models in the literature. We also introduce two new regimes, +i$w$CDM and fi$w$CDM, where the conditions from Tables~\ref{tab:Com_PEC} and~\ref{tab:Com_AE_BR} can be used to set priors and avoid the most severe theoretical pathologies faced by these models.  

Beyond the foundation laid in this paper and the two companion papers, several important directions remain open. First, the analytical results presented here allow each of the models to be constrained with cosmological data at the background level, using the different regimes outlined above. Future work will also incorporate the perturbation equations for each of the interactions, enabling comparisons with structure growth and CMB data. In addition, broader interaction functions with stronger field-theoretical motivation deserve further exploration. Taken together, these developments will make it possible to evaluate more fully the potential of IDE models to address cosmological tensions and to confront the wealth of upcoming precision data.

\begin{acknowledgments}
EDV is supported by a Royal Society Dorothy Hodgkin Research Fellowship. This article is based upon work from the COST Action CA21136 - ``Addressing observational tensions in cosmology with systematics and fundamental physics (CosmoVerse)'', supported by COST - ``European Cooperation in Science and Technology''.
\end{acknowledgments}

\bibliographystyle{apsrev4-2}
\bibliography{References, biblio}  

\begin{thebibliography}{252}%
\makeatletter
\providecommand \@ifxundefined [1]{%
 \@ifx{#1\undefined}
}%
\providecommand \@ifnum [1]{%
 \ifnum #1\expandafter \@firstoftwo
 \else \expandafter \@secondoftwo
 \fi
}%
\providecommand \@ifx [1]{%
 \ifx #1\expandafter \@firstoftwo
 \else \expandafter \@secondoftwo
 \fi
}%
\providecommand \natexlab [1]{#1}%
\providecommand \enquote  [1]{``#1''}%
\providecommand \bibnamefont  [1]{#1}%
\providecommand \bibfnamefont [1]{#1}%
\providecommand \citenamefont [1]{#1}%
\providecommand \href@noop [0]{\@secondoftwo}%
\providecommand \href [0]{\begingroup \@sanitize@url \@href}%
\providecommand \@href[1]{\@@startlink{#1}\@@href}%
\providecommand \@@href[1]{\endgroup#1\@@endlink}%
\providecommand \@sanitize@url [0]{\catcode `\\12\catcode `\$12\catcode `\&12\catcode `\#12\catcode `\^12\catcode `\_12\catcode `\%12\relax}%
\providecommand \@@startlink[1]{}%
\providecommand \@@endlink[0]{}%
\providecommand \url  [0]{\begingroup\@sanitize@url \@url }%
\providecommand \@url [1]{\endgroup\@href {#1}{\urlprefix }}%
\providecommand \urlprefix  [0]{URL }%
\providecommand \Eprint [0]{\href }%
\providecommand \doibase [0]{https://doi.org/}%
\providecommand \selectlanguage [0]{\@gobble}%
\providecommand \bibinfo  [0]{\@secondoftwo}%
\providecommand \bibfield  [0]{\@secondoftwo}%
\providecommand \translation [1]{[#1]}%
\providecommand \BibitemOpen [0]{}%
\providecommand \bibitemStop [0]{}%
\providecommand \bibitemNoStop [0]{.\EOS\space}%
\providecommand \EOS [0]{\spacefactor3000\relax}%
\providecommand \BibitemShut  [1]{\csname bibitem#1\endcsname}%
\let\auto@bib@innerbib\@empty
\bibitem [{\citenamefont {van~der Westhuizen}\ \emph {et~al.}(2025{\natexlab{a}})\citenamefont {van~der Westhuizen}, \citenamefont {Abebe},\ and\ \citenamefont {Di~Valentino}}]{vanderWesthuizen:2025I}%
  \BibitemOpen
  \bibfield  {author} {\bibinfo {author} {\bibfnamefont {M.}~\bibnamefont {van~der Westhuizen}}, \bibinfo {author} {\bibfnamefont {A.}~\bibnamefont {Abebe}},\ and\ \bibinfo {author} {\bibfnamefont {E.}~\bibnamefont {Di~Valentino}},\ }\href@noop {} {\bibinfo {title} {{I. Linear Interacting Dark Energy: Analytical Solutions and Theoretical Pathologies}}} (\bibinfo {year} {2025}{\natexlab{a}}),\ \Eprint {https://arxiv.org/abs/2509.04495} {arXiv:2509.04495 [gr-qc]} \BibitemShut {NoStop}%
\bibitem [{\citenamefont {van~der Westhuizen}\ \emph {et~al.}(2025{\natexlab{b}})\citenamefont {van~der Westhuizen}, \citenamefont {Abebe},\ and\ \citenamefont {Di~Valentino}}]{vanderWesthuizen:2025II}%
  \BibitemOpen
  \bibfield  {author} {\bibinfo {author} {\bibfnamefont {M.}~\bibnamefont {van~der Westhuizen}}, \bibinfo {author} {\bibfnamefont {A.}~\bibnamefont {Abebe}},\ and\ \bibinfo {author} {\bibfnamefont {E.}~\bibnamefont {Di~Valentino}},\ }\href@noop {} {\bibinfo {title} {{II. Non-Linear Interacting Dark Energy: Analytical Solutions and Theoretical Pathologies}}} (\bibinfo {year} {2025}{\natexlab{b}}),\ \Eprint {https://arxiv.org/abs/2509.04494} {arXiv:2509.04494 [gr-qc]} \BibitemShut {NoStop}%
\bibitem [{\citenamefont {Weinberg}(1989)}]{Weinberg:1988cp}%
  \BibitemOpen
  \bibfield  {author} {\bibinfo {author} {\bibfnamefont {S.}~\bibnamefont {Weinberg}},\ }\href {https://doi.org/10.1103/RevModPhys.61.1} {\bibfield  {journal} {\bibinfo  {journal} {Rev. Mod. Phys.}\ }\textbf {\bibinfo {volume} {61}},\ \bibinfo {pages} {1} (\bibinfo {year} {1989})}\BibitemShut {NoStop}%
\bibitem [{\citenamefont {Amendola}(2000)}]{Amendola_2000}%
  \BibitemOpen
  \bibfield  {author} {\bibinfo {author} {\bibfnamefont {L.}~\bibnamefont {Amendola}},\ }\href {https://doi.org/10.1103/PhysRevD.62.043511} {\bibfield  {journal} {\bibinfo  {journal} {Phys. Rev. D}\ }\textbf {\bibinfo {volume} {62}},\ \bibinfo {pages} {043511} (\bibinfo {year} {2000})},\ \Eprint {https://arxiv.org/abs/astro-ph/9908023} {arXiv:astro-ph/9908023} \BibitemShut {NoStop}%
\bibitem [{\citenamefont {Zimdahl}\ and\ \citenamefont {Pavon}(2001)}]{Zimdahl_2001}%
  \BibitemOpen
  \bibfield  {author} {\bibinfo {author} {\bibfnamefont {W.}~\bibnamefont {Zimdahl}}\ and\ \bibinfo {author} {\bibfnamefont {D.}~\bibnamefont {Pavon}},\ }\href {https://doi.org/10.1016/S0370-2693(01)01174-1} {\bibfield  {journal} {\bibinfo  {journal} {Phys. Lett. B}\ }\textbf {\bibinfo {volume} {521}},\ \bibinfo {pages} {133} (\bibinfo {year} {2001})},\ \Eprint {https://arxiv.org/abs/astro-ph/0105479} {arXiv:astro-ph/0105479} \BibitemShut {NoStop}%
\bibitem [{\citenamefont {Chimento}\ \emph {et~al.}(2003)\citenamefont {Chimento}, \citenamefont {Jakubi}, \citenamefont {Pavon},\ and\ \citenamefont {Zimdahl}}]{Chimento_2003}%
  \BibitemOpen
  \bibfield  {author} {\bibinfo {author} {\bibfnamefont {L.~P.}\ \bibnamefont {Chimento}}, \bibinfo {author} {\bibfnamefont {A.~S.}\ \bibnamefont {Jakubi}}, \bibinfo {author} {\bibfnamefont {D.}~\bibnamefont {Pavon}},\ and\ \bibinfo {author} {\bibfnamefont {W.}~\bibnamefont {Zimdahl}},\ }\href {https://doi.org/10.1103/PhysRevD.67.083513} {\bibfield  {journal} {\bibinfo  {journal} {Phys. Rev. D}\ }\textbf {\bibinfo {volume} {67}},\ \bibinfo {pages} {083513} (\bibinfo {year} {2003})},\ \Eprint {https://arxiv.org/abs/astro-ph/0303145} {arXiv:astro-ph/0303145} \BibitemShut {NoStop}%
\bibitem [{\citenamefont {Farrar}\ and\ \citenamefont {Peebles}(2004)}]{Farrar_2004}%
  \BibitemOpen
  \bibfield  {author} {\bibinfo {author} {\bibfnamefont {G.~R.}\ \bibnamefont {Farrar}}\ and\ \bibinfo {author} {\bibfnamefont {P.~J.~E.}\ \bibnamefont {Peebles}},\ }\href {https://doi.org/10.1086/381728} {\bibfield  {journal} {\bibinfo  {journal} {Astrophys. J.}\ }\textbf {\bibinfo {volume} {604}},\ \bibinfo {pages} {1} (\bibinfo {year} {2004})},\ \Eprint {https://arxiv.org/abs/astro-ph/0307316} {arXiv:astro-ph/0307316} \BibitemShut {NoStop}%
\bibitem [{\citenamefont {Wang}\ and\ \citenamefont {Meng}(2005)}]{Wang_2004}%
  \BibitemOpen
  \bibfield  {author} {\bibinfo {author} {\bibfnamefont {P.}~\bibnamefont {Wang}}\ and\ \bibinfo {author} {\bibfnamefont {X.-H.}\ \bibnamefont {Meng}},\ }\href {https://doi.org/10.1088/0264-9381/22/2/003} {\bibfield  {journal} {\bibinfo  {journal} {Class. Quant. Grav.}\ }\textbf {\bibinfo {volume} {22}},\ \bibinfo {pages} {283} (\bibinfo {year} {2005})},\ \Eprint {https://arxiv.org/abs/astro-ph/0408495} {arXiv:astro-ph/0408495} \BibitemShut {NoStop}%
\bibitem [{\citenamefont {Olivares}\ \emph {et~al.}(2006)\citenamefont {Olivares}, \citenamefont {Atrio-Barandela},\ and\ \citenamefont {Pavon}}]{Olivares_2006}%
  \BibitemOpen
  \bibfield  {author} {\bibinfo {author} {\bibfnamefont {G.}~\bibnamefont {Olivares}}, \bibinfo {author} {\bibfnamefont {F.}~\bibnamefont {Atrio-Barandela}},\ and\ \bibinfo {author} {\bibfnamefont {D.}~\bibnamefont {Pavon}},\ }\href {https://doi.org/10.1103/PhysRevD.74.043521} {\bibfield  {journal} {\bibinfo  {journal} {Phys. Rev. D}\ }\textbf {\bibinfo {volume} {74}},\ \bibinfo {pages} {043521} (\bibinfo {year} {2006})},\ \Eprint {https://arxiv.org/abs/astro-ph/0607604} {arXiv:astro-ph/0607604} \BibitemShut {NoStop}%
\bibitem [{\citenamefont {Sadjadi}\ and\ \citenamefont {Alimohammadi}(2006)}]{Sadjadi_2006}%
  \BibitemOpen
  \bibfield  {author} {\bibinfo {author} {\bibfnamefont {H.~M.}\ \bibnamefont {Sadjadi}}\ and\ \bibinfo {author} {\bibfnamefont {M.}~\bibnamefont {Alimohammadi}},\ }\href {https://doi.org/10.1103/PhysRevD.74.103007} {\bibfield  {journal} {\bibinfo  {journal} {Phys. Rev. D}\ }\textbf {\bibinfo {volume} {74}},\ \bibinfo {pages} {103007} (\bibinfo {year} {2006})},\ \Eprint {https://arxiv.org/abs/gr-qc/0610080} {arXiv:gr-qc/0610080} \BibitemShut {NoStop}%
\bibitem [{\citenamefont {Quartin}\ \emph {et~al.}(2008)\citenamefont {Quartin}, \citenamefont {Calvao}, \citenamefont {Joras}, \citenamefont {Reis},\ and\ \citenamefont {Waga}}]{Quartin_2008}%
  \BibitemOpen
  \bibfield  {author} {\bibinfo {author} {\bibfnamefont {M.}~\bibnamefont {Quartin}}, \bibinfo {author} {\bibfnamefont {M.~O.}\ \bibnamefont {Calvao}}, \bibinfo {author} {\bibfnamefont {S.~E.}\ \bibnamefont {Joras}}, \bibinfo {author} {\bibfnamefont {R.~R.~R.}\ \bibnamefont {Reis}},\ and\ \bibinfo {author} {\bibfnamefont {I.}~\bibnamefont {Waga}},\ }\href {https://doi.org/10.1088/1475-7516/2008/05/007} {\bibfield  {journal} {\bibinfo  {journal} {JCAP}\ }\textbf {\bibinfo {volume} {05}},\ \bibinfo {pages} {007}},\ \Eprint {https://arxiv.org/abs/0802.0546} {arXiv:0802.0546 [astro-ph]} \BibitemShut {NoStop}%
\bibitem [{\citenamefont {del Campo}\ \emph {et~al.}(2009)\citenamefont {del Campo}, \citenamefont {Herrera},\ and\ \citenamefont {Pavon}}]{Campo_2009}%
  \BibitemOpen
  \bibfield  {author} {\bibinfo {author} {\bibfnamefont {S.}~\bibnamefont {del Campo}}, \bibinfo {author} {\bibfnamefont {R.}~\bibnamefont {Herrera}},\ and\ \bibinfo {author} {\bibfnamefont {D.}~\bibnamefont {Pavon}},\ }\href {https://doi.org/10.1088/1475-7516/2009/01/020} {\bibfield  {journal} {\bibinfo  {journal} {JCAP}\ }\textbf {\bibinfo {volume} {01}},\ \bibinfo {pages} {020}},\ \Eprint {https://arxiv.org/abs/0812.2210} {arXiv:0812.2210 [gr-qc]} \BibitemShut {NoStop}%
\bibitem [{\citenamefont {Caldera-Cabral}\ \emph {et~al.}(2009{\natexlab{a}})\citenamefont {Caldera-Cabral}, \citenamefont {Maartens},\ and\ \citenamefont {Urena-Lopez}}]{Caldera_Cabral_2009_DSA}%
  \BibitemOpen
  \bibfield  {author} {\bibinfo {author} {\bibfnamefont {G.}~\bibnamefont {Caldera-Cabral}}, \bibinfo {author} {\bibfnamefont {R.}~\bibnamefont {Maartens}},\ and\ \bibinfo {author} {\bibfnamefont {L.~A.}\ \bibnamefont {Urena-Lopez}},\ }\href {https://doi.org/10.1103/PhysRevD.79.063518} {\bibfield  {journal} {\bibinfo  {journal} {Phys. Rev. D}\ }\textbf {\bibinfo {volume} {79}},\ \bibinfo {pages} {063518} (\bibinfo {year} {2009}{\natexlab{a}})},\ \Eprint {https://arxiv.org/abs/0812.1827} {arXiv:0812.1827 [gr-qc]} \BibitemShut {NoStop}%
\bibitem [{\citenamefont {He}\ \emph {et~al.}(2011)\citenamefont {He}, \citenamefont {Wang},\ and\ \citenamefont {Abdalla}}]{He_2011}%
  \BibitemOpen
  \bibfield  {author} {\bibinfo {author} {\bibfnamefont {J.-H.}\ \bibnamefont {He}}, \bibinfo {author} {\bibfnamefont {B.}~\bibnamefont {Wang}},\ and\ \bibinfo {author} {\bibfnamefont {E.}~\bibnamefont {Abdalla}},\ }\href {https://doi.org/10.1103/PhysRevD.83.063515} {\bibfield  {journal} {\bibinfo  {journal} {Phys. Rev. D}\ }\textbf {\bibinfo {volume} {83}},\ \bibinfo {pages} {063515} (\bibinfo {year} {2011})},\ \Eprint {https://arxiv.org/abs/1012.3904} {arXiv:1012.3904 [astro-ph.CO]} \BibitemShut {NoStop}%
\bibitem [{\citenamefont {del Campo}\ \emph {et~al.}(2015)\citenamefont {del Campo}, \citenamefont {Herrera},\ and\ \citenamefont {Pav{\'o}n}}]{delcampo2015interactiondarksector}%
  \BibitemOpen
  \bibfield  {author} {\bibinfo {author} {\bibfnamefont {S.}~\bibnamefont {del Campo}}, \bibinfo {author} {\bibfnamefont {R.}~\bibnamefont {Herrera}},\ and\ \bibinfo {author} {\bibfnamefont {D.}~\bibnamefont {Pav{\'o}n}},\ }\href {https://doi.org/10.1103/PhysRevD.91.123539} {\bibinfo {title} {{Interaction in the dark sector}}} (\bibinfo {year} {2015}),\ \Eprint {https://arxiv.org/abs/1507.00187} {arXiv:1507.00187 [gr-qc]} \BibitemShut {NoStop}%
\bibitem [{\citenamefont {von Marttens}\ \emph {et~al.}(2019)\citenamefont {von Marttens}, \citenamefont {Casarini}, \citenamefont {Mota},\ and\ \citenamefont {Zimdahl}}]{von_Marttens_2019}%
  \BibitemOpen
  \bibfield  {author} {\bibinfo {author} {\bibfnamefont {R.}~\bibnamefont {von Marttens}}, \bibinfo {author} {\bibfnamefont {L.}~\bibnamefont {Casarini}}, \bibinfo {author} {\bibfnamefont {D.~F.}\ \bibnamefont {Mota}},\ and\ \bibinfo {author} {\bibfnamefont {W.}~\bibnamefont {Zimdahl}},\ }\href {https://doi.org/10.1016/j.dark.2018.10.007} {\bibfield  {journal} {\bibinfo  {journal} {Phys. Dark Univ.}\ }\textbf {\bibinfo {volume} {23}},\ \bibinfo {pages} {100248} (\bibinfo {year} {2019})},\ \Eprint {https://arxiv.org/abs/1807.11380} {arXiv:1807.11380 [astro-ph.CO]} \BibitemShut {NoStop}%
\bibitem [{\citenamefont {Kumar}\ and\ \citenamefont {Nunes}(2016)}]{Kumar:2016zpg}%
  \BibitemOpen
  \bibfield  {author} {\bibinfo {author} {\bibfnamefont {S.}~\bibnamefont {Kumar}}\ and\ \bibinfo {author} {\bibfnamefont {R.~C.}\ \bibnamefont {Nunes}},\ }\href {https://doi.org/10.1103/PhysRevD.94.123511} {\bibfield  {journal} {\bibinfo  {journal} {Phys. Rev. D}\ }\textbf {\bibinfo {volume} {94}},\ \bibinfo {pages} {123511} (\bibinfo {year} {2016})},\ \Eprint {https://arxiv.org/abs/1608.02454} {arXiv:1608.02454 [astro-ph.CO]} \BibitemShut {NoStop}%
\bibitem [{\citenamefont {Murgia}\ \emph {et~al.}(2016)\citenamefont {Murgia}, \citenamefont {Gariazzo},\ and\ \citenamefont {Fornengo}}]{Murgia:2016ccp}%
  \BibitemOpen
  \bibfield  {author} {\bibinfo {author} {\bibfnamefont {R.}~\bibnamefont {Murgia}}, \bibinfo {author} {\bibfnamefont {S.}~\bibnamefont {Gariazzo}},\ and\ \bibinfo {author} {\bibfnamefont {N.}~\bibnamefont {Fornengo}},\ }\href {https://doi.org/10.1088/1475-7516/2016/04/014} {\bibfield  {journal} {\bibinfo  {journal} {JCAP}\ }\textbf {\bibinfo {volume} {04}},\ \bibinfo {pages} {014}},\ \Eprint {https://arxiv.org/abs/1602.01765} {arXiv:1602.01765 [astro-ph.CO]} \BibitemShut {NoStop}%
\bibitem [{\citenamefont {Kumar}\ and\ \citenamefont {Nunes}(2017)}]{Kumar:2017dnp}%
  \BibitemOpen
  \bibfield  {author} {\bibinfo {author} {\bibfnamefont {S.}~\bibnamefont {Kumar}}\ and\ \bibinfo {author} {\bibfnamefont {R.~C.}\ \bibnamefont {Nunes}},\ }\href {https://doi.org/10.1103/PhysRevD.96.103511} {\bibfield  {journal} {\bibinfo  {journal} {Phys. Rev. D}\ }\textbf {\bibinfo {volume} {96}},\ \bibinfo {pages} {103511} (\bibinfo {year} {2017})},\ \Eprint {https://arxiv.org/abs/1702.02143} {arXiv:1702.02143 [astro-ph.CO]} \BibitemShut {NoStop}%
\bibitem [{\citenamefont {Di~Valentino}\ \emph {et~al.}(2017{\natexlab{a}})\citenamefont {Di~Valentino}, \citenamefont {Melchiorri},\ and\ \citenamefont {Mena}}]{DiValentino:2017iww}%
  \BibitemOpen
  \bibfield  {author} {\bibinfo {author} {\bibfnamefont {E.}~\bibnamefont {Di~Valentino}}, \bibinfo {author} {\bibfnamefont {A.}~\bibnamefont {Melchiorri}},\ and\ \bibinfo {author} {\bibfnamefont {O.}~\bibnamefont {Mena}},\ }\href {https://doi.org/10.1103/PhysRevD.96.043503} {\bibfield  {journal} {\bibinfo  {journal} {Phys. Rev. D}\ }\textbf {\bibinfo {volume} {96}},\ \bibinfo {pages} {043503} (\bibinfo {year} {2017}{\natexlab{a}})},\ \Eprint {https://arxiv.org/abs/1704.08342} {arXiv:1704.08342 [astro-ph.CO]} \BibitemShut {NoStop}%
\bibitem [{\citenamefont {Kumar}(2021)}]{Kumar:2021eev}%
  \BibitemOpen
  \bibfield  {author} {\bibinfo {author} {\bibfnamefont {S.}~\bibnamefont {Kumar}},\ }\href {https://doi.org/10.1016/j.dark.2021.100862} {\bibfield  {journal} {\bibinfo  {journal} {Phys. Dark Univ.}\ }\textbf {\bibinfo {volume} {33}},\ \bibinfo {pages} {100862} (\bibinfo {year} {2021})},\ \Eprint {https://arxiv.org/abs/2102.12902} {arXiv:2102.12902 [astro-ph.CO]} \BibitemShut {NoStop}%
\bibitem [{\citenamefont {Pan}\ and\ \citenamefont {Yang}(2023)}]{Pan:2023mie}%
  \BibitemOpen
  \bibfield  {author} {\bibinfo {author} {\bibfnamefont {S.}~\bibnamefont {Pan}}\ and\ \bibinfo {author} {\bibfnamefont {W.}~\bibnamefont {Yang}},\ }\href {https://doi.org/10.1007/978-981-99-0177-7_29} {\bibinfo {title} {{On the interacting dark energy scenarios - the case for Hubble constant tension}}} (\bibinfo {year} {2023}),\ \Eprint {https://arxiv.org/abs/2310.07260} {arXiv:2310.07260 [astro-ph.CO]} \BibitemShut {NoStop}%
\bibitem [{\citenamefont {Benisty}\ \emph {et~al.}(2024)\citenamefont {Benisty}, \citenamefont {Pan}, \citenamefont {Staicova}, \citenamefont {Di~Valentino},\ and\ \citenamefont {Nunes}}]{Benisty:2024lmj}%
  \BibitemOpen
  \bibfield  {author} {\bibinfo {author} {\bibfnamefont {D.}~\bibnamefont {Benisty}}, \bibinfo {author} {\bibfnamefont {S.}~\bibnamefont {Pan}}, \bibinfo {author} {\bibfnamefont {D.}~\bibnamefont {Staicova}}, \bibinfo {author} {\bibfnamefont {E.}~\bibnamefont {Di~Valentino}},\ and\ \bibinfo {author} {\bibfnamefont {R.~C.}\ \bibnamefont {Nunes}},\ }\href {https://doi.org/10.1051/0004-6361/202449883} {\bibfield  {journal} {\bibinfo  {journal} {Astron. Astrophys.}\ }\textbf {\bibinfo {volume} {688}},\ \bibinfo {pages} {A156} (\bibinfo {year} {2024})},\ \Eprint {https://arxiv.org/abs/2403.00056} {arXiv:2403.00056 [astro-ph.CO]} \BibitemShut {NoStop}%
\bibitem [{\citenamefont {Yang}\ \emph {et~al.}(2020{\natexlab{a}})\citenamefont {Yang}, \citenamefont {Di~Valentino}, \citenamefont {Mena}, \citenamefont {Pan},\ and\ \citenamefont {Nunes}}]{Yang:2020uga}%
  \BibitemOpen
  \bibfield  {author} {\bibinfo {author} {\bibfnamefont {W.}~\bibnamefont {Yang}}, \bibinfo {author} {\bibfnamefont {E.}~\bibnamefont {Di~Valentino}}, \bibinfo {author} {\bibfnamefont {O.}~\bibnamefont {Mena}}, \bibinfo {author} {\bibfnamefont {S.}~\bibnamefont {Pan}},\ and\ \bibinfo {author} {\bibfnamefont {R.~C.}\ \bibnamefont {Nunes}},\ }\href {https://doi.org/10.1103/PhysRevD.101.083509} {\bibfield  {journal} {\bibinfo  {journal} {Phys. Rev. D}\ }\textbf {\bibinfo {volume} {101}},\ \bibinfo {pages} {083509} (\bibinfo {year} {2020}{\natexlab{a}})},\ \Eprint {https://arxiv.org/abs/2001.10852} {arXiv:2001.10852 [astro-ph.CO]} \BibitemShut {NoStop}%
\bibitem [{\citenamefont {Forconi}\ \emph {et~al.}(2024{\natexlab{a}})\citenamefont {Forconi}, \citenamefont {Giar\`e}, \citenamefont {Mena}, \citenamefont {Ruchika}, \citenamefont {Di~Valentino}, \citenamefont {Melchiorri},\ and\ \citenamefont {Nunes}}]{Forconi:2023hsj}%
  \BibitemOpen
  \bibfield  {author} {\bibinfo {author} {\bibfnamefont {M.}~\bibnamefont {Forconi}}, \bibinfo {author} {\bibfnamefont {W.}~\bibnamefont {Giar\`e}}, \bibinfo {author} {\bibfnamefont {O.}~\bibnamefont {Mena}}, \bibinfo {author} {\bibnamefont {Ruchika}}, \bibinfo {author} {\bibfnamefont {E.}~\bibnamefont {Di~Valentino}}, \bibinfo {author} {\bibfnamefont {A.}~\bibnamefont {Melchiorri}},\ and\ \bibinfo {author} {\bibfnamefont {R.~C.}\ \bibnamefont {Nunes}},\ }\href {https://doi.org/10.1088/1475-7516/2024/05/097} {\bibfield  {journal} {\bibinfo  {journal} {JCAP}\ }\textbf {\bibinfo {volume} {05}},\ \bibinfo {pages} {097}},\ \Eprint {https://arxiv.org/abs/2312.11074} {arXiv:2312.11074 [astro-ph.CO]} \BibitemShut {NoStop}%
\bibitem [{\citenamefont {Pourtsidou}\ and\ \citenamefont {Tram}(2016)}]{Pourtsidou:2016ico}%
  \BibitemOpen
  \bibfield  {author} {\bibinfo {author} {\bibfnamefont {A.}~\bibnamefont {Pourtsidou}}\ and\ \bibinfo {author} {\bibfnamefont {T.}~\bibnamefont {Tram}},\ }\href {https://doi.org/10.1103/PhysRevD.94.043518} {\bibfield  {journal} {\bibinfo  {journal} {Phys. Rev. D}\ }\textbf {\bibinfo {volume} {94}},\ \bibinfo {pages} {043518} (\bibinfo {year} {2016})},\ \Eprint {https://arxiv.org/abs/1604.04222} {arXiv:1604.04222 [astro-ph.CO]} \BibitemShut {NoStop}%
\bibitem [{\citenamefont {Di~Valentino}(2021)}]{DiValentino:2020vnx}%
  \BibitemOpen
  \bibfield  {author} {\bibinfo {author} {\bibfnamefont {E.}~\bibnamefont {Di~Valentino}},\ }\href {https://doi.org/10.1093/mnras/stab187} {\bibfield  {journal} {\bibinfo  {journal} {Mon. Not. Roy. Astron. Soc.}\ }\textbf {\bibinfo {volume} {502}},\ \bibinfo {pages} {2065} (\bibinfo {year} {2021})},\ \Eprint {https://arxiv.org/abs/2011.00246} {arXiv:2011.00246 [astro-ph.CO]} \BibitemShut {NoStop}%
\bibitem [{\citenamefont {Di~Valentino}\ and\ \citenamefont {Mena}(2020)}]{DiValentino:2020leo}%
  \BibitemOpen
  \bibfield  {author} {\bibinfo {author} {\bibfnamefont {E.}~\bibnamefont {Di~Valentino}}\ and\ \bibinfo {author} {\bibfnamefont {O.}~\bibnamefont {Mena}},\ }\href {https://doi.org/10.1093/mnrasl/slaa175} {\bibfield  {journal} {\bibinfo  {journal} {Mon. Not. Roy. Astron. Soc.}\ }\textbf {\bibinfo {volume} {500}},\ \bibinfo {pages} {L22} (\bibinfo {year} {2020})},\ \Eprint {https://arxiv.org/abs/2009.12620} {arXiv:2009.12620 [astro-ph.CO]} \BibitemShut {NoStop}%
\bibitem [{\citenamefont {Nunes}\ and\ \citenamefont {Di~Valentino}(2021)}]{Nunes:2021zzi}%
  \BibitemOpen
  \bibfield  {author} {\bibinfo {author} {\bibfnamefont {R.~C.}\ \bibnamefont {Nunes}}\ and\ \bibinfo {author} {\bibfnamefont {E.}~\bibnamefont {Di~Valentino}},\ }\href {https://doi.org/10.1103/PhysRevD.104.063529} {\bibfield  {journal} {\bibinfo  {journal} {Phys. Rev. D}\ }\textbf {\bibinfo {volume} {104}},\ \bibinfo {pages} {063529} (\bibinfo {year} {2021})},\ \Eprint {https://arxiv.org/abs/2107.09151} {arXiv:2107.09151 [astro-ph.CO]} \BibitemShut {NoStop}%
\bibitem [{\citenamefont {Yang}\ \emph {et~al.}(2018)\citenamefont {Yang}, \citenamefont {Mukherjee}, \citenamefont {Di~Valentino},\ and\ \citenamefont {Pan}}]{Yang:2018uae}%
  \BibitemOpen
  \bibfield  {author} {\bibinfo {author} {\bibfnamefont {W.}~\bibnamefont {Yang}}, \bibinfo {author} {\bibfnamefont {A.}~\bibnamefont {Mukherjee}}, \bibinfo {author} {\bibfnamefont {E.}~\bibnamefont {Di~Valentino}},\ and\ \bibinfo {author} {\bibfnamefont {S.}~\bibnamefont {Pan}},\ }\href {https://doi.org/10.1103/PhysRevD.98.123527} {\bibfield  {journal} {\bibinfo  {journal} {Phys. Rev. D}\ }\textbf {\bibinfo {volume} {98}},\ \bibinfo {pages} {123527} (\bibinfo {year} {2018})},\ \Eprint {https://arxiv.org/abs/1809.06883} {arXiv:1809.06883 [astro-ph.CO]} \BibitemShut {NoStop}%
\bibitem [{\citenamefont {von Marttens}\ \emph {et~al.}(2020{\natexlab{a}})\citenamefont {von Marttens}, \citenamefont {Lombriser}, \citenamefont {Kunz}, \citenamefont {Marra}, \citenamefont {Casarini},\ and\ \citenamefont {Alcaniz}}]{vonMarttens:2019ixw}%
  \BibitemOpen
  \bibfield  {author} {\bibinfo {author} {\bibfnamefont {R.}~\bibnamefont {von Marttens}}, \bibinfo {author} {\bibfnamefont {L.}~\bibnamefont {Lombriser}}, \bibinfo {author} {\bibfnamefont {M.}~\bibnamefont {Kunz}}, \bibinfo {author} {\bibfnamefont {V.}~\bibnamefont {Marra}}, \bibinfo {author} {\bibfnamefont {L.}~\bibnamefont {Casarini}},\ and\ \bibinfo {author} {\bibfnamefont {J.}~\bibnamefont {Alcaniz}},\ }\href {https://doi.org/10.1016/j.dark.2020.100490} {\bibfield  {journal} {\bibinfo  {journal} {Phys. Dark Univ.}\ }\textbf {\bibinfo {volume} {28}},\ \bibinfo {pages} {100490} (\bibinfo {year} {2020}{\natexlab{a}})},\ \Eprint {https://arxiv.org/abs/1911.02618} {arXiv:1911.02618 [astro-ph.CO]} \BibitemShut {NoStop}%
\bibitem [{\citenamefont {Lucca}\ and\ \citenamefont {Hooper}(2020{\natexlab{a}})}]{Lucca:2020zjb}%
  \BibitemOpen
  \bibfield  {author} {\bibinfo {author} {\bibfnamefont {M.}~\bibnamefont {Lucca}}\ and\ \bibinfo {author} {\bibfnamefont {D.~C.}\ \bibnamefont {Hooper}},\ }\href {https://doi.org/10.1103/PhysRevD.102.123502} {\bibfield  {journal} {\bibinfo  {journal} {Phys. Rev. D}\ }\textbf {\bibinfo {volume} {102}},\ \bibinfo {pages} {123502} (\bibinfo {year} {2020}{\natexlab{a}})},\ \Eprint {https://arxiv.org/abs/2002.06127} {arXiv:2002.06127 [astro-ph.CO]} \BibitemShut {NoStop}%
\bibitem [{\citenamefont {Zhai}\ \emph {et~al.}(2023{\natexlab{a}})\citenamefont {Zhai}, \citenamefont {Giar\`e}, \citenamefont {van~de Bruck}, \citenamefont {Di~Valentino}, \citenamefont {Mena},\ and\ \citenamefont {Nunes}}]{Zhai:2023yny}%
  \BibitemOpen
  \bibfield  {author} {\bibinfo {author} {\bibfnamefont {Y.}~\bibnamefont {Zhai}}, \bibinfo {author} {\bibfnamefont {W.}~\bibnamefont {Giar\`e}}, \bibinfo {author} {\bibfnamefont {C.}~\bibnamefont {van~de Bruck}}, \bibinfo {author} {\bibfnamefont {E.}~\bibnamefont {Di~Valentino}}, \bibinfo {author} {\bibfnamefont {O.}~\bibnamefont {Mena}},\ and\ \bibinfo {author} {\bibfnamefont {R.~C.}\ \bibnamefont {Nunes}},\ }\href {https://doi.org/10.1088/1475-7516/2023/07/032} {\bibfield  {journal} {\bibinfo  {journal} {JCAP}\ }\textbf {\bibinfo {volume} {07}},\ \bibinfo {pages} {032}},\ \Eprint {https://arxiv.org/abs/2303.08201} {arXiv:2303.08201 [astro-ph.CO]} \BibitemShut {NoStop}%
\bibitem [{\citenamefont {Bernui}\ \emph {et~al.}(2023{\natexlab{a}})\citenamefont {Bernui}, \citenamefont {Di~Valentino}, \citenamefont {Giar\`e}, \citenamefont {Kumar},\ and\ \citenamefont {Nunes}}]{Bernui:2023byc}%
  \BibitemOpen
  \bibfield  {author} {\bibinfo {author} {\bibfnamefont {A.}~\bibnamefont {Bernui}}, \bibinfo {author} {\bibfnamefont {E.}~\bibnamefont {Di~Valentino}}, \bibinfo {author} {\bibfnamefont {W.}~\bibnamefont {Giar\`e}}, \bibinfo {author} {\bibfnamefont {S.}~\bibnamefont {Kumar}},\ and\ \bibinfo {author} {\bibfnamefont {R.~C.}\ \bibnamefont {Nunes}},\ }\href {https://doi.org/10.1103/PhysRevD.107.103531} {\bibfield  {journal} {\bibinfo  {journal} {Phys. Rev. D}\ }\textbf {\bibinfo {volume} {107}},\ \bibinfo {pages} {103531} (\bibinfo {year} {2023}{\natexlab{a}})},\ \Eprint {https://arxiv.org/abs/2301.06097} {arXiv:2301.06097 [astro-ph.CO]} \BibitemShut {NoStop}%
\bibitem [{\citenamefont {Hoerning}\ \emph {et~al.}(2023)\citenamefont {Hoerning}, \citenamefont {Landim}, \citenamefont {Ponte}, \citenamefont {Rolim}, \citenamefont {Abdalla},\ and\ \citenamefont {Abdalla}}]{Hoerning:2023hks}%
  \BibitemOpen
  \bibfield  {author} {\bibinfo {author} {\bibfnamefont {G.~A.}\ \bibnamefont {Hoerning}}, \bibinfo {author} {\bibfnamefont {R.~G.}\ \bibnamefont {Landim}}, \bibinfo {author} {\bibfnamefont {L.~O.}\ \bibnamefont {Ponte}}, \bibinfo {author} {\bibfnamefont {R.~P.}\ \bibnamefont {Rolim}}, \bibinfo {author} {\bibfnamefont {F.~B.}\ \bibnamefont {Abdalla}},\ and\ \bibinfo {author} {\bibfnamefont {E.}~\bibnamefont {Abdalla}},\ }\href@noop {} {\bibinfo {title} {{Constraints on interacting dark energy revisited: implications for the Hubble tension}}} (\bibinfo {year} {2023}),\ \Eprint {https://arxiv.org/abs/2308.05807} {arXiv:2308.05807 [astro-ph.CO]} \BibitemShut {NoStop}%
\bibitem [{\citenamefont {Giar\`e}\ \emph {et~al.}(2024{\natexlab{a}})\citenamefont {Giar\`e}, \citenamefont {Zhai}, \citenamefont {Pan}, \citenamefont {Di~Valentino}, \citenamefont {Nunes},\ and\ \citenamefont {van~de Bruck}}]{Giare:2024ytc}%
  \BibitemOpen
  \bibfield  {author} {\bibinfo {author} {\bibfnamefont {W.}~\bibnamefont {Giar\`e}}, \bibinfo {author} {\bibfnamefont {Y.}~\bibnamefont {Zhai}}, \bibinfo {author} {\bibfnamefont {S.}~\bibnamefont {Pan}}, \bibinfo {author} {\bibfnamefont {E.}~\bibnamefont {Di~Valentino}}, \bibinfo {author} {\bibfnamefont {R.~C.}\ \bibnamefont {Nunes}},\ and\ \bibinfo {author} {\bibfnamefont {C.}~\bibnamefont {van~de Bruck}},\ }\href {https://doi.org/10.1103/PhysRevD.110.063527} {\bibfield  {journal} {\bibinfo  {journal} {Phys. Rev. D}\ }\textbf {\bibinfo {volume} {110}},\ \bibinfo {pages} {063527} (\bibinfo {year} {2024}{\natexlab{a}})},\ \Eprint {https://arxiv.org/abs/2404.02110} {arXiv:2404.02110 [astro-ph.CO]} \BibitemShut {NoStop}%
\bibitem [{\citenamefont {Escamilla}\ \emph {et~al.}(2023)\citenamefont {Escamilla}, \citenamefont {Akarsu}, \citenamefont {Di~Valentino},\ and\ \citenamefont {Vazquez}}]{Escamilla:2023shf}%
  \BibitemOpen
  \bibfield  {author} {\bibinfo {author} {\bibfnamefont {L.~A.}\ \bibnamefont {Escamilla}}, \bibinfo {author} {\bibfnamefont {O.}~\bibnamefont {Akarsu}}, \bibinfo {author} {\bibfnamefont {E.}~\bibnamefont {Di~Valentino}},\ and\ \bibinfo {author} {\bibfnamefont {J.~A.}\ \bibnamefont {Vazquez}},\ }\href {https://doi.org/10.1088/1475-7516/2023/11/051} {\bibfield  {journal} {\bibinfo  {journal} {JCAP}\ }\textbf {\bibinfo {volume} {11}},\ \bibinfo {pages} {051}},\ \Eprint {https://arxiv.org/abs/2305.16290} {arXiv:2305.16290 [astro-ph.CO]} \BibitemShut {NoStop}%
\bibitem [{\citenamefont {van~der Westhuizen}\ and\ \citenamefont {Abebe}(2024)}]{vanderWesthuizen:2023hcl}%
  \BibitemOpen
  \bibfield  {author} {\bibinfo {author} {\bibfnamefont {M.~A.}\ \bibnamefont {van~der Westhuizen}}\ and\ \bibinfo {author} {\bibfnamefont {A.}~\bibnamefont {Abebe}},\ }\href {https://doi.org/10.1088/1475-7516/2024/01/048} {\bibfield  {journal} {\bibinfo  {journal} {JCAP}\ }\textbf {\bibinfo {volume} {01}},\ \bibinfo {pages} {048}},\ \Eprint {https://arxiv.org/abs/2302.11949} {arXiv:2302.11949 [gr-qc]} \BibitemShut {NoStop}%
\bibitem [{\citenamefont {Silva}\ \emph {et~al.}(2024{\natexlab{a}})\citenamefont {Silva}, \citenamefont {Z\'u\~niga Bola\~no}, \citenamefont {Nunes},\ and\ \citenamefont {Di~Valentino}}]{Silva:2024ift}%
  \BibitemOpen
  \bibfield  {author} {\bibinfo {author} {\bibfnamefont {E.}~\bibnamefont {Silva}}, \bibinfo {author} {\bibfnamefont {U.}~\bibnamefont {Z\'u\~niga Bola\~no}}, \bibinfo {author} {\bibfnamefont {R.~C.}\ \bibnamefont {Nunes}},\ and\ \bibinfo {author} {\bibfnamefont {E.}~\bibnamefont {Di~Valentino}},\ }\href@noop {} {\bibinfo {title} {{Non-Linear Matter Power Spectrum Modeling in Interacting Dark Energy Cosmologies}}} (\bibinfo {year} {2024}{\natexlab{a}}),\ \Eprint {https://arxiv.org/abs/2403.19590} {arXiv:2403.19590 [astro-ph.CO]} \BibitemShut {NoStop}%
\bibitem [{\citenamefont {Di~Valentino}\ \emph {et~al.}(2020{\natexlab{a}})\citenamefont {Di~Valentino}, \citenamefont {Melchiorri}, \citenamefont {Mena},\ and\ \citenamefont {Vagnozzi}}]{DiValentino:2019ffd}%
  \BibitemOpen
  \bibfield  {author} {\bibinfo {author} {\bibfnamefont {E.}~\bibnamefont {Di~Valentino}}, \bibinfo {author} {\bibfnamefont {A.}~\bibnamefont {Melchiorri}}, \bibinfo {author} {\bibfnamefont {O.}~\bibnamefont {Mena}},\ and\ \bibinfo {author} {\bibfnamefont {S.}~\bibnamefont {Vagnozzi}},\ }\href {https://doi.org/10.1016/j.dark.2020.100666} {\bibfield  {journal} {\bibinfo  {journal} {Phys. Dark Univ.}\ }\textbf {\bibinfo {volume} {30}},\ \bibinfo {pages} {100666} (\bibinfo {year} {2020}{\natexlab{a}})},\ \Eprint {https://arxiv.org/abs/1908.04281} {arXiv:1908.04281 [astro-ph.CO]} \BibitemShut {NoStop}%
\bibitem [{\citenamefont {Li}\ \emph {et~al.}(2024{\natexlab{a}})\citenamefont {Li}, \citenamefont {Wu}, \citenamefont {Du}, \citenamefont {Jin}, \citenamefont {Li}, \citenamefont {Zhang},\ and\ \citenamefont {Zhang}}]{Li:2024qso}%
  \BibitemOpen
  \bibfield  {author} {\bibinfo {author} {\bibfnamefont {T.-N.}\ \bibnamefont {Li}}, \bibinfo {author} {\bibfnamefont {P.-J.}\ \bibnamefont {Wu}}, \bibinfo {author} {\bibfnamefont {G.-H.}\ \bibnamefont {Du}}, \bibinfo {author} {\bibfnamefont {S.-J.}\ \bibnamefont {Jin}}, \bibinfo {author} {\bibfnamefont {H.-L.}\ \bibnamefont {Li}}, \bibinfo {author} {\bibfnamefont {J.-F.}\ \bibnamefont {Zhang}},\ and\ \bibinfo {author} {\bibfnamefont {X.}~\bibnamefont {Zhang}},\ }\href@noop {} {\bibinfo {title} {{Constraints on interacting dark energy models from the DESI BAO and DES supernovae data}}} (\bibinfo {year} {2024}{\natexlab{a}}),\ \Eprint {https://arxiv.org/abs/2407.14934} {arXiv:2407.14934 [astro-ph.CO]} \BibitemShut {NoStop}%
\bibitem [{\citenamefont {Pooya}(2024{\natexlab{a}})}]{Pooya:2024wsq}%
  \BibitemOpen
  \bibfield  {author} {\bibinfo {author} {\bibfnamefont {N.~N.}\ \bibnamefont {Pooya}},\ }\href {https://doi.org/10.1103/PhysRevD.110.043510} {\bibfield  {journal} {\bibinfo  {journal} {Phys. Rev. D}\ }\textbf {\bibinfo {volume} {110}},\ \bibinfo {pages} {043510} (\bibinfo {year} {2024}{\natexlab{a}})},\ \Eprint {https://arxiv.org/abs/2407.03766} {arXiv:2407.03766 [astro-ph.CO]} \BibitemShut {NoStop}%
\bibitem [{\citenamefont {Halder}\ \emph {et~al.}(2024{\natexlab{a}})\citenamefont {Halder}, \citenamefont {de~Haro}, \citenamefont {Saha},\ and\ \citenamefont {Pan}}]{Halder:2024uao}%
  \BibitemOpen
  \bibfield  {author} {\bibinfo {author} {\bibfnamefont {S.}~\bibnamefont {Halder}}, \bibinfo {author} {\bibfnamefont {J.}~\bibnamefont {de~Haro}}, \bibinfo {author} {\bibfnamefont {T.}~\bibnamefont {Saha}},\ and\ \bibinfo {author} {\bibfnamefont {S.}~\bibnamefont {Pan}},\ }\href {https://doi.org/10.1103/PhysRevD.109.083522} {\bibfield  {journal} {\bibinfo  {journal} {Phys. Rev. D}\ }\textbf {\bibinfo {volume} {109}},\ \bibinfo {pages} {083522} (\bibinfo {year} {2024}{\natexlab{a}})},\ \Eprint {https://arxiv.org/abs/2403.01397} {arXiv:2403.01397 [gr-qc]} \BibitemShut {NoStop}%
\bibitem [{\citenamefont {Castello}\ \emph {et~al.}(2024)\citenamefont {Castello}, \citenamefont {Mancarella}, \citenamefont {Grimm}, \citenamefont {Sobral-Blanco}, \citenamefont {Tutusaus},\ and\ \citenamefont {Bonvin}}]{Castello:2023zjr}%
  \BibitemOpen
  \bibfield  {author} {\bibinfo {author} {\bibfnamefont {S.}~\bibnamefont {Castello}}, \bibinfo {author} {\bibfnamefont {M.}~\bibnamefont {Mancarella}}, \bibinfo {author} {\bibfnamefont {N.}~\bibnamefont {Grimm}}, \bibinfo {author} {\bibfnamefont {D.}~\bibnamefont {Sobral-Blanco}}, \bibinfo {author} {\bibfnamefont {I.}~\bibnamefont {Tutusaus}},\ and\ \bibinfo {author} {\bibfnamefont {C.}~\bibnamefont {Bonvin}},\ }\href {https://doi.org/10.1088/1475-7516/2024/05/003} {\bibfield  {journal} {\bibinfo  {journal} {JCAP}\ }\textbf {\bibinfo {volume} {05}},\ \bibinfo {pages} {003}},\ \Eprint {https://arxiv.org/abs/2311.14425} {arXiv:2311.14425 [astro-ph.CO]} \BibitemShut {NoStop}%
\bibitem [{\citenamefont {Yao}\ and\ \citenamefont {Meng}(2023)}]{Yao:2023jau}%
  \BibitemOpen
  \bibfield  {author} {\bibinfo {author} {\bibfnamefont {Y.-H.}\ \bibnamefont {Yao}}\ and\ \bibinfo {author} {\bibfnamefont {X.-H.}\ \bibnamefont {Meng}},\ }\href {https://doi.org/10.1016/j.dark.2022.101165} {\bibfield  {journal} {\bibinfo  {journal} {Phys. Dark Univ.}\ }\textbf {\bibinfo {volume} {39}},\ \bibinfo {pages} {101165} (\bibinfo {year} {2023})}\BibitemShut {NoStop}%
\bibitem [{\citenamefont {Mishra}\ \emph {et~al.}(2023)\citenamefont {Mishra}, \citenamefont {Pacif}, \citenamefont {Kumar},\ and\ \citenamefont {Bamba}}]{Mishra:2023ueo}%
  \BibitemOpen
  \bibfield  {author} {\bibinfo {author} {\bibfnamefont {K.~R.}\ \bibnamefont {Mishra}}, \bibinfo {author} {\bibfnamefont {S.~K.~J.}\ \bibnamefont {Pacif}}, \bibinfo {author} {\bibfnamefont {R.}~\bibnamefont {Kumar}},\ and\ \bibinfo {author} {\bibfnamefont {K.}~\bibnamefont {Bamba}},\ }\href {https://doi.org/10.1016/j.dark.2023.101211} {\bibfield  {journal} {\bibinfo  {journal} {Phys. Dark Univ.}\ }\textbf {\bibinfo {volume} {40}},\ \bibinfo {pages} {101211} (\bibinfo {year} {2023})},\ \Eprint {https://arxiv.org/abs/2301.08743} {arXiv:2301.08743 [gr-qc]} \BibitemShut {NoStop}%
\bibitem [{\citenamefont {Nunes}\ \emph {et~al.}(2016)\citenamefont {Nunes}, \citenamefont {Pan},\ and\ \citenamefont {Saridakis}}]{Nunes:2016dlj}%
  \BibitemOpen
  \bibfield  {author} {\bibinfo {author} {\bibfnamefont {R.~C.}\ \bibnamefont {Nunes}}, \bibinfo {author} {\bibfnamefont {S.}~\bibnamefont {Pan}},\ and\ \bibinfo {author} {\bibfnamefont {E.~N.}\ \bibnamefont {Saridakis}},\ }\href {https://doi.org/10.1103/PhysRevD.94.023508} {\bibfield  {journal} {\bibinfo  {journal} {Phys. Rev. D}\ }\textbf {\bibinfo {volume} {94}},\ \bibinfo {pages} {023508} (\bibinfo {year} {2016})},\ \Eprint {https://arxiv.org/abs/1605.01712} {arXiv:1605.01712 [astro-ph.CO]} \BibitemShut {NoStop}%
\bibitem [{\citenamefont {Silva}\ \emph {et~al.}(2025{\natexlab{a}})\citenamefont {Silva}, \citenamefont {Sabogal}, \citenamefont {Souza}, \citenamefont {Nunes}, \citenamefont {Di~Valentino},\ and\ \citenamefont {Kumar}}]{Silva:2025hxw}%
  \BibitemOpen
  \bibfield  {author} {\bibinfo {author} {\bibfnamefont {E.}~\bibnamefont {Silva}}, \bibinfo {author} {\bibfnamefont {M.~A.}\ \bibnamefont {Sabogal}}, \bibinfo {author} {\bibfnamefont {M.~S.}\ \bibnamefont {Souza}}, \bibinfo {author} {\bibfnamefont {R.~C.}\ \bibnamefont {Nunes}}, \bibinfo {author} {\bibfnamefont {E.}~\bibnamefont {Di~Valentino}},\ and\ \bibinfo {author} {\bibfnamefont {S.}~\bibnamefont {Kumar}},\ }\href@noop {} {\bibinfo {title} {{New Constraints on Interacting Dark Energy from DESI DR2 BAO Observations}}} (\bibinfo {year} {2025}{\natexlab{a}}),\ \Eprint {https://arxiv.org/abs/2503.23225} {arXiv:2503.23225 [astro-ph.CO]} \BibitemShut {NoStop}%
\bibitem [{\citenamefont {Zheng}\ \emph {et~al.}(2017)\citenamefont {Zheng}, \citenamefont {Biesiada}, \citenamefont {Cao}, \citenamefont {Qi},\ and\ \citenamefont {Zhu}}]{Zheng_2017}%
  \BibitemOpen
  \bibfield  {author} {\bibinfo {author} {\bibfnamefont {X.}~\bibnamefont {Zheng}}, \bibinfo {author} {\bibfnamefont {M.}~\bibnamefont {Biesiada}}, \bibinfo {author} {\bibfnamefont {S.}~\bibnamefont {Cao}}, \bibinfo {author} {\bibfnamefont {J.}~\bibnamefont {Qi}},\ and\ \bibinfo {author} {\bibfnamefont {Z.-H.}\ \bibnamefont {Zhu}},\ }\href {https://doi.org/10.1088/1475-7516/2017/10/030} {\bibfield  {journal} {\bibinfo  {journal} {JCAP}\ }\textbf {\bibinfo {volume} {10}},\ \bibinfo {pages} {030}},\ \Eprint {https://arxiv.org/abs/1705.06204} {arXiv:1705.06204 [astro-ph.CO]} \BibitemShut {NoStop}%
\bibitem [{\citenamefont {Kumar}\ \emph {et~al.}(2019)\citenamefont {Kumar}, \citenamefont {Nunes},\ and\ \citenamefont {Yadav}}]{Kumar_2019}%
  \BibitemOpen
  \bibfield  {author} {\bibinfo {author} {\bibfnamefont {S.}~\bibnamefont {Kumar}}, \bibinfo {author} {\bibfnamefont {R.~C.}\ \bibnamefont {Nunes}},\ and\ \bibinfo {author} {\bibfnamefont {S.~K.}\ \bibnamefont {Yadav}},\ }\href {https://doi.org/10.1140/epjc/s10052-019-7087-7} {\bibfield  {journal} {\bibinfo  {journal} {Eur. Phys. J. C}\ }\textbf {\bibinfo {volume} {79}},\ \bibinfo {pages} {576} (\bibinfo {year} {2019})},\ \Eprint {https://arxiv.org/abs/1903.04865} {arXiv:1903.04865 [astro-ph.CO]} \BibitemShut {NoStop}%
\bibitem [{\citenamefont {Anchordoqui}\ \emph {et~al.}(2021)\citenamefont {Anchordoqui}, \citenamefont {Di~Valentino}, \citenamefont {Pan},\ and\ \citenamefont {Yang}}]{Anchordoqui_2021}%
  \BibitemOpen
  \bibfield  {author} {\bibinfo {author} {\bibfnamefont {L.~A.}\ \bibnamefont {Anchordoqui}}, \bibinfo {author} {\bibfnamefont {E.}~\bibnamefont {Di~Valentino}}, \bibinfo {author} {\bibfnamefont {S.}~\bibnamefont {Pan}},\ and\ \bibinfo {author} {\bibfnamefont {W.}~\bibnamefont {Yang}},\ }\href {https://doi.org/10.1016/j.jheap.2021.08.001} {\bibfield  {journal} {\bibinfo  {journal} {JHEAp}\ }\textbf {\bibinfo {volume} {32}},\ \bibinfo {pages} {28} (\bibinfo {year} {2021})},\ \Eprint {https://arxiv.org/abs/2107.13932} {arXiv:2107.13932 [astro-ph.CO]} \BibitemShut {NoStop}%
\bibitem [{\citenamefont {Pan}\ \emph {et~al.}(2019)\citenamefont {Pan}, \citenamefont {Yang}, \citenamefont {Singha},\ and\ \citenamefont {Saridakis}}]{Pan_2019}%
  \BibitemOpen
  \bibfield  {author} {\bibinfo {author} {\bibfnamefont {S.}~\bibnamefont {Pan}}, \bibinfo {author} {\bibfnamefont {W.}~\bibnamefont {Yang}}, \bibinfo {author} {\bibfnamefont {C.}~\bibnamefont {Singha}},\ and\ \bibinfo {author} {\bibfnamefont {E.~N.}\ \bibnamefont {Saridakis}},\ }\href {https://doi.org/10.1103/PhysRevD.100.083539} {\bibfield  {journal} {\bibinfo  {journal} {Phys. Rev. D}\ }\textbf {\bibinfo {volume} {100}},\ \bibinfo {pages} {083539} (\bibinfo {year} {2019})},\ \Eprint {https://arxiv.org/abs/1903.10969} {arXiv:1903.10969 [astro-ph.CO]} \BibitemShut {NoStop}%
\bibitem [{\citenamefont {Guo}\ \emph {et~al.}(2021)\citenamefont {Guo}, \citenamefont {Feng}, \citenamefont {Yao},\ and\ \citenamefont {Chen}}]{Guo_2021}%
  \BibitemOpen
  \bibfield  {author} {\bibinfo {author} {\bibfnamefont {R.-Y.}\ \bibnamefont {Guo}}, \bibinfo {author} {\bibfnamefont {L.}~\bibnamefont {Feng}}, \bibinfo {author} {\bibfnamefont {T.-Y.}\ \bibnamefont {Yao}},\ and\ \bibinfo {author} {\bibfnamefont {X.-Y.}\ \bibnamefont {Chen}},\ }\href {https://doi.org/10.1088/1475-7516/2021/12/036} {\bibfield  {journal} {\bibinfo  {journal} {JCAP}\ }\textbf {\bibinfo {volume} {12}}\bibfield  {number} {\bibinfo  {number} { (12)},\ \bibinfo {pages} {036}},\ }\Eprint {https://arxiv.org/abs/2110.02536} {arXiv:2110.02536 [gr-qc]} \BibitemShut {NoStop}%
\bibitem [{\citenamefont {Gao}\ \emph {et~al.}(2021)\citenamefont {Gao}, \citenamefont {Zhao}, \citenamefont {Xue},\ and\ \citenamefont {Zhang}}]{Gao:2021xnk}%
  \BibitemOpen
  \bibfield  {author} {\bibinfo {author} {\bibfnamefont {L.-Y.}\ \bibnamefont {Gao}}, \bibinfo {author} {\bibfnamefont {Z.-W.}\ \bibnamefont {Zhao}}, \bibinfo {author} {\bibfnamefont {S.-S.}\ \bibnamefont {Xue}},\ and\ \bibinfo {author} {\bibfnamefont {X.}~\bibnamefont {Zhang}},\ }\href {https://doi.org/10.1088/1475-7516/2021/07/005} {\bibfield  {journal} {\bibinfo  {journal} {JCAP}\ }\textbf {\bibinfo {volume} {07}},\ \bibinfo {pages} {005}},\ \Eprint {https://arxiv.org/abs/2101.10714} {arXiv:2101.10714 [astro-ph.CO]} \BibitemShut {NoStop}%
\bibitem [{\citenamefont {Di~Valentino}\ \emph {et~al.}(2021{\natexlab{a}})\citenamefont {Di~Valentino}, \citenamefont {Mena}, \citenamefont {Pan}, \citenamefont {Visinelli}, \citenamefont {Yang}, \citenamefont {Melchiorri}, \citenamefont {Mota}, \citenamefont {Riess},\ and\ \citenamefont {Silk}}]{Di_Valentino_2021_H0_review}%
  \BibitemOpen
  \bibfield  {author} {\bibinfo {author} {\bibfnamefont {E.}~\bibnamefont {Di~Valentino}}, \bibinfo {author} {\bibfnamefont {O.}~\bibnamefont {Mena}}, \bibinfo {author} {\bibfnamefont {S.}~\bibnamefont {Pan}}, \bibinfo {author} {\bibfnamefont {L.}~\bibnamefont {Visinelli}}, \bibinfo {author} {\bibfnamefont {W.}~\bibnamefont {Yang}}, \bibinfo {author} {\bibfnamefont {A.}~\bibnamefont {Melchiorri}}, \bibinfo {author} {\bibfnamefont {D.~F.}\ \bibnamefont {Mota}}, \bibinfo {author} {\bibfnamefont {A.~G.}\ \bibnamefont {Riess}},\ and\ \bibinfo {author} {\bibfnamefont {J.}~\bibnamefont {Silk}},\ }\href {https://doi.org/10.1088/1361-6382/ac086d} {\bibfield  {journal} {\bibinfo  {journal} {Class. Quant. Grav.}\ }\textbf {\bibinfo {volume} {38}},\ \bibinfo {pages} {153001} (\bibinfo {year} {2021}{\natexlab{a}})},\ \Eprint {https://arxiv.org/abs/2103.01183} {arXiv:2103.01183 [astro-ph.CO]} \BibitemShut {NoStop}%
\bibitem [{\citenamefont {Gariazzo}\ \emph {et~al.}(2022)\citenamefont {Gariazzo}, \citenamefont {Di~Valentino}, \citenamefont {Mena},\ and\ \citenamefont {Nunes}}]{Gariazzo_2022}%
  \BibitemOpen
  \bibfield  {author} {\bibinfo {author} {\bibfnamefont {S.}~\bibnamefont {Gariazzo}}, \bibinfo {author} {\bibfnamefont {E.}~\bibnamefont {Di~Valentino}}, \bibinfo {author} {\bibfnamefont {O.}~\bibnamefont {Mena}},\ and\ \bibinfo {author} {\bibfnamefont {R.~C.}\ \bibnamefont {Nunes}},\ }\href {https://doi.org/10.1103/PhysRevD.106.023530} {\bibfield  {journal} {\bibinfo  {journal} {Phys. Rev. D}\ }\textbf {\bibinfo {volume} {106}},\ \bibinfo {pages} {023530} (\bibinfo {year} {2022})},\ \Eprint {https://arxiv.org/abs/2111.03152} {arXiv:2111.03152 [astro-ph.CO]} \BibitemShut {NoStop}%
\bibitem [{\citenamefont {Wang}\ \emph {et~al.}(2022)\citenamefont {Wang}, \citenamefont {Zhang}, \citenamefont {He}, \citenamefont {Zhang},\ and\ \citenamefont {Zhang}}]{Wang_2022}%
  \BibitemOpen
  \bibfield  {author} {\bibinfo {author} {\bibfnamefont {L.-F.}\ \bibnamefont {Wang}}, \bibinfo {author} {\bibfnamefont {J.-H.}\ \bibnamefont {Zhang}}, \bibinfo {author} {\bibfnamefont {D.-Z.}\ \bibnamefont {He}}, \bibinfo {author} {\bibfnamefont {J.-F.}\ \bibnamefont {Zhang}},\ and\ \bibinfo {author} {\bibfnamefont {X.}~\bibnamefont {Zhang}},\ }\href {https://doi.org/10.1093/mnras/stac1468} {\bibfield  {journal} {\bibinfo  {journal} {Mon. Not. Roy. Astron. Soc.}\ }\textbf {\bibinfo {volume} {514}},\ \bibinfo {pages} {1433} (\bibinfo {year} {2022})},\ \Eprint {https://arxiv.org/abs/2102.09331} {arXiv:2102.09331 [astro-ph.CO]} \BibitemShut {NoStop}%
\bibitem [{\citenamefont {Califano}\ \emph {et~al.}(2023)\citenamefont {Califano}, \citenamefont {de~Martino}, \citenamefont {Vernieri},\ and\ \citenamefont {Capozziello}}]{Califano_2023}%
  \BibitemOpen
  \bibfield  {author} {\bibinfo {author} {\bibfnamefont {M.}~\bibnamefont {Califano}}, \bibinfo {author} {\bibfnamefont {I.}~\bibnamefont {de~Martino}}, \bibinfo {author} {\bibfnamefont {D.}~\bibnamefont {Vernieri}},\ and\ \bibinfo {author} {\bibfnamefont {S.}~\bibnamefont {Capozziello}},\ }\href {https://doi.org/10.1103/PhysRevD.107.123519} {\bibfield  {journal} {\bibinfo  {journal} {Phys. Rev. D}\ }\textbf {\bibinfo {volume} {107}},\ \bibinfo {pages} {123519} (\bibinfo {year} {2023})},\ \Eprint {https://arxiv.org/abs/2208.13999} {arXiv:2208.13999 [astro-ph.CO]} \BibitemShut {NoStop}%
\bibitem [{\citenamefont {Pan}\ and\ \citenamefont {Yang}(2024)}]{Pan_2024}%
  \BibitemOpen
  \bibfield  {author} {\bibinfo {author} {\bibfnamefont {S.}~\bibnamefont {Pan}}\ and\ \bibinfo {author} {\bibfnamefont {W.}~\bibnamefont {Yang}},\ }\bibinfo {title} {On the interacting dark energy scenarios—the case for hubble constant tension},\ in\ \href {https://doi.org/10.1007/978-981-99-0177-7_29} {\emph {\bibinfo {booktitle} {The Hubble Constant Tension}}}\ (\bibinfo  {publisher} {Springer Nature Singapore},\ \bibinfo {year} {2024})\ p.\ \bibinfo {pages} {531–551}\BibitemShut {NoStop}%
\bibitem [{\citenamefont {Liu}\ \emph {et~al.}(2023)\citenamefont {Liu}, \citenamefont {Zhou}, \citenamefont {Mu},\ and\ \citenamefont {Xu}}]{Liu_2023}%
  \BibitemOpen
  \bibfield  {author} {\bibinfo {author} {\bibfnamefont {G.}~\bibnamefont {Liu}}, \bibinfo {author} {\bibfnamefont {Z.}~\bibnamefont {Zhou}}, \bibinfo {author} {\bibfnamefont {Y.}~\bibnamefont {Mu}},\ and\ \bibinfo {author} {\bibfnamefont {L.}~\bibnamefont {Xu}},\ }\href {https://doi.org/10.1103/PhysRevD.108.083523} {\bibfield  {journal} {\bibinfo  {journal} {Phys. Rev. D}\ }\textbf {\bibinfo {volume} {108}},\ \bibinfo {pages} {083523} (\bibinfo {year} {2023})},\ \Eprint {https://arxiv.org/abs/2307.07228} {arXiv:2307.07228 [astro-ph.CO]} \BibitemShut {NoStop}%
\bibitem [{\citenamefont {Liu}\ \emph {et~al.}(2024)\citenamefont {Liu}, \citenamefont {Zhou}, \citenamefont {Mu},\ and\ \citenamefont {Xu}}]{Liu_2024}%
  \BibitemOpen
  \bibfield  {author} {\bibinfo {author} {\bibfnamefont {G.}~\bibnamefont {Liu}}, \bibinfo {author} {\bibfnamefont {Z.}~\bibnamefont {Zhou}}, \bibinfo {author} {\bibfnamefont {Y.}~\bibnamefont {Mu}},\ and\ \bibinfo {author} {\bibfnamefont {L.}~\bibnamefont {Xu}},\ }\href {https://doi.org/10.1093/mnras/stae661} {\bibfield  {journal} {\bibinfo  {journal} {Mon. Not. Roy. Astron. Soc.}\ }\textbf {\bibinfo {volume} {529}},\ \bibinfo {pages} {1852} (\bibinfo {year} {2024})},\ \Eprint {https://arxiv.org/abs/2308.07069} {arXiv:2308.07069 [astro-ph.CO]} \BibitemShut {NoStop}%
\bibitem [{\citenamefont {Sabogal}\ \emph {et~al.}(2025)\citenamefont {Sabogal}, \citenamefont {Silva}, \citenamefont {Nunes}, \citenamefont {Kumar},\ and\ \citenamefont {Di~Valentino}}]{Sabogal_2025}%
  \BibitemOpen
  \bibfield  {author} {\bibinfo {author} {\bibfnamefont {M.~A.}\ \bibnamefont {Sabogal}}, \bibinfo {author} {\bibfnamefont {E.}~\bibnamefont {Silva}}, \bibinfo {author} {\bibfnamefont {R.~C.}\ \bibnamefont {Nunes}}, \bibinfo {author} {\bibfnamefont {S.}~\bibnamefont {Kumar}},\ and\ \bibinfo {author} {\bibfnamefont {E.}~\bibnamefont {Di~Valentino}},\ }\href {https://doi.org/10.1103/PhysRevD.111.043531} {\bibfield  {journal} {\bibinfo  {journal} {Phys. Rev. D}\ }\textbf {\bibinfo {volume} {111}},\ \bibinfo {pages} {043531} (\bibinfo {year} {2025})},\ \Eprint {https://arxiv.org/abs/2501.10323} {arXiv:2501.10323 [astro-ph.CO]} \BibitemShut {NoStop}%
\bibitem [{\citenamefont {Yang}\ \emph {et~al.}(2025{\natexlab{a}})\citenamefont {Yang}, \citenamefont {Zhang}, \citenamefont {Mena}, \citenamefont {Pan},\ and\ \citenamefont {Di~Valentino}}]{Yang:2025uyv}%
  \BibitemOpen
  \bibfield  {author} {\bibinfo {author} {\bibfnamefont {W.}~\bibnamefont {Yang}}, \bibinfo {author} {\bibfnamefont {S.}~\bibnamefont {Zhang}}, \bibinfo {author} {\bibfnamefont {O.}~\bibnamefont {Mena}}, \bibinfo {author} {\bibfnamefont {S.}~\bibnamefont {Pan}},\ and\ \bibinfo {author} {\bibfnamefont {E.}~\bibnamefont {Di~Valentino}},\ }\href@noop {} {\bibinfo {title} {{Dark Energy Is Not That Into You: Variable Couplings after DESI DR2 BAO}}} (\bibinfo {year} {2025}{\natexlab{a}}),\ \Eprint {https://arxiv.org/abs/2508.19109} {arXiv:2508.19109 [astro-ph.CO]} \BibitemShut {NoStop}%
\bibitem [{\citenamefont {Vagnozzi}(2023)}]{Vagnozzi:2023nrq}%
  \BibitemOpen
  \bibfield  {author} {\bibinfo {author} {\bibfnamefont {S.}~\bibnamefont {Vagnozzi}},\ }\href {https://doi.org/10.3390/universe9090393} {\bibfield  {journal} {\bibinfo  {journal} {Universe}\ }\textbf {\bibinfo {volume} {9}},\ \bibinfo {pages} {393} (\bibinfo {year} {2023})},\ \Eprint {https://arxiv.org/abs/2308.16628} {arXiv:2308.16628 [astro-ph.CO]} \BibitemShut {NoStop}%
\bibitem [{\citenamefont {Di~Valentino}\ \emph {et~al.}(2020{\natexlab{b}})\citenamefont {Di~Valentino}, \citenamefont {Melchiorri}, \citenamefont {Mena},\ and\ \citenamefont {Vagnozzi}}]{Di_Valentino_2020_rhode}%
  \BibitemOpen
  \bibfield  {author} {\bibinfo {author} {\bibfnamefont {E.}~\bibnamefont {Di~Valentino}}, \bibinfo {author} {\bibfnamefont {A.}~\bibnamefont {Melchiorri}}, \bibinfo {author} {\bibfnamefont {O.}~\bibnamefont {Mena}},\ and\ \bibinfo {author} {\bibfnamefont {S.}~\bibnamefont {Vagnozzi}},\ }\href {https://doi.org/10.1016/j.dark.2020.100666} {\bibfield  {journal} {\bibinfo  {journal} {Phys. Dark Univ.}\ }\textbf {\bibinfo {volume} {30}},\ \bibinfo {pages} {100666} (\bibinfo {year} {2020}{\natexlab{b}})},\ \Eprint {https://arxiv.org/abs/1908.04281} {arXiv:1908.04281 [astro-ph.CO]} \BibitemShut {NoStop}%
\bibitem [{\citenamefont {Lucca}(2021)}]{lucca2021darkenergydarkmatterinteractions}%
  \BibitemOpen
  \bibfield  {author} {\bibinfo {author} {\bibfnamefont {M.}~\bibnamefont {Lucca}},\ }\href {https://doi.org/10.1016/j.dark.2021.100899} {\bibfield  {journal} {\bibinfo  {journal} {Phys. Dark Univ.}\ }\textbf {\bibinfo {volume} {34}},\ \bibinfo {pages} {100899} (\bibinfo {year} {2021})},\ \Eprint {https://arxiv.org/abs/2105.09249} {arXiv:2105.09249 [astro-ph.CO]} \BibitemShut {NoStop}%
\bibitem [{\citenamefont {Sabogal}\ \emph {et~al.}(2024)\citenamefont {Sabogal}, \citenamefont {Silva}, \citenamefont {Nunes}, \citenamefont {Kumar}, \citenamefont {Di~Valentino},\ and\ \citenamefont {Giarè}}]{Sabogal_2024}%
  \BibitemOpen
  \bibfield  {author} {\bibinfo {author} {\bibfnamefont {M.~A.}\ \bibnamefont {Sabogal}}, \bibinfo {author} {\bibfnamefont {E.}~\bibnamefont {Silva}}, \bibinfo {author} {\bibfnamefont {R.~C.}\ \bibnamefont {Nunes}}, \bibinfo {author} {\bibfnamefont {S.}~\bibnamefont {Kumar}}, \bibinfo {author} {\bibfnamefont {E.}~\bibnamefont {Di~Valentino}},\ and\ \bibinfo {author} {\bibfnamefont {W.}~\bibnamefont {Giarè}},\ }\bibfield  {journal} {\bibinfo  {journal} {Physical Review D}\ }\textbf {\bibinfo {volume} {110}},\ \href {https://doi.org/10.1103/physrevd.110.123508} {10.1103/physrevd.110.123508} (\bibinfo {year} {2024})\BibitemShut {NoStop}%
\bibitem [{\citenamefont {Yang}\ \emph {et~al.}(2025{\natexlab{b}})\citenamefont {Yang}, \citenamefont {Pan}, \citenamefont {Valentino}, \citenamefont {Mena}, \citenamefont {Mota},\ and\ \citenamefont {Chakraborty}}]{yang2025probingcoldnaturedark}%
  \BibitemOpen
  \bibfield  {author} {\bibinfo {author} {\bibfnamefont {W.}~\bibnamefont {Yang}}, \bibinfo {author} {\bibfnamefont {S.}~\bibnamefont {Pan}}, \bibinfo {author} {\bibfnamefont {E.~D.}\ \bibnamefont {Valentino}}, \bibinfo {author} {\bibfnamefont {O.}~\bibnamefont {Mena}}, \bibinfo {author} {\bibfnamefont {D.~F.}\ \bibnamefont {Mota}},\ and\ \bibinfo {author} {\bibfnamefont {S.}~\bibnamefont {Chakraborty}},\ }\href {https://arxiv.org/abs/2504.11973} {\bibinfo {title} {Probing the cold nature of dark matter}} (\bibinfo {year} {2025}{\natexlab{b}}),\ \Eprint {https://arxiv.org/abs/2504.11973} {arXiv:2504.11973 [astro-ph.CO]} \BibitemShut {NoStop}%
\bibitem [{\citenamefont {Liu}\ \emph {et~al.}(2025)\citenamefont {Liu}, \citenamefont {Wu}, \citenamefont {Pan},\ and\ \citenamefont {Yang}}]{Liu_2025}%
  \BibitemOpen
  \bibfield  {author} {\bibinfo {author} {\bibfnamefont {W.}~\bibnamefont {Liu}}, \bibinfo {author} {\bibfnamefont {Y.}~\bibnamefont {Wu}}, \bibinfo {author} {\bibfnamefont {S.}~\bibnamefont {Pan}},\ and\ \bibinfo {author} {\bibfnamefont {W.}~\bibnamefont {Yang}},\ }\href {https://doi.org/10.1016/j.jheap.2025.100403} {\bibfield  {journal} {\bibinfo  {journal} {Journal of High Energy Astrophysics}\ }\textbf {\bibinfo {volume} {47}},\ \bibinfo {pages} {100403} (\bibinfo {year} {2025})}\BibitemShut {NoStop}%
\bibitem [{\citenamefont {Abdul~Karim}\ \emph {et~al.}(2025)\citenamefont {Abdul~Karim} \emph {et~al.}}]{DESI:2025zgx}%
  \BibitemOpen
  \bibfield  {author} {\bibinfo {author} {\bibfnamefont {M.}~\bibnamefont {Abdul~Karim}} \emph {et~al.} (\bibinfo {collaboration} {DESI}),\ }\href@noop {} {\bibinfo {title} {{DESI DR2 Results II: Measurements of Baryon Acoustic Oscillations and Cosmological Constraints}}} (\bibinfo {year} {2025}),\ \Eprint {https://arxiv.org/abs/2503.14738} {arXiv:2503.14738 [astro-ph.CO]} \BibitemShut {NoStop}%
\bibitem [{\citenamefont {Lodha}\ \emph {et~al.}(2025{\natexlab{a}})\citenamefont {Lodha} \emph {et~al.}}]{DESI:2025fii}%
  \BibitemOpen
  \bibfield  {author} {\bibinfo {author} {\bibfnamefont {K.}~\bibnamefont {Lodha}} \emph {et~al.} (\bibinfo {collaboration} {DESI}),\ }\href@noop {} {\bibinfo {title} {{Extended Dark Energy analysis using DESI DR2 BAO measurements}}} (\bibinfo {year} {2025}{\natexlab{a}}),\ \Eprint {https://arxiv.org/abs/2503.14743} {arXiv:2503.14743 [astro-ph.CO]} \BibitemShut {NoStop}%
\bibitem [{\citenamefont {Andrade}\ \emph {et~al.}(2025)\citenamefont {Andrade} \emph {et~al.}}]{DESI:2025qqy}%
  \BibitemOpen
  \bibfield  {author} {\bibinfo {author} {\bibfnamefont {U.}~\bibnamefont {Andrade}} \emph {et~al.} (\bibinfo {collaboration} {DESI}),\ }\href@noop {} {\bibinfo {title} {{Validation of the DESI DR2 Measurements of Baryon Acoustic Oscillations from Galaxies and Quasars}}} (\bibinfo {year} {2025}),\ \Eprint {https://arxiv.org/abs/2503.14742} {arXiv:2503.14742 [astro-ph.CO]} \BibitemShut {NoStop}%
\bibitem [{\citenamefont {Gu}\ \emph {et~al.}(2025)\citenamefont {Gu} \emph {et~al.}}]{DESI:2025wyn}%
  \BibitemOpen
  \bibfield  {author} {\bibinfo {author} {\bibfnamefont {G.}~\bibnamefont {Gu}} \emph {et~al.} (\bibinfo {collaboration} {DESI}),\ }\href@noop {} {\bibinfo {title} {{Dynamical Dark Energy in light of the DESI DR2 Baryonic Acoustic Oscillations Measurements}}} (\bibinfo {year} {2025}),\ \Eprint {https://arxiv.org/abs/2504.06118} {arXiv:2504.06118 [astro-ph.CO]} \BibitemShut {NoStop}%
\bibitem [{\citenamefont {Adame}\ \emph {et~al.}(2024)\citenamefont {Adame} \emph {et~al.}}]{DESI:2024mwx}%
  \BibitemOpen
  \bibfield  {author} {\bibinfo {author} {\bibfnamefont {A.~G.}\ \bibnamefont {Adame}} \emph {et~al.} (\bibinfo {collaboration} {DESI}),\ }\href@noop {} {\bibinfo {title} {{DESI 2024 VI: Cosmological Constraints from the Measurements of Baryon Acoustic Oscillations}}} (\bibinfo {year} {2024}),\ \Eprint {https://arxiv.org/abs/2404.03002} {arXiv:2404.03002 [astro-ph.CO]} \BibitemShut {NoStop}%
\bibitem [{\citenamefont {Cort\^es}\ and\ \citenamefont {Liddle}(2024)}]{Cortes:2024lgw}%
  \BibitemOpen
  \bibfield  {author} {\bibinfo {author} {\bibfnamefont {M.}~\bibnamefont {Cort\^es}}\ and\ \bibinfo {author} {\bibfnamefont {A.~R.}\ \bibnamefont {Liddle}},\ }\href {https://doi.org/10.1088/1475-7516/2024/12/007} {\bibfield  {journal} {\bibinfo  {journal} {JCAP}\ }\textbf {\bibinfo {volume} {12}},\ \bibinfo {pages} {007}},\ \Eprint {https://arxiv.org/abs/2404.08056} {arXiv:2404.08056 [astro-ph.CO]} \BibitemShut {NoStop}%
\bibitem [{\citenamefont {Shlivko}\ and\ \citenamefont {Steinhardt}(2024)}]{Shlivko:2024llw}%
  \BibitemOpen
  \bibfield  {author} {\bibinfo {author} {\bibfnamefont {D.}~\bibnamefont {Shlivko}}\ and\ \bibinfo {author} {\bibfnamefont {P.~J.}\ \bibnamefont {Steinhardt}},\ }\href {https://doi.org/10.1016/j.physletb.2024.138826} {\bibfield  {journal} {\bibinfo  {journal} {Phys. Lett. B}\ }\textbf {\bibinfo {volume} {855}},\ \bibinfo {pages} {138826} (\bibinfo {year} {2024})},\ \Eprint {https://arxiv.org/abs/2405.03933} {arXiv:2405.03933 [astro-ph.CO]} \BibitemShut {NoStop}%
\bibitem [{\citenamefont {Luongo}\ and\ \citenamefont {Muccino}(2024)}]{Luongo:2024fww}%
  \BibitemOpen
  \bibfield  {author} {\bibinfo {author} {\bibfnamefont {O.}~\bibnamefont {Luongo}}\ and\ \bibinfo {author} {\bibfnamefont {M.}~\bibnamefont {Muccino}},\ }\href {https://doi.org/10.1051/0004-6361/202450512} {\bibfield  {journal} {\bibinfo  {journal} {Astron. Astrophys.}\ }\textbf {\bibinfo {volume} {690}},\ \bibinfo {pages} {A40} (\bibinfo {year} {2024})},\ \Eprint {https://arxiv.org/abs/2404.07070} {arXiv:2404.07070 [astro-ph.CO]} \BibitemShut {NoStop}%
\bibitem [{\citenamefont {Yin}(2024)}]{Yin:2024hba}%
  \BibitemOpen
  \bibfield  {author} {\bibinfo {author} {\bibfnamefont {W.}~\bibnamefont {Yin}},\ }\href {https://doi.org/10.1007/JHEP05(2024)327} {\bibfield  {journal} {\bibinfo  {journal} {JHEP}\ }\textbf {\bibinfo {volume} {05}},\ \bibinfo {pages} {327}},\ \Eprint {https://arxiv.org/abs/2404.06444} {arXiv:2404.06444 [hep-ph]} \BibitemShut {NoStop}%
\bibitem [{\citenamefont {Gialamas}\ \emph {et~al.}(2024)\citenamefont {Gialamas}, \citenamefont {H\"utsi}, \citenamefont {Kannike}, \citenamefont {Racioppi}, \citenamefont {Raidal}, \citenamefont {Vasar},\ and\ \citenamefont {Veerm\"ae}}]{Gialamas:2024lyw}%
  \BibitemOpen
  \bibfield  {author} {\bibinfo {author} {\bibfnamefont {I.~D.}\ \bibnamefont {Gialamas}}, \bibinfo {author} {\bibfnamefont {G.}~\bibnamefont {H\"utsi}}, \bibinfo {author} {\bibfnamefont {K.}~\bibnamefont {Kannike}}, \bibinfo {author} {\bibfnamefont {A.}~\bibnamefont {Racioppi}}, \bibinfo {author} {\bibfnamefont {M.}~\bibnamefont {Raidal}}, \bibinfo {author} {\bibfnamefont {M.}~\bibnamefont {Vasar}},\ and\ \bibinfo {author} {\bibfnamefont {H.}~\bibnamefont {Veerm\"ae}},\ }\href@noop {} {\bibinfo {title} {{Interpreting DESI 2024 BAO: late-time dynamical dark energy or a local effect?}}} (\bibinfo {year} {2024}),\ \Eprint {https://arxiv.org/abs/2406.07533} {arXiv:2406.07533 [astro-ph.CO]} \BibitemShut {NoStop}%
\bibitem [{\citenamefont {Dinda}(2024)}]{Dinda:2024kjf}%
  \BibitemOpen
  \bibfield  {author} {\bibinfo {author} {\bibfnamefont {B.~R.}\ \bibnamefont {Dinda}},\ }\href {https://doi.org/10.1088/1475-7516/2024/09/062} {\bibfield  {journal} {\bibinfo  {journal} {JCAP}\ }\textbf {\bibinfo {volume} {09}},\ \bibinfo {pages} {062}},\ \Eprint {https://arxiv.org/abs/2405.06618} {arXiv:2405.06618 [astro-ph.CO]} \BibitemShut {NoStop}%
\bibitem [{\citenamefont {Najafi}\ \emph {et~al.}(2024)\citenamefont {Najafi}, \citenamefont {Pan}, \citenamefont {Di~Valentino},\ and\ \citenamefont {Firouzjaee}}]{Najafi:2024qzm}%
  \BibitemOpen
  \bibfield  {author} {\bibinfo {author} {\bibfnamefont {M.}~\bibnamefont {Najafi}}, \bibinfo {author} {\bibfnamefont {S.}~\bibnamefont {Pan}}, \bibinfo {author} {\bibfnamefont {E.}~\bibnamefont {Di~Valentino}},\ and\ \bibinfo {author} {\bibfnamefont {J.~T.}\ \bibnamefont {Firouzjaee}},\ }\href {https://doi.org/10.1016/j.dark.2024.101539} {\bibfield  {journal} {\bibinfo  {journal} {Phys. Dark Univ.}\ }\textbf {\bibinfo {volume} {45}},\ \bibinfo {pages} {101539} (\bibinfo {year} {2024})},\ \Eprint {https://arxiv.org/abs/2407.14939} {arXiv:2407.14939 [astro-ph.CO]} \BibitemShut {NoStop}%
\bibitem [{\citenamefont {Wang}\ and\ \citenamefont {Piao}(2024)}]{Wang:2024dka}%
  \BibitemOpen
  \bibfield  {author} {\bibinfo {author} {\bibfnamefont {H.}~\bibnamefont {Wang}}\ and\ \bibinfo {author} {\bibfnamefont {Y.-S.}\ \bibnamefont {Piao}},\ }\href@noop {} {\bibinfo {title} {{Dark energy in light of recent DESI BAO and Hubble tension}}} (\bibinfo {year} {2024}),\ \Eprint {https://arxiv.org/abs/2404.18579} {arXiv:2404.18579 [astro-ph.CO]} \BibitemShut {NoStop}%
\bibitem [{\citenamefont {Ye}\ \emph {et~al.}(2024)\citenamefont {Ye}, \citenamefont {Martinelli}, \citenamefont {Hu},\ and\ \citenamefont {Silvestri}}]{Ye:2024ywg}%
  \BibitemOpen
  \bibfield  {author} {\bibinfo {author} {\bibfnamefont {G.}~\bibnamefont {Ye}}, \bibinfo {author} {\bibfnamefont {M.}~\bibnamefont {Martinelli}}, \bibinfo {author} {\bibfnamefont {B.}~\bibnamefont {Hu}},\ and\ \bibinfo {author} {\bibfnamefont {A.}~\bibnamefont {Silvestri}},\ }\href@noop {} {\bibinfo {title} {{Non-minimally coupled gravity as a physically viable fit to DESI 2024 BAO}}} (\bibinfo {year} {2024}),\ \Eprint {https://arxiv.org/abs/2407.15832} {arXiv:2407.15832 [astro-ph.CO]} \BibitemShut {NoStop}%
\bibitem [{\citenamefont {Tada}\ and\ \citenamefont {Terada}(2024)}]{Tada:2024znt}%
  \BibitemOpen
  \bibfield  {author} {\bibinfo {author} {\bibfnamefont {Y.}~\bibnamefont {Tada}}\ and\ \bibinfo {author} {\bibfnamefont {T.}~\bibnamefont {Terada}},\ }\href {https://doi.org/10.1103/PhysRevD.109.L121305} {\bibfield  {journal} {\bibinfo  {journal} {Phys. Rev. D}\ }\textbf {\bibinfo {volume} {109}},\ \bibinfo {pages} {L121305} (\bibinfo {year} {2024})},\ \Eprint {https://arxiv.org/abs/2404.05722} {arXiv:2404.05722 [astro-ph.CO]} \BibitemShut {NoStop}%
\bibitem [{\citenamefont {Carloni}\ \emph {et~al.}(2025)\citenamefont {Carloni}, \citenamefont {Luongo},\ and\ \citenamefont {Muccino}}]{Carloni:2024zpl}%
  \BibitemOpen
  \bibfield  {author} {\bibinfo {author} {\bibfnamefont {Y.}~\bibnamefont {Carloni}}, \bibinfo {author} {\bibfnamefont {O.}~\bibnamefont {Luongo}},\ and\ \bibinfo {author} {\bibfnamefont {M.}~\bibnamefont {Muccino}},\ }\href {https://doi.org/10.1103/PhysRevD.111.023512} {\bibfield  {journal} {\bibinfo  {journal} {Phys. Rev. D}\ }\textbf {\bibinfo {volume} {111}},\ \bibinfo {pages} {023512} (\bibinfo {year} {2025})},\ \Eprint {https://arxiv.org/abs/2404.12068} {arXiv:2404.12068 [astro-ph.CO]} \BibitemShut {NoStop}%
\bibitem [{\citenamefont {Park}\ \emph {et~al.}(2024{\natexlab{a}})\citenamefont {Park}, \citenamefont {de~Cruz~P\'erez},\ and\ \citenamefont {Ratra}}]{Chan-GyungPark:2024mlx}%
  \BibitemOpen
  \bibfield  {author} {\bibinfo {author} {\bibfnamefont {C.-G.}\ \bibnamefont {Park}}, \bibinfo {author} {\bibfnamefont {J.}~\bibnamefont {de~Cruz~P\'erez}},\ and\ \bibinfo {author} {\bibfnamefont {B.}~\bibnamefont {Ratra}},\ }\href {https://doi.org/10.1103/PhysRevD.110.123533} {\bibfield  {journal} {\bibinfo  {journal} {Phys. Rev. D}\ }\textbf {\bibinfo {volume} {110}},\ \bibinfo {pages} {123533} (\bibinfo {year} {2024}{\natexlab{a}})},\ \Eprint {https://arxiv.org/abs/2405.00502} {arXiv:2405.00502 [astro-ph.CO]} \BibitemShut {NoStop}%
\bibitem [{\citenamefont {Lodha}\ \emph {et~al.}(2025{\natexlab{b}})\citenamefont {Lodha} \emph {et~al.}}]{DESI:2024kob}%
  \BibitemOpen
  \bibfield  {author} {\bibinfo {author} {\bibfnamefont {K.}~\bibnamefont {Lodha}} \emph {et~al.} (\bibinfo {collaboration} {DESI}),\ }\href {https://doi.org/10.1103/PhysRevD.111.023532} {\bibfield  {journal} {\bibinfo  {journal} {Phys. Rev. D}\ }\textbf {\bibinfo {volume} {111}},\ \bibinfo {pages} {023532} (\bibinfo {year} {2025}{\natexlab{b}})},\ \Eprint {https://arxiv.org/abs/2405.13588} {arXiv:2405.13588 [astro-ph.CO]} \BibitemShut {NoStop}%
\bibitem [{\citenamefont {Ramadan}\ \emph {et~al.}(2024)\citenamefont {Ramadan}, \citenamefont {Sakstein},\ and\ \citenamefont {Rubin}}]{Ramadan:2024kmn}%
  \BibitemOpen
  \bibfield  {author} {\bibinfo {author} {\bibfnamefont {O.~F.}\ \bibnamefont {Ramadan}}, \bibinfo {author} {\bibfnamefont {J.}~\bibnamefont {Sakstein}},\ and\ \bibinfo {author} {\bibfnamefont {D.}~\bibnamefont {Rubin}},\ }\href {https://doi.org/10.1103/PhysRevD.110.L041303} {\bibfield  {journal} {\bibinfo  {journal} {Phys. Rev. D}\ }\textbf {\bibinfo {volume} {110}},\ \bibinfo {pages} {L041303} (\bibinfo {year} {2024})},\ \Eprint {https://arxiv.org/abs/2405.18747} {arXiv:2405.18747 [astro-ph.CO]} \BibitemShut {NoStop}%
\bibitem [{\citenamefont {Notari}\ \emph {et~al.}(2024{\natexlab{a}})\citenamefont {Notari}, \citenamefont {Redi},\ and\ \citenamefont {Tesi}}]{Notari:2024rti}%
  \BibitemOpen
  \bibfield  {author} {\bibinfo {author} {\bibfnamefont {A.}~\bibnamefont {Notari}}, \bibinfo {author} {\bibfnamefont {M.}~\bibnamefont {Redi}},\ and\ \bibinfo {author} {\bibfnamefont {A.}~\bibnamefont {Tesi}},\ }\href {https://doi.org/10.1088/1475-7516/2024/11/025} {\bibfield  {journal} {\bibinfo  {journal} {JCAP}\ }\textbf {\bibinfo {volume} {11}},\ \bibinfo {pages} {025}},\ \Eprint {https://arxiv.org/abs/2406.08459} {arXiv:2406.08459 [astro-ph.CO]} \BibitemShut {NoStop}%
\bibitem [{\citenamefont {Orchard}\ and\ \citenamefont {C\'ardenas}(2024)}]{Orchard:2024bve}%
  \BibitemOpen
  \bibfield  {author} {\bibinfo {author} {\bibfnamefont {L.}~\bibnamefont {Orchard}}\ and\ \bibinfo {author} {\bibfnamefont {V.~H.}\ \bibnamefont {C\'ardenas}},\ }\href {https://doi.org/10.1016/j.dark.2024.101678} {\bibfield  {journal} {\bibinfo  {journal} {Phys. Dark Univ.}\ }\textbf {\bibinfo {volume} {46}},\ \bibinfo {pages} {101678} (\bibinfo {year} {2024})},\ \Eprint {https://arxiv.org/abs/2407.05579} {arXiv:2407.05579 [astro-ph.CO]} \BibitemShut {NoStop}%
\bibitem [{\citenamefont {Hern\'andez-Almada}\ \emph {et~al.}(2024)\citenamefont {Hern\'andez-Almada}, \citenamefont {Mendoza-Mart\'\i{}nez}, \citenamefont {Garc\'\i{}a-Aspeitia},\ and\ \citenamefont {Motta}}]{Hernandez-Almada:2024ost}%
  \BibitemOpen
  \bibfield  {author} {\bibinfo {author} {\bibfnamefont {A.}~\bibnamefont {Hern\'andez-Almada}}, \bibinfo {author} {\bibfnamefont {M.~L.}\ \bibnamefont {Mendoza-Mart\'\i{}nez}}, \bibinfo {author} {\bibfnamefont {M.~A.}\ \bibnamefont {Garc\'\i{}a-Aspeitia}},\ and\ \bibinfo {author} {\bibfnamefont {V.}~\bibnamefont {Motta}},\ }\href {https://doi.org/10.1016/j.dark.2024.101668} {\bibfield  {journal} {\bibinfo  {journal} {Phys. Dark Univ.}\ }\textbf {\bibinfo {volume} {46}},\ \bibinfo {pages} {101668} (\bibinfo {year} {2024})},\ \Eprint {https://arxiv.org/abs/2407.09430} {arXiv:2407.09430 [astro-ph.CO]} \BibitemShut {NoStop}%
\bibitem [{\citenamefont {Pourojaghi}\ \emph {et~al.}(2024)\citenamefont {Pourojaghi}, \citenamefont {Malekjani},\ and\ \citenamefont {Davari}}]{Pourojaghi:2024tmw}%
  \BibitemOpen
  \bibfield  {author} {\bibinfo {author} {\bibfnamefont {S.}~\bibnamefont {Pourojaghi}}, \bibinfo {author} {\bibfnamefont {M.}~\bibnamefont {Malekjani}},\ and\ \bibinfo {author} {\bibfnamefont {Z.}~\bibnamefont {Davari}},\ }\href@noop {} {\bibinfo {title} {{Cosmological constraints on dark energy parametrizations after DESI 2024: Persistent deviation from standard $\Lambda$CDM cosmology}}} (\bibinfo {year} {2024}),\ \Eprint {https://arxiv.org/abs/2407.09767} {arXiv:2407.09767 [astro-ph.CO]} \BibitemShut {NoStop}%
\bibitem [{\citenamefont {Giar\`e}\ \emph {et~al.}(2024{\natexlab{b}})\citenamefont {Giar\`e}, \citenamefont {Najafi}, \citenamefont {Pan}, \citenamefont {Di~Valentino},\ and\ \citenamefont {Firouzjaee}}]{Giare:2024gpk}%
  \BibitemOpen
  \bibfield  {author} {\bibinfo {author} {\bibfnamefont {W.}~\bibnamefont {Giar\`e}}, \bibinfo {author} {\bibfnamefont {M.}~\bibnamefont {Najafi}}, \bibinfo {author} {\bibfnamefont {S.}~\bibnamefont {Pan}}, \bibinfo {author} {\bibfnamefont {E.}~\bibnamefont {Di~Valentino}},\ and\ \bibinfo {author} {\bibfnamefont {J.~T.}\ \bibnamefont {Firouzjaee}},\ }\href {https://doi.org/10.1088/1475-7516/2024/10/035} {\bibfield  {journal} {\bibinfo  {journal} {JCAP}\ }\textbf {\bibinfo {volume} {10}},\ \bibinfo {pages} {035}},\ \Eprint {https://arxiv.org/abs/2407.16689} {arXiv:2407.16689 [astro-ph.CO]} \BibitemShut {NoStop}%
\bibitem [{\citenamefont {Rebou\c{c}as}\ \emph {et~al.}(2025)\citenamefont {Rebou\c{c}as}, \citenamefont {de~Souza}, \citenamefont {Zhong}, \citenamefont {Miranda},\ and\ \citenamefont {Rosenfeld}}]{Reboucas:2024smm}%
  \BibitemOpen
  \bibfield  {author} {\bibinfo {author} {\bibfnamefont {J.~a.}\ \bibnamefont {Rebou\c{c}as}}, \bibinfo {author} {\bibfnamefont {D.~H.~F.}\ \bibnamefont {de~Souza}}, \bibinfo {author} {\bibfnamefont {K.}~\bibnamefont {Zhong}}, \bibinfo {author} {\bibfnamefont {V.}~\bibnamefont {Miranda}},\ and\ \bibinfo {author} {\bibfnamefont {R.}~\bibnamefont {Rosenfeld}},\ }\href {https://doi.org/10.1088/1475-7516/2025/02/024} {\bibfield  {journal} {\bibinfo  {journal} {JCAP}\ }\textbf {\bibinfo {volume} {02}},\ \bibinfo {pages} {024}},\ \Eprint {https://arxiv.org/abs/2408.14628} {arXiv:2408.14628 [astro-ph.CO]} \BibitemShut {NoStop}%
\bibitem [{\citenamefont {Giar\`e}(2024)}]{Giare:2024ocw}%
  \BibitemOpen
  \bibfield  {author} {\bibinfo {author} {\bibfnamefont {W.}~\bibnamefont {Giar\`e}},\ }\href@noop {} {\bibinfo {title} {{Dynamical Dark Energy Beyond Planck? Constraints from multiple CMB probes, DESI BAO and Type-Ia Supernovae}}} (\bibinfo {year} {2024}),\ \Eprint {https://arxiv.org/abs/2409.17074} {arXiv:2409.17074 [astro-ph.CO]} \BibitemShut {NoStop}%
\bibitem [{\citenamefont {Park}\ \emph {et~al.}(2024{\natexlab{b}})\citenamefont {Park}, \citenamefont {de~Cruz~Perez},\ and\ \citenamefont {Ratra}}]{Chan-GyungPark:2024brx}%
  \BibitemOpen
  \bibfield  {author} {\bibinfo {author} {\bibfnamefont {C.-G.}\ \bibnamefont {Park}}, \bibinfo {author} {\bibfnamefont {J.}~\bibnamefont {de~Cruz~Perez}},\ and\ \bibinfo {author} {\bibfnamefont {B.}~\bibnamefont {Ratra}},\ }\href@noop {} {\bibinfo {title} {{Is the $w_0w_a$CDM cosmological parameterization evidence for dark energy dynamics partially caused by the excess smoothing of Planck CMB anisotropy data?}}} (\bibinfo {year} {2024}{\natexlab{b}}),\ \Eprint {https://arxiv.org/abs/2410.13627} {arXiv:2410.13627 [astro-ph.CO]} \BibitemShut {NoStop}%
\bibitem [{\citenamefont {Menci}\ \emph {et~al.}(2024)\citenamefont {Menci}, \citenamefont {Sen},\ and\ \citenamefont {Castellano}}]{Menci:2024hop}%
  \BibitemOpen
  \bibfield  {author} {\bibinfo {author} {\bibfnamefont {N.}~\bibnamefont {Menci}}, \bibinfo {author} {\bibfnamefont {A.~A.}\ \bibnamefont {Sen}},\ and\ \bibinfo {author} {\bibfnamefont {M.}~\bibnamefont {Castellano}},\ }\href {https://doi.org/10.3847/1538-4357/ad8d5b} {\bibfield  {journal} {\bibinfo  {journal} {Astrophys. J.}\ }\textbf {\bibinfo {volume} {976}},\ \bibinfo {pages} {227} (\bibinfo {year} {2024})},\ \Eprint {https://arxiv.org/abs/2410.22940} {arXiv:2410.22940 [astro-ph.CO]} \BibitemShut {NoStop}%
\bibitem [{\citenamefont {Li}\ \emph {et~al.}(2024{\natexlab{b}})\citenamefont {Li}, \citenamefont {Li}, \citenamefont {Du}, \citenamefont {Wu}, \citenamefont {Feng}, \citenamefont {Zhang},\ and\ \citenamefont {Zhang}}]{Li:2024qus}%
  \BibitemOpen
  \bibfield  {author} {\bibinfo {author} {\bibfnamefont {T.-N.}\ \bibnamefont {Li}}, \bibinfo {author} {\bibfnamefont {Y.-H.}\ \bibnamefont {Li}}, \bibinfo {author} {\bibfnamefont {G.-H.}\ \bibnamefont {Du}}, \bibinfo {author} {\bibfnamefont {P.-J.}\ \bibnamefont {Wu}}, \bibinfo {author} {\bibfnamefont {L.}~\bibnamefont {Feng}}, \bibinfo {author} {\bibfnamefont {J.-F.}\ \bibnamefont {Zhang}},\ and\ \bibinfo {author} {\bibfnamefont {X.}~\bibnamefont {Zhang}},\ }\href@noop {} {\bibinfo {title} {{Revisiting holographic dark energy after DESI 2024}}} (\bibinfo {year} {2024}{\natexlab{b}}),\ \Eprint {https://arxiv.org/abs/2411.08639} {arXiv:2411.08639 [astro-ph.CO]} \BibitemShut {NoStop}%
\bibitem [{\citenamefont {Li}\ and\ \citenamefont {Wang}(2024)}]{Li:2024hrv}%
  \BibitemOpen
  \bibfield  {author} {\bibinfo {author} {\bibfnamefont {J.-X.}\ \bibnamefont {Li}}\ and\ \bibinfo {author} {\bibfnamefont {S.}~\bibnamefont {Wang}},\ }\href@noop {} {\bibinfo {title} {{A comprehensive numerical study on four categories of holographic dark energy models}}} (\bibinfo {year} {2024}),\ \Eprint {https://arxiv.org/abs/2412.09064} {arXiv:2412.09064 [astro-ph.CO]} \BibitemShut {NoStop}%
\bibitem [{\citenamefont {Notari}\ \emph {et~al.}(2024{\natexlab{b}})\citenamefont {Notari}, \citenamefont {Redi},\ and\ \citenamefont {Tesi}}]{Notari:2024zmi}%
  \BibitemOpen
  \bibfield  {author} {\bibinfo {author} {\bibfnamefont {A.}~\bibnamefont {Notari}}, \bibinfo {author} {\bibfnamefont {M.}~\bibnamefont {Redi}},\ and\ \bibinfo {author} {\bibfnamefont {A.}~\bibnamefont {Tesi}},\ }\href@noop {} {\bibinfo {title} {{BAO vs. SN evidence for evolving dark energy}}} (\bibinfo {year} {2024}{\natexlab{b}}),\ \Eprint {https://arxiv.org/abs/2411.11685} {arXiv:2411.11685 [astro-ph.CO]} \BibitemShut {NoStop}%
\bibitem [{\citenamefont {Gao}\ \emph {et~al.}(2025)\citenamefont {Gao}, \citenamefont {Peng}, \citenamefont {Gao},\ and\ \citenamefont {Gong}}]{Gao:2024ily}%
  \BibitemOpen
  \bibfield  {author} {\bibinfo {author} {\bibfnamefont {Q.}~\bibnamefont {Gao}}, \bibinfo {author} {\bibfnamefont {Z.}~\bibnamefont {Peng}}, \bibinfo {author} {\bibfnamefont {S.}~\bibnamefont {Gao}},\ and\ \bibinfo {author} {\bibfnamefont {Y.}~\bibnamefont {Gong}},\ }\href {https://doi.org/10.3390/universe11010010} {\bibfield  {journal} {\bibinfo  {journal} {Universe}\ }\textbf {\bibinfo {volume} {11}},\ \bibinfo {pages} {10} (\bibinfo {year} {2025})},\ \Eprint {https://arxiv.org/abs/2411.16046} {arXiv:2411.16046 [astro-ph.CO]} \BibitemShut {NoStop}%
\bibitem [{\citenamefont {Fikri}\ \emph {et~al.}(2024)\citenamefont {Fikri}, \citenamefont {ElKhateeb}, \citenamefont {Lashin},\ and\ \citenamefont {El~Hanafy}}]{Fikri:2024klc}%
  \BibitemOpen
  \bibfield  {author} {\bibinfo {author} {\bibfnamefont {R.}~\bibnamefont {Fikri}}, \bibinfo {author} {\bibfnamefont {E.}~\bibnamefont {ElKhateeb}}, \bibinfo {author} {\bibfnamefont {E.~S.}\ \bibnamefont {Lashin}},\ and\ \bibinfo {author} {\bibfnamefont {W.}~\bibnamefont {El~Hanafy}},\ }\href@noop {} {\bibinfo {title} {{A preference for dynamical phantom dark energy using one-parameter model with Planck, DESI DR1 BAO and SN data}}} (\bibinfo {year} {2024}),\ \Eprint {https://arxiv.org/abs/2411.19362} {arXiv:2411.19362 [astro-ph.CO]} \BibitemShut {NoStop}%
\bibitem [{\citenamefont {Jiang}\ \emph {et~al.}(2024)\citenamefont {Jiang}, \citenamefont {Pedrotti}, \citenamefont {da~Costa},\ and\ \citenamefont {Vagnozzi}}]{Jiang:2024xnu}%
  \BibitemOpen
  \bibfield  {author} {\bibinfo {author} {\bibfnamefont {J.-Q.}\ \bibnamefont {Jiang}}, \bibinfo {author} {\bibfnamefont {D.}~\bibnamefont {Pedrotti}}, \bibinfo {author} {\bibfnamefont {S.~S.}\ \bibnamefont {da~Costa}},\ and\ \bibinfo {author} {\bibfnamefont {S.}~\bibnamefont {Vagnozzi}},\ }\href@noop {} {\bibinfo {title} {{Non-parametric late-time expansion history reconstruction and implications for the Hubble tension in light of DESI}}} (\bibinfo {year} {2024}),\ \Eprint {https://arxiv.org/abs/2408.02365} {arXiv:2408.02365 [astro-ph.CO]} \BibitemShut {NoStop}%
\bibitem [{\citenamefont {Zheng}\ \emph {et~al.}(2024)\citenamefont {Zheng}, \citenamefont {Qiang},\ and\ \citenamefont {You}}]{Zheng:2024qzi}%
  \BibitemOpen
  \bibfield  {author} {\bibinfo {author} {\bibfnamefont {J.}~\bibnamefont {Zheng}}, \bibinfo {author} {\bibfnamefont {D.-C.}\ \bibnamefont {Qiang}},\ and\ \bibinfo {author} {\bibfnamefont {Z.-Q.}\ \bibnamefont {You}},\ }\href@noop {} {\bibinfo {title} {{Cosmological constraints on dark energy models using DESI BAO 2024}}} (\bibinfo {year} {2024}),\ \Eprint {https://arxiv.org/abs/2412.04830} {arXiv:2412.04830 [astro-ph.CO]} \BibitemShut {NoStop}%
\bibitem [{\citenamefont {G\'omez-Valent}\ and\ \citenamefont {Sol\`a~Peracaula}(2025)}]{Gomez-Valent:2024ejh}%
  \BibitemOpen
  \bibfield  {author} {\bibinfo {author} {\bibfnamefont {A.}~\bibnamefont {G\'omez-Valent}}\ and\ \bibinfo {author} {\bibfnamefont {J.}~\bibnamefont {Sol\`a~Peracaula}},\ }\href {https://doi.org/10.1016/j.physletb.2025.139391} {\bibfield  {journal} {\bibinfo  {journal} {Phys. Lett. B}\ }\textbf {\bibinfo {volume} {864}},\ \bibinfo {pages} {139391} (\bibinfo {year} {2025})},\ \Eprint {https://arxiv.org/abs/2412.15124} {arXiv:2412.15124 [astro-ph.CO]} \BibitemShut {NoStop}%
\bibitem [{\citenamefont {Roy~Choudhury}\ and\ \citenamefont {Okumura}(2024)}]{RoyChoudhury:2024wri}%
  \BibitemOpen
  \bibfield  {author} {\bibinfo {author} {\bibfnamefont {S.}~\bibnamefont {Roy~Choudhury}}\ and\ \bibinfo {author} {\bibfnamefont {T.}~\bibnamefont {Okumura}},\ }\href {https://doi.org/10.3847/2041-8213/ad8c26} {\bibfield  {journal} {\bibinfo  {journal} {Astrophys. J. Lett.}\ }\textbf {\bibinfo {volume} {976}},\ \bibinfo {pages} {L11} (\bibinfo {year} {2024})},\ \Eprint {https://arxiv.org/abs/2409.13022} {arXiv:2409.13022 [astro-ph.CO]} \BibitemShut {NoStop}%
\bibitem [{\citenamefont {Lewis}\ and\ \citenamefont {Chamberlain}(2024)}]{Lewis:2024cqj}%
  \BibitemOpen
  \bibfield  {author} {\bibinfo {author} {\bibfnamefont {A.}~\bibnamefont {Lewis}}\ and\ \bibinfo {author} {\bibfnamefont {E.}~\bibnamefont {Chamberlain}},\ }\href@noop {} {\bibinfo {title} {{Understanding acoustic scale observations: the one-sided fight against $\Lambda$}}} (\bibinfo {year} {2024}),\ \Eprint {https://arxiv.org/abs/2412.13894} {arXiv:2412.13894 [astro-ph.CO]} \BibitemShut {NoStop}%
\bibitem [{\citenamefont {Wolf}\ \emph {et~al.}(2024)\citenamefont {Wolf}, \citenamefont {Garc{\'\i}a-Garc{\'\i}a}, \citenamefont {Bartlett},\ and\ \citenamefont {Ferreira}}]{Wolf:2024eph}%
  \BibitemOpen
  \bibfield  {author} {\bibinfo {author} {\bibfnamefont {W.~J.}\ \bibnamefont {Wolf}}, \bibinfo {author} {\bibfnamefont {C.}~\bibnamefont {Garc{\'\i}a-Garc{\'\i}a}}, \bibinfo {author} {\bibfnamefont {D.~J.}\ \bibnamefont {Bartlett}},\ and\ \bibinfo {author} {\bibfnamefont {P.~G.}\ \bibnamefont {Ferreira}},\ }\href {https://doi.org/10.1103/PhysRevD.110.083528} {\bibfield  {journal} {\bibinfo  {journal} {Phys. Rev. D}\ }\textbf {\bibinfo {volume} {110}},\ \bibinfo {pages} {083528} (\bibinfo {year} {2024})},\ \Eprint {https://arxiv.org/abs/2408.17318} {arXiv:2408.17318 [astro-ph.CO]} \BibitemShut {NoStop}%
\bibitem [{\citenamefont {Wolf}\ \emph {et~al.}(2025{\natexlab{a}})\citenamefont {Wolf}, \citenamefont {Ferreira},\ and\ \citenamefont {Garc{\'\i}a-Garc{\'\i}a}}]{Wolf:2024stt}%
  \BibitemOpen
  \bibfield  {author} {\bibinfo {author} {\bibfnamefont {W.~J.}\ \bibnamefont {Wolf}}, \bibinfo {author} {\bibfnamefont {P.~G.}\ \bibnamefont {Ferreira}},\ and\ \bibinfo {author} {\bibfnamefont {C.}~\bibnamefont {Garc{\'\i}a-Garc{\'\i}a}},\ }\href {https://doi.org/10.1103/PhysRevD.111.L041303} {\bibfield  {journal} {\bibinfo  {journal} {Phys. Rev. D}\ }\textbf {\bibinfo {volume} {111}},\ \bibinfo {pages} {L041303} (\bibinfo {year} {2025}{\natexlab{a}})},\ \Eprint {https://arxiv.org/abs/2409.17019} {arXiv:2409.17019 [astro-ph.CO]} \BibitemShut {NoStop}%
\bibitem [{\citenamefont {Wolf}\ \emph {et~al.}(2025{\natexlab{b}})\citenamefont {Wolf}, \citenamefont {Garc{\'\i}a-Garc{\'\i}a}, \citenamefont {Anton},\ and\ \citenamefont {Ferreira}}]{Wolf:2025jed}%
  \BibitemOpen
  \bibfield  {author} {\bibinfo {author} {\bibfnamefont {W.~J.}\ \bibnamefont {Wolf}}, \bibinfo {author} {\bibfnamefont {C.}~\bibnamefont {Garc{\'\i}a-Garc{\'\i}a}}, \bibinfo {author} {\bibfnamefont {T.}~\bibnamefont {Anton}},\ and\ \bibinfo {author} {\bibfnamefont {P.~G.}\ \bibnamefont {Ferreira}},\ }\href {https://doi.org/10.1103/jysf-k72m} {\bibfield  {journal} {\bibinfo  {journal} {Phys. Rev. Lett.}\ }\textbf {\bibinfo {volume} {135}},\ \bibinfo {pages} {081001} (\bibinfo {year} {2025}{\natexlab{b}})},\ \Eprint {https://arxiv.org/abs/2504.07679} {arXiv:2504.07679 [astro-ph.CO]} \BibitemShut {NoStop}%
\bibitem [{\citenamefont {Wolf}\ \emph {et~al.}(2025{\natexlab{c}})\citenamefont {Wolf}, \citenamefont {Garc\'\i{}a-Garc\'\i{}a},\ and\ \citenamefont {Ferreira}}]{Wolf:2025jlc}%
  \BibitemOpen
  \bibfield  {author} {\bibinfo {author} {\bibfnamefont {W.~J.}\ \bibnamefont {Wolf}}, \bibinfo {author} {\bibfnamefont {C.}~\bibnamefont {Garc\'\i{}a-Garc\'\i{}a}},\ and\ \bibinfo {author} {\bibfnamefont {P.~G.}\ \bibnamefont {Ferreira}},\ }\href@noop {} {\bibinfo {title} {{Robustness of Dark Energy Phenomenology Across Different Parameterizations}}} (\bibinfo {year} {2025}{\natexlab{c}}),\ \Eprint {https://arxiv.org/abs/2502.04929} {arXiv:2502.04929 [astro-ph.CO]} \BibitemShut {NoStop}%
\bibitem [{\citenamefont {Shajib}\ and\ \citenamefont {Frieman}(2025)}]{Shajib:2025tpd}%
  \BibitemOpen
  \bibfield  {author} {\bibinfo {author} {\bibfnamefont {A.~J.}\ \bibnamefont {Shajib}}\ and\ \bibinfo {author} {\bibfnamefont {J.~A.}\ \bibnamefont {Frieman}},\ }\href@noop {} {\bibinfo {title} {{Evolving dark energy models: Current and forecast constraints}}} (\bibinfo {year} {2025}),\ \Eprint {https://arxiv.org/abs/2502.06929} {arXiv:2502.06929 [astro-ph.CO]} \BibitemShut {NoStop}%
\bibitem [{\citenamefont {Giar\`e}\ \emph {et~al.}(2025)\citenamefont {Giar\`e}, \citenamefont {Mahassen}, \citenamefont {Di~Valentino},\ and\ \citenamefont {Pan}}]{Giare:2025pzu}%
  \BibitemOpen
  \bibfield  {author} {\bibinfo {author} {\bibfnamefont {W.}~\bibnamefont {Giar\`e}}, \bibinfo {author} {\bibfnamefont {T.}~\bibnamefont {Mahassen}}, \bibinfo {author} {\bibfnamefont {E.}~\bibnamefont {Di~Valentino}},\ and\ \bibinfo {author} {\bibfnamefont {S.}~\bibnamefont {Pan}},\ }\href {https://doi.org/10.1016/j.dark.2025.101906} {\bibfield  {journal} {\bibinfo  {journal} {Phys. Dark Univ.}\ }\textbf {\bibinfo {volume} {48}},\ \bibinfo {pages} {101906} (\bibinfo {year} {2025})},\ \Eprint {https://arxiv.org/abs/2502.10264} {arXiv:2502.10264 [astro-ph.CO]} \BibitemShut {NoStop}%
\bibitem [{\citenamefont {Chaussidon}\ \emph {et~al.}(2025)\citenamefont {Chaussidon} \emph {et~al.}}]{Chaussidon:2025npr}%
  \BibitemOpen
  \bibfield  {author} {\bibinfo {author} {\bibfnamefont {E.}~\bibnamefont {Chaussidon}} \emph {et~al.},\ }\href@noop {} {\bibinfo {title} {{Early time solution as an alternative to the late time evolving dark energy with DESI DR2 BAO}}} (\bibinfo {year} {2025}),\ \Eprint {https://arxiv.org/abs/2503.24343} {arXiv:2503.24343 [astro-ph.CO]} \BibitemShut {NoStop}%
\bibitem [{\citenamefont {Kessler}\ \emph {et~al.}(2025)\citenamefont {Kessler}, \citenamefont {Escamilla}, \citenamefont {Pan},\ and\ \citenamefont {Di~Valentino}}]{Kessler:2025kju}%
  \BibitemOpen
  \bibfield  {author} {\bibinfo {author} {\bibfnamefont {D.~A.}\ \bibnamefont {Kessler}}, \bibinfo {author} {\bibfnamefont {L.~A.}\ \bibnamefont {Escamilla}}, \bibinfo {author} {\bibfnamefont {S.}~\bibnamefont {Pan}},\ and\ \bibinfo {author} {\bibfnamefont {E.}~\bibnamefont {Di~Valentino}},\ }\href@noop {} {\bibinfo {title} {{One-parameter dynamical dark energy: Hints for oscillations}}} (\bibinfo {year} {2025}),\ \Eprint {https://arxiv.org/abs/2504.00776} {arXiv:2504.00776 [astro-ph.CO]} \BibitemShut {NoStop}%
\bibitem [{\citenamefont {Pang}\ \emph {et~al.}(2025)\citenamefont {Pang}, \citenamefont {Zhang},\ and\ \citenamefont {Huang}}]{Pang:2025lvh}%
  \BibitemOpen
  \bibfield  {author} {\bibinfo {author} {\bibfnamefont {Y.-H.}\ \bibnamefont {Pang}}, \bibinfo {author} {\bibfnamefont {X.}~\bibnamefont {Zhang}},\ and\ \bibinfo {author} {\bibfnamefont {Q.-G.}\ \bibnamefont {Huang}},\ }\href@noop {} {\bibinfo {title} {{The Impact of the Hubble Tension on the Evidence for Dynamical Dark Energy}}} (\bibinfo {year} {2025}),\ \Eprint {https://arxiv.org/abs/2503.21600} {arXiv:2503.21600 [astro-ph.CO]} \BibitemShut {NoStop}%
\bibitem [{\citenamefont {Roy~Choudhury}(2025)}]{RoyChoudhury:2025dhe}%
  \BibitemOpen
  \bibfield  {author} {\bibinfo {author} {\bibfnamefont {S.}~\bibnamefont {Roy~Choudhury}},\ }\href@noop {} {\bibinfo {title} {{Cosmology in Extended Parameter Space with DESI DR2 BAO: A 2$\sigma$+ Detection of Non-zero Neutrino Masses with an Update on Dynamical Dark Energy and Lensing Anomaly}}} (\bibinfo {year} {2025}),\ \Eprint {https://arxiv.org/abs/2504.15340} {arXiv:2504.15340 [astro-ph.CO]} \BibitemShut {NoStop}%
\bibitem [{\citenamefont {Scherer}\ \emph {et~al.}(2025)\citenamefont {Scherer}, \citenamefont {Sabogal}, \citenamefont {Nunes},\ and\ \citenamefont {De~Felice}}]{Scherer:2025esj}%
  \BibitemOpen
  \bibfield  {author} {\bibinfo {author} {\bibfnamefont {M.}~\bibnamefont {Scherer}}, \bibinfo {author} {\bibfnamefont {M.~A.}\ \bibnamefont {Sabogal}}, \bibinfo {author} {\bibfnamefont {R.~C.}\ \bibnamefont {Nunes}},\ and\ \bibinfo {author} {\bibfnamefont {A.}~\bibnamefont {De~Felice}},\ }\href@noop {} {\bibinfo {title} {{Challenging $\Lambda$CDM: 5$\sigma$ Evidence for a Dynamical Dark Energy Late-Time Transition}}} (\bibinfo {year} {2025}),\ \Eprint {https://arxiv.org/abs/2504.20664} {arXiv:2504.20664 [astro-ph.CO]} \BibitemShut {NoStop}%
\bibitem [{\citenamefont {Teixeira}\ \emph {et~al.}(2025)\citenamefont {Teixeira}, \citenamefont {Giar{\`e}}, \citenamefont {Hogg}, \citenamefont {Montandon}, \citenamefont {Poudou},\ and\ \citenamefont {Poulin}}]{Teixeira:2025czm}%
  \BibitemOpen
  \bibfield  {author} {\bibinfo {author} {\bibfnamefont {E.~M.}\ \bibnamefont {Teixeira}}, \bibinfo {author} {\bibfnamefont {W.}~\bibnamefont {Giar{\`e}}}, \bibinfo {author} {\bibfnamefont {N.~B.}\ \bibnamefont {Hogg}}, \bibinfo {author} {\bibfnamefont {T.}~\bibnamefont {Montandon}}, \bibinfo {author} {\bibfnamefont {A.}~\bibnamefont {Poudou}},\ and\ \bibinfo {author} {\bibfnamefont {V.}~\bibnamefont {Poulin}},\ }\href@noop {} {\bibinfo {title} {{Implications of distance duality violation for the $H_0$ tension and evolving dark energy}}} (\bibinfo {year} {2025}),\ \Eprint {https://arxiv.org/abs/2504.10464} {arXiv:2504.10464 [astro-ph.CO]} \BibitemShut {NoStop}%
\bibitem [{\citenamefont {Specogna}\ \emph {et~al.}(2025)\citenamefont {Specogna}, \citenamefont {Adil}, \citenamefont {Ozulker}, \citenamefont {Di~Valentino}, \citenamefont {Nunes}, \citenamefont {Akarsu},\ and\ \citenamefont {Sen}}]{Specogna:2025guo}%
  \BibitemOpen
  \bibfield  {author} {\bibinfo {author} {\bibfnamefont {E.}~\bibnamefont {Specogna}}, \bibinfo {author} {\bibfnamefont {S.~A.}\ \bibnamefont {Adil}}, \bibinfo {author} {\bibfnamefont {E.}~\bibnamefont {Ozulker}}, \bibinfo {author} {\bibfnamefont {E.}~\bibnamefont {Di~Valentino}}, \bibinfo {author} {\bibfnamefont {R.~C.}\ \bibnamefont {Nunes}}, \bibinfo {author} {\bibfnamefont {O.}~\bibnamefont {Akarsu}},\ and\ \bibinfo {author} {\bibfnamefont {A.~A.}\ \bibnamefont {Sen}},\ }\href@noop {} {\bibinfo {title} {{Updated Constraints on Omnipotent Dark Energy: A Comprehensive Analysis with CMB and BAO Data}}} (\bibinfo {year} {2025}),\ \Eprint {https://arxiv.org/abs/2504.17859} {arXiv:2504.17859 [gr-qc]} \BibitemShut {NoStop}%
\bibitem [{\citenamefont {Cheng}\ \emph {et~al.}(2025{\natexlab{a}})\citenamefont {Cheng}, \citenamefont {Di~Valentino}, \citenamefont {Escamilla}, \citenamefont {Sen},\ and\ \citenamefont {Visinelli}}]{Cheng:2025lod}%
  \BibitemOpen
  \bibfield  {author} {\bibinfo {author} {\bibfnamefont {H.}~\bibnamefont {Cheng}}, \bibinfo {author} {\bibfnamefont {E.}~\bibnamefont {Di~Valentino}}, \bibinfo {author} {\bibfnamefont {L.~A.}\ \bibnamefont {Escamilla}}, \bibinfo {author} {\bibfnamefont {A.~A.}\ \bibnamefont {Sen}},\ and\ \bibinfo {author} {\bibfnamefont {L.}~\bibnamefont {Visinelli}},\ }\href@noop {} {\bibinfo {title} {{Pressure Parametrization of Dark Energy: First and Second-Order Constraints with Latest Cosmological Data}}} (\bibinfo {year} {2025}{\natexlab{a}}),\ \Eprint {https://arxiv.org/abs/2505.02932} {arXiv:2505.02932 [astro-ph.CO]} \BibitemShut {NoStop}%
\bibitem [{\citenamefont {Cheng}\ \emph {et~al.}(2025{\natexlab{b}})\citenamefont {Cheng}, \citenamefont {Di~Valentino},\ and\ \citenamefont {Visinelli}}]{Cheng:2025hug}%
  \BibitemOpen
  \bibfield  {author} {\bibinfo {author} {\bibfnamefont {H.}~\bibnamefont {Cheng}}, \bibinfo {author} {\bibfnamefont {E.}~\bibnamefont {Di~Valentino}},\ and\ \bibinfo {author} {\bibfnamefont {L.}~\bibnamefont {Visinelli}},\ }\href@noop {} {\bibinfo {title} {{Cosmic Strings as Dynamical Dark Energy: Novel Constraints}}} (\bibinfo {year} {2025}{\natexlab{b}}),\ \Eprint {https://arxiv.org/abs/2505.22066} {arXiv:2505.22066 [astro-ph.CO]} \BibitemShut {NoStop}%
\bibitem [{\citenamefont {{\"O}z{\"u}lker}\ \emph {et~al.}(2025)\citenamefont {{\"O}z{\"u}lker}, \citenamefont {Di~Valentino},\ and\ \citenamefont {Giar{\`e}}}]{Ozulker:2025ehg}%
  \BibitemOpen
  \bibfield  {author} {\bibinfo {author} {\bibfnamefont {E.}~\bibnamefont {{\"O}z{\"u}lker}}, \bibinfo {author} {\bibfnamefont {E.}~\bibnamefont {Di~Valentino}},\ and\ \bibinfo {author} {\bibfnamefont {W.}~\bibnamefont {Giar{\`e}}},\ }\href@noop {} {\bibinfo {title} {{Dark Energy Crosses the Line: Quantifying and Testing the Evidence for Phantom Crossing}}} (\bibinfo {year} {2025}),\ \Eprint {https://arxiv.org/abs/2506.19053} {arXiv:2506.19053 [astro-ph.CO]} \BibitemShut {NoStop}%
\bibitem [{\citenamefont {Gialamas}\ \emph {et~al.}(2025)\citenamefont {Gialamas}, \citenamefont {H{\"u}tsi}, \citenamefont {Raidal}, \citenamefont {Urrutia}, \citenamefont {Vasar},\ and\ \citenamefont {Veerm{\"a}e}}]{Gialamas:2025pwv}%
  \BibitemOpen
  \bibfield  {author} {\bibinfo {author} {\bibfnamefont {I.~D.}\ \bibnamefont {Gialamas}}, \bibinfo {author} {\bibfnamefont {G.}~\bibnamefont {H{\"u}tsi}}, \bibinfo {author} {\bibfnamefont {M.}~\bibnamefont {Raidal}}, \bibinfo {author} {\bibfnamefont {J.}~\bibnamefont {Urrutia}}, \bibinfo {author} {\bibfnamefont {M.}~\bibnamefont {Vasar}},\ and\ \bibinfo {author} {\bibfnamefont {H.}~\bibnamefont {Veerm{\"a}e}},\ }\href@noop {} {\bibinfo {title} {{Quintessence and phantoms in light of DESI 2025}}} (\bibinfo {year} {2025}),\ \Eprint {https://arxiv.org/abs/2506.21542} {arXiv:2506.21542 [astro-ph.CO]} \BibitemShut {NoStop}%
\bibitem [{\citenamefont {Lee}\ \emph {et~al.}(2025)\citenamefont {Lee}, \citenamefont {Yang}, \citenamefont {Di~Valentino}, \citenamefont {Pan},\ and\ \citenamefont {van~de Bruck}}]{Lee:2025pzo}%
  \BibitemOpen
  \bibfield  {author} {\bibinfo {author} {\bibfnamefont {D.~H.}\ \bibnamefont {Lee}}, \bibinfo {author} {\bibfnamefont {W.}~\bibnamefont {Yang}}, \bibinfo {author} {\bibfnamefont {E.}~\bibnamefont {Di~Valentino}}, \bibinfo {author} {\bibfnamefont {S.}~\bibnamefont {Pan}},\ and\ \bibinfo {author} {\bibfnamefont {C.}~\bibnamefont {van~de Bruck}},\ }\href@noop {} {\bibinfo {title} {{The Shape of Dark Energy: Constraining Its Evolution with a General Parametrization}}} (\bibinfo {year} {2025}),\ \Eprint {https://arxiv.org/abs/2507.11432} {arXiv:2507.11432 [astro-ph.CO]} \BibitemShut {NoStop}%
\bibitem [{\citenamefont {Chimento}(2012)}]{Chimento_2012}%
  \BibitemOpen
  \bibfield  {author} {\bibinfo {author} {\bibfnamefont {L.~P.}\ \bibnamefont {Chimento}},\ }in\ \href {https://doi.org/10.1063/1.4756808} {\emph {\bibinfo {booktitle} {AIP Conference Proceedings}}}\ (\bibinfo  {publisher} {AIP},\ \bibinfo {year} {2012})\ p.\ \bibinfo {pages} {30–38}\BibitemShut {NoStop}%
\bibitem [{\citenamefont {Pan}\ \emph {et~al.}(2015)\citenamefont {Pan}, \citenamefont {Bhattacharya},\ and\ \citenamefont {Chakraborty}}]{Pan_2015}%
  \BibitemOpen
  \bibfield  {author} {\bibinfo {author} {\bibfnamefont {S.}~\bibnamefont {Pan}}, \bibinfo {author} {\bibfnamefont {S.}~\bibnamefont {Bhattacharya}},\ and\ \bibinfo {author} {\bibfnamefont {S.}~\bibnamefont {Chakraborty}},\ }\href {https://doi.org/10.1093/mnras/stv1495} {\bibfield  {journal} {\bibinfo  {journal} {Monthly Notices of the Royal Astronomical Society}\ }\textbf {\bibinfo {volume} {452}},\ \bibinfo {pages} {3038–3046} (\bibinfo {year} {2015})}\BibitemShut {NoStop}%
\bibitem [{\citenamefont {Pan}\ and\ \citenamefont {Sharov}(2017)}]{Pan_2017}%
  \BibitemOpen
  \bibfield  {author} {\bibinfo {author} {\bibfnamefont {S.}~\bibnamefont {Pan}}\ and\ \bibinfo {author} {\bibfnamefont {G.~S.}\ \bibnamefont {Sharov}},\ }\href {https://doi.org/10.1093/mnras/stx2278} {\bibfield  {journal} {\bibinfo  {journal} {Monthly Notices of the Royal Astronomical Society}\ }\textbf {\bibinfo {volume} {472}},\ \bibinfo {pages} {4736–4749} (\bibinfo {year} {2017})}\BibitemShut {NoStop}%
\bibitem [{\citenamefont {He}\ and\ \citenamefont {Wang}(2008)}]{He_2008}%
  \BibitemOpen
  \bibfield  {author} {\bibinfo {author} {\bibfnamefont {J.-H.}\ \bibnamefont {He}}\ and\ \bibinfo {author} {\bibfnamefont {B.}~\bibnamefont {Wang}},\ }\href {https://doi.org/10.1088/1475-7516/2008/06/010} {\bibfield  {journal} {\bibinfo  {journal} {Journal of Cosmology and Astroparticle Physics}\ }\textbf {\bibinfo {volume} {2008}}\bibinfo  {number} { (06)},\ \bibinfo {pages} {010}}\BibitemShut {NoStop}%
\bibitem [{\citenamefont {Caldera-Cabral}\ \emph {et~al.}(2009{\natexlab{b}})\citenamefont {Caldera-Cabral}, \citenamefont {Maartens},\ and\ \citenamefont {Schaefer}}]{Caldera_Cabral_2009_structure}%
  \BibitemOpen
\bibfield  {number} {  }\bibfield  {author} {\bibinfo {author} {\bibfnamefont {G.}~\bibnamefont {Caldera-Cabral}}, \bibinfo {author} {\bibfnamefont {R.}~\bibnamefont {Maartens}},\ and\ \bibinfo {author} {\bibfnamefont {B.~M.}\ \bibnamefont {Schaefer}},\ }\href {https://doi.org/10.1088/1475-7516/2009/07/027} {\bibfield  {journal} {\bibinfo  {journal} {Journal of Cosmology and Astroparticle Physics}\ }\textbf {\bibinfo {volume} {2009}}\bibinfo  {number} { (07)},\ \bibinfo {pages} {027–027}}\BibitemShut {NoStop}%
\bibitem [{\citenamefont {Pan}\ \emph {et~al.}(2020{\natexlab{a}})\citenamefont {Pan}, \citenamefont {de~Haro}, \citenamefont {Yang},\ and\ \citenamefont {Amorós}}]{Pan_2020}%
  \BibitemOpen
\bibfield  {number} {  }\bibfield  {author} {\bibinfo {author} {\bibfnamefont {S.}~\bibnamefont {Pan}}, \bibinfo {author} {\bibfnamefont {J.}~\bibnamefont {de~Haro}}, \bibinfo {author} {\bibfnamefont {W.}~\bibnamefont {Yang}},\ and\ \bibinfo {author} {\bibfnamefont {J.}~\bibnamefont {Amorós}},\ }\bibfield  {journal} {\bibinfo  {journal} {Physical Review D}\ }\textbf {\bibinfo {volume} {101}},\ \href {https://doi.org/10.1103/physrevd.101.123506} {10.1103/physrevd.101.123506} (\bibinfo {year} {2020}{\natexlab{a}})\BibitemShut {NoStop}%
\bibitem [{\citenamefont {Nojiri}\ \emph {et~al.}(2005)\citenamefont {Nojiri}, \citenamefont {Odintsov},\ and\ \citenamefont {Tsujikawa}}]{Nojiri_2005_IDE}%
  \BibitemOpen
  \bibfield  {author} {\bibinfo {author} {\bibfnamefont {S.}~\bibnamefont {Nojiri}}, \bibinfo {author} {\bibfnamefont {S.~D.}\ \bibnamefont {Odintsov}},\ and\ \bibinfo {author} {\bibfnamefont {S.}~\bibnamefont {Tsujikawa}},\ }\href {https://doi.org/10.1103/PhysRevD.71.063004} {\bibfield  {journal} {\bibinfo  {journal} {Physical Review D}\ }\textbf {\bibinfo {volume} {71}},\ \bibinfo {pages} {063004} (\bibinfo {year} {2005})}\BibitemShut {NoStop}%
\bibitem [{\citenamefont {de~Haro}\ \emph {et~al.}(2023)\citenamefont {de~Haro}, \citenamefont {Nojiri}, \citenamefont {Odintsov}, \citenamefont {Oikonomou},\ and\ \citenamefont {Pan}}]{de_Haro_2023}%
  \BibitemOpen
  \bibfield  {author} {\bibinfo {author} {\bibfnamefont {J.}~\bibnamefont {de~Haro}}, \bibinfo {author} {\bibfnamefont {S.}~\bibnamefont {Nojiri}}, \bibinfo {author} {\bibfnamefont {S.}~\bibnamefont {Odintsov}}, \bibinfo {author} {\bibfnamefont {V.}~\bibnamefont {Oikonomou}},\ and\ \bibinfo {author} {\bibfnamefont {S.}~\bibnamefont {Pan}},\ }\href {https://doi.org/10.1016/j.physrep.2023.09.003} {\bibfield  {journal} {\bibinfo  {journal} {Physics Reports}\ }\textbf {\bibinfo {volume} {1034}},\ \bibinfo {pages} {1–114} (\bibinfo {year} {2023})}\BibitemShut {NoStop}%
\bibitem [{\citenamefont {Valiviita}\ \emph {et~al.}(2008)\citenamefont {Valiviita}, \citenamefont {Majerotto},\ and\ \citenamefont {Maartens}}]{Valiviita:2008iv}%
  \BibitemOpen
  \bibfield  {author} {\bibinfo {author} {\bibfnamefont {J.}~\bibnamefont {Valiviita}}, \bibinfo {author} {\bibfnamefont {E.}~\bibnamefont {Majerotto}},\ and\ \bibinfo {author} {\bibfnamefont {R.}~\bibnamefont {Maartens}},\ }\href {https://doi.org/10.1088/1475-7516/2008/07/020} {\bibfield  {journal} {\bibinfo  {journal} {JCAP}\ }\textbf {\bibinfo {volume} {07}},\ \bibinfo {pages} {020}},\ \Eprint {https://arxiv.org/abs/0804.0232} {arXiv:0804.0232 [astro-ph]} \BibitemShut {NoStop}%
\bibitem [{\citenamefont {He}\ \emph {et~al.}(2009)\citenamefont {He}, \citenamefont {Wang},\ and\ \citenamefont {Abdalla}}]{He_2009}%
  \BibitemOpen
  \bibfield  {author} {\bibinfo {author} {\bibfnamefont {J.-H.}\ \bibnamefont {He}}, \bibinfo {author} {\bibfnamefont {B.}~\bibnamefont {Wang}},\ and\ \bibinfo {author} {\bibfnamefont {E.}~\bibnamefont {Abdalla}},\ }\href {https://doi.org/10.1016/j.physletb.2008.11.062} {\bibfield  {journal} {\bibinfo  {journal} {Physics Letters B}\ }\textbf {\bibinfo {volume} {671}},\ \bibinfo {pages} {139–145} (\bibinfo {year} {2009})}\BibitemShut {NoStop}%
\bibitem [{\citenamefont {Costa}\ \emph {et~al.}(2014)\citenamefont {Costa}, \citenamefont {Xu}, \citenamefont {Wang}, \citenamefont {Ferreira},\ and\ \citenamefont {Abdalla}}]{Costa_2014}%
  \BibitemOpen
  \bibfield  {author} {\bibinfo {author} {\bibfnamefont {A.~A.}\ \bibnamefont {Costa}}, \bibinfo {author} {\bibfnamefont {X.-D.}\ \bibnamefont {Xu}}, \bibinfo {author} {\bibfnamefont {B.}~\bibnamefont {Wang}}, \bibinfo {author} {\bibfnamefont {E.~G.}\ \bibnamefont {Ferreira}},\ and\ \bibinfo {author} {\bibfnamefont {E.}~\bibnamefont {Abdalla}},\ }\bibfield  {journal} {\bibinfo  {journal} {Physical Review D}\ }\textbf {\bibinfo {volume} {89}},\ \href {https://doi.org/10.1103/physrevd.89.103531} {10.1103/physrevd.89.103531} (\bibinfo {year} {2014})\BibitemShut {NoStop}%
\bibitem [{\citenamefont {Eingorn}\ and\ \citenamefont {Kiefer}(2015)}]{Eingorn_2015}%
  \BibitemOpen
  \bibfield  {author} {\bibinfo {author} {\bibfnamefont {M.}~\bibnamefont {Eingorn}}\ and\ \bibinfo {author} {\bibfnamefont {C.}~\bibnamefont {Kiefer}},\ }\href {https://doi.org/10.1088/1475-7516/2015/07/036} {\bibfield  {journal} {\bibinfo  {journal} {Journal of Cosmology and Astroparticle Physics}\ }\textbf {\bibinfo {volume} {2015}}\bibinfo  {number} { (07)},\ \bibinfo {pages} {036–036}}\BibitemShut {NoStop}%
\bibitem [{\citenamefont {Costa}\ \emph {et~al.}(2017)\citenamefont {Costa}, \citenamefont {Xu}, \citenamefont {Wang},\ and\ \citenamefont {Abdalla}}]{Costa_2017}%
  \BibitemOpen
\bibfield  {number} {  }\bibfield  {author} {\bibinfo {author} {\bibfnamefont {A.~A.}\ \bibnamefont {Costa}}, \bibinfo {author} {\bibfnamefont {X.-D.}\ \bibnamefont {Xu}}, \bibinfo {author} {\bibfnamefont {B.}~\bibnamefont {Wang}},\ and\ \bibinfo {author} {\bibfnamefont {E.}~\bibnamefont {Abdalla}},\ }\href {https://doi.org/10.1088/1475-7516/2017/01/028} {\bibfield  {journal} {\bibinfo  {journal} {Journal of Cosmology and Astroparticle Physics}\ }\textbf {\bibinfo {volume} {2017}}\bibinfo  {number} { (01)},\ \bibinfo {pages} {028–028}}\BibitemShut {NoStop}%
\bibitem [{\citenamefont {Carrasco}\ \emph {et~al.}(2023)\citenamefont {Carrasco}, \citenamefont {Rincon}, \citenamefont {Saavedra},\ and\ \citenamefont {Videla}}]{carrasco2023discriminatinginteractingdarkenergy}%
  \BibitemOpen
\bibfield  {number} {  }\bibfield  {author} {\bibinfo {author} {\bibfnamefont {R.}~\bibnamefont {Carrasco}}, \bibinfo {author} {\bibfnamefont {A.}~\bibnamefont {Rincon}}, \bibinfo {author} {\bibfnamefont {J.}~\bibnamefont {Saavedra}},\ and\ \bibinfo {author} {\bibfnamefont {N.}~\bibnamefont {Videla}},\ }\href {https://arxiv.org/abs/2310.04324} {\bibinfo {title} {Discriminating interacting dark energy models using statefinder diagnostic}} (\bibinfo {year} {2023}),\ \Eprint {https://arxiv.org/abs/2310.04324} {arXiv:2310.04324 [gr-qc]} \BibitemShut {NoStop}%
\bibitem [{\citenamefont {Arevalo}\ \emph {et~al.}(2017)\citenamefont {Arevalo}, \citenamefont {Cid},\ and\ \citenamefont {Moya}}]{Arevalo_2017}%
  \BibitemOpen
  \bibfield  {author} {\bibinfo {author} {\bibfnamefont {F.}~\bibnamefont {Arevalo}}, \bibinfo {author} {\bibfnamefont {A.}~\bibnamefont {Cid}},\ and\ \bibinfo {author} {\bibfnamefont {J.}~\bibnamefont {Moya}},\ }\bibfield  {journal} {\bibinfo  {journal} {The European Physical Journal C}\ }\textbf {\bibinfo {volume} {77}},\ \href {https://doi.org/10.1140/epjc/s10052-017-5128-7} {10.1140/epjc/s10052-017-5128-7} (\bibinfo {year} {2017})\BibitemShut {NoStop}%
\bibitem [{\citenamefont {Cid}\ \emph {et~al.}(2019)\citenamefont {Cid}, \citenamefont {Santos}, \citenamefont {Pigozzo}, \citenamefont {Ferreira},\ and\ \citenamefont {Alcaniz}}]{Cid_2019}%
  \BibitemOpen
  \bibfield  {author} {\bibinfo {author} {\bibfnamefont {A.}~\bibnamefont {Cid}}, \bibinfo {author} {\bibfnamefont {B.}~\bibnamefont {Santos}}, \bibinfo {author} {\bibfnamefont {C.}~\bibnamefont {Pigozzo}}, \bibinfo {author} {\bibfnamefont {T.}~\bibnamefont {Ferreira}},\ and\ \bibinfo {author} {\bibfnamefont {J.}~\bibnamefont {Alcaniz}},\ }\href {https://doi.org/10.1088/1475-7516/2019/03/030} {\bibfield  {journal} {\bibinfo  {journal} {Journal of Cosmology and Astroparticle Physics}\ }\textbf {\bibinfo {volume} {2019}}\bibinfo  {number} { (03)},\ \bibinfo {pages} {030–030}}\BibitemShut {NoStop}%
\bibitem [{\citenamefont {Lepe}\ and\ \citenamefont {Peña}(2016)}]{Lepe_2016}%
  \BibitemOpen
\bibfield  {number} {  }\bibfield  {author} {\bibinfo {author} {\bibfnamefont {S.}~\bibnamefont {Lepe}}\ and\ \bibinfo {author} {\bibfnamefont {F.}~\bibnamefont {Peña}},\ }\bibfield  {journal} {\bibinfo  {journal} {The European Physical Journal C}\ }\textbf {\bibinfo {volume} {76}},\ \href {https://doi.org/10.1140/epjc/s10052-016-4347-7} {10.1140/epjc/s10052-016-4347-7} (\bibinfo {year} {2016})\BibitemShut {NoStop}%
\bibitem [{\citenamefont {Costa}\ \emph {et~al.}(2019)\citenamefont {Costa}, \citenamefont {Marcondes}, \citenamefont {Landim}, \citenamefont {Abdalla}, \citenamefont {Abramo}, \citenamefont {Xavier}, \citenamefont {Orsi}, \citenamefont {Devi}, \citenamefont {Cenarro}, \citenamefont {Cristóbal-Hornillos}, \citenamefont {Dupke}, \citenamefont {Ederoclite}, \citenamefont {Marín-Franch}, \citenamefont {Oliveira}, \citenamefont {Vázquez~Ramió}, \citenamefont {Taylor},\ and\ \citenamefont {Varela}}]{Costa_2019}%
  \BibitemOpen
  \bibfield  {author} {\bibinfo {author} {\bibfnamefont {A.~A.}\ \bibnamefont {Costa}}, \bibinfo {author} {\bibfnamefont {R.~J.~F.}\ \bibnamefont {Marcondes}}, \bibinfo {author} {\bibfnamefont {R.~G.}\ \bibnamefont {Landim}}, \bibinfo {author} {\bibfnamefont {E.}~\bibnamefont {Abdalla}}, \bibinfo {author} {\bibfnamefont {L.~R.}\ \bibnamefont {Abramo}}, \bibinfo {author} {\bibfnamefont {H.~S.}\ \bibnamefont {Xavier}}, \bibinfo {author} {\bibfnamefont {A.~A.}\ \bibnamefont {Orsi}}, \bibinfo {author} {\bibfnamefont {N.~C.}\ \bibnamefont {Devi}}, \bibinfo {author} {\bibfnamefont {A.~J.}\ \bibnamefont {Cenarro}}, \bibinfo {author} {\bibfnamefont {D.}~\bibnamefont {Cristóbal-Hornillos}}, \bibinfo {author} {\bibfnamefont {R.~A.}\ \bibnamefont {Dupke}}, \bibinfo {author} {\bibfnamefont {A.}~\bibnamefont {Ederoclite}}, \bibinfo {author} {\bibfnamefont {A.}~\bibnamefont {Marín-Franch}}, \bibinfo {author} {\bibfnamefont {C.~M.}\ \bibnamefont {Oliveira}}, \bibinfo {author} {\bibfnamefont {H.}~\bibnamefont
  {Vázquez~Ramió}}, \bibinfo {author} {\bibfnamefont {K.}~\bibnamefont {Taylor}},\ and\ \bibinfo {author} {\bibfnamefont {J.}~\bibnamefont {Varela}},\ }\href {https://doi.org/10.1093/mnras/stz1675} {\bibfield  {journal} {\bibinfo  {journal} {Monthly Notices of the Royal Astronomical Society}\ }\textbf {\bibinfo {volume} {488}},\ \bibinfo {pages} {78–88} (\bibinfo {year} {2019})}\BibitemShut {NoStop}%
\bibitem [{\citenamefont {An}\ \emph {et~al.}(2019)\citenamefont {An}, \citenamefont {Xu}, \citenamefont {Zhang}, \citenamefont {Wang},\ and\ \citenamefont {Yue}}]{An_2019}%
  \BibitemOpen
  \bibfield  {author} {\bibinfo {author} {\bibfnamefont {R.}~\bibnamefont {An}}, \bibinfo {author} {\bibfnamefont {X.}~\bibnamefont {Xu}}, \bibinfo {author} {\bibfnamefont {J.}~\bibnamefont {Zhang}}, \bibinfo {author} {\bibfnamefont {B.}~\bibnamefont {Wang}},\ and\ \bibinfo {author} {\bibfnamefont {B.}~\bibnamefont {Yue}},\ }\href {https://doi.org/10.1088/1475-7516/2019/01/034} {\bibfield  {journal} {\bibinfo  {journal} {Journal of Cosmology and Astroparticle Physics}\ }\textbf {\bibinfo {volume} {2019}}\bibinfo  {number} { (01)},\ \bibinfo {pages} {034–034}}\BibitemShut {NoStop}%
\bibitem [{\citenamefont {Mahata}\ and\ \citenamefont {Chakraborty}(2015)}]{Mahata_2015}%
  \BibitemOpen
\bibfield  {number} {  }\bibfield  {author} {\bibinfo {author} {\bibfnamefont {N.}~\bibnamefont {Mahata}}\ and\ \bibinfo {author} {\bibfnamefont {S.}~\bibnamefont {Chakraborty}},\ }\href {https://doi.org/10.1142/s0217732315500091} {\bibfield  {journal} {\bibinfo  {journal} {Modern Physics Letters A}\ }\textbf {\bibinfo {volume} {30}},\ \bibinfo {pages} {1550009} (\bibinfo {year} {2015})}\BibitemShut {NoStop}%
\bibitem [{\citenamefont {Olivares}\ \emph {et~al.}(2005)\citenamefont {Olivares}, \citenamefont {Atrio-Barandela},\ and\ \citenamefont {Pavón}}]{Olivares_2005}%
  \BibitemOpen
  \bibfield  {author} {\bibinfo {author} {\bibfnamefont {G.}~\bibnamefont {Olivares}}, \bibinfo {author} {\bibfnamefont {F.}~\bibnamefont {Atrio-Barandela}},\ and\ \bibinfo {author} {\bibfnamefont {D.}~\bibnamefont {Pavón}},\ }\bibfield  {journal} {\bibinfo  {journal} {Physical Review D}\ }\textbf {\bibinfo {volume} {71}},\ \href {https://doi.org/10.1103/physrevd.71.063523} {10.1103/physrevd.71.063523} (\bibinfo {year} {2005})\BibitemShut {NoStop}%
\bibitem [{\citenamefont {Califano}\ \emph {et~al.}(2024)\citenamefont {Califano}, \citenamefont {De~Martino},\ and\ \citenamefont {Lazkoz}}]{Califano_2024}%
  \BibitemOpen
  \bibfield  {author} {\bibinfo {author} {\bibfnamefont {M.}~\bibnamefont {Califano}}, \bibinfo {author} {\bibfnamefont {I.}~\bibnamefont {De~Martino}},\ and\ \bibinfo {author} {\bibfnamefont {R.}~\bibnamefont {Lazkoz}},\ }\bibfield  {journal} {\bibinfo  {journal} {Physical Review D}\ }\textbf {\bibinfo {volume} {110}},\ \href {https://doi.org/10.1103/physrevd.110.083519} {10.1103/physrevd.110.083519} (\bibinfo {year} {2024})\BibitemShut {NoStop}%
\bibitem [{\citenamefont {Arévalo}\ and\ \citenamefont {Cid}(2022)}]{Ar_valo_2022}%
  \BibitemOpen
  \bibfield  {author} {\bibinfo {author} {\bibfnamefont {F.}~\bibnamefont {Arévalo}}\ and\ \bibinfo {author} {\bibfnamefont {A.}~\bibnamefont {Cid}},\ }\bibfield  {journal} {\bibinfo  {journal} {The European Physical Journal C}\ }\textbf {\bibinfo {volume} {82}},\ \href {https://doi.org/10.1140/epjc/s10052-022-10898-6} {10.1140/epjc/s10052-022-10898-6} (\bibinfo {year} {2022})\BibitemShut {NoStop}%
\bibitem [{\citenamefont {Wang}\ \emph {et~al.}(2016)\citenamefont {Wang}, \citenamefont {Abdalla}, \citenamefont {Atrio-Barandela},\ and\ \citenamefont {Pavon}}]{Wang:2016lxa}%
  \BibitemOpen
  \bibfield  {author} {\bibinfo {author} {\bibfnamefont {B.}~\bibnamefont {Wang}}, \bibinfo {author} {\bibfnamefont {E.}~\bibnamefont {Abdalla}}, \bibinfo {author} {\bibfnamefont {F.}~\bibnamefont {Atrio-Barandela}},\ and\ \bibinfo {author} {\bibfnamefont {D.}~\bibnamefont {Pavon}},\ }\href {https://doi.org/10.1088/0034-4885/79/9/096901} {\bibfield  {journal} {\bibinfo  {journal} {Rept. Prog. Phys.}\ }\textbf {\bibinfo {volume} {79}},\ \bibinfo {pages} {096901} (\bibinfo {year} {2016})},\ \Eprint {https://arxiv.org/abs/1603.08299} {arXiv:1603.08299 [astro-ph.CO]} \BibitemShut {NoStop}%
\bibitem [{\citenamefont {Feng}\ and\ \citenamefont {Zhang}(2016)}]{Feng_2016}%
  \BibitemOpen
  \bibfield  {author} {\bibinfo {author} {\bibfnamefont {L.}~\bibnamefont {Feng}}\ and\ \bibinfo {author} {\bibfnamefont {X.}~\bibnamefont {Zhang}},\ }\href {https://doi.org/10.1088/1475-7516/2016/08/072} {\bibfield  {journal} {\bibinfo  {journal} {Journal of Cosmology and Astroparticle Physics}\ }\textbf {\bibinfo {volume} {2016}}\bibinfo  {number} { (08)},\ \bibinfo {pages} {072–072}}\BibitemShut {NoStop}%
\bibitem [{\citenamefont {Huang}\ \emph {et~al.}(2019)\citenamefont {Huang}, \citenamefont {Huang}, \citenamefont {Chen}, \citenamefont {Zhang},\ and\ \citenamefont {Tu}}]{Huang_2019}%
  \BibitemOpen
\bibfield  {number} {  }\bibfield  {author} {\bibinfo {author} {\bibfnamefont {Q.}~\bibnamefont {Huang}}, \bibinfo {author} {\bibfnamefont {H.}~\bibnamefont {Huang}}, \bibinfo {author} {\bibfnamefont {J.}~\bibnamefont {Chen}}, \bibinfo {author} {\bibfnamefont {L.}~\bibnamefont {Zhang}},\ and\ \bibinfo {author} {\bibfnamefont {F.}~\bibnamefont {Tu}},\ }\href {https://doi.org/10.1088/1361-6382/ab3504} {\bibfield  {journal} {\bibinfo  {journal} {Classical and Quantum Gravity}\ }\textbf {\bibinfo {volume} {36}},\ \bibinfo {pages} {175001} (\bibinfo {year} {2019})}\BibitemShut {NoStop}%
\bibitem [{\citenamefont {Abdollahi~Zadeh}\ \emph {et~al.}(2019)\citenamefont {Abdollahi~Zadeh}, \citenamefont {Sheykhi},\ and\ \citenamefont {Moradpour}}]{AbdollahiZadeh:2019lsx}%
  \BibitemOpen
  \bibfield  {author} {\bibinfo {author} {\bibfnamefont {M.}~\bibnamefont {Abdollahi~Zadeh}}, \bibinfo {author} {\bibfnamefont {A.}~\bibnamefont {Sheykhi}},\ and\ \bibinfo {author} {\bibfnamefont {H.}~\bibnamefont {Moradpour}},\ }\href {https://doi.org/10.1007/s10714-018-2497-7} {\bibfield  {journal} {\bibinfo  {journal} {Gen. Rel. Grav.}\ }\textbf {\bibinfo {volume} {51}},\ \bibinfo {pages} {12} (\bibinfo {year} {2019})}\BibitemShut {NoStop}%
\bibitem [{\citenamefont {Pan}\ \emph {et~al.}(2020{\natexlab{b}})\citenamefont {Pan}, \citenamefont {Sharov},\ and\ \citenamefont {Yang}}]{Pan_2020_Field}%
  \BibitemOpen
  \bibfield  {author} {\bibinfo {author} {\bibfnamefont {S.}~\bibnamefont {Pan}}, \bibinfo {author} {\bibfnamefont {G.~S.}\ \bibnamefont {Sharov}},\ and\ \bibinfo {author} {\bibfnamefont {W.}~\bibnamefont {Yang}},\ }\bibfield  {journal} {\bibinfo  {journal} {Physical Review D}\ }\textbf {\bibinfo {volume} {101}},\ \href {https://doi.org/10.1103/physrevd.101.103533} {10.1103/physrevd.101.103533} (\bibinfo {year} {2020}{\natexlab{b}})\BibitemShut {NoStop}%
\bibitem [{\citenamefont {An}\ \emph {et~al.}(2017)\citenamefont {An}, \citenamefont {Feng},\ and\ \citenamefont {Wang}}]{An_2017}%
  \BibitemOpen
  \bibfield  {author} {\bibinfo {author} {\bibfnamefont {R.}~\bibnamefont {An}}, \bibinfo {author} {\bibfnamefont {C.}~\bibnamefont {Feng}},\ and\ \bibinfo {author} {\bibfnamefont {B.}~\bibnamefont {Wang}},\ }\href {https://doi.org/10.1088/1475-7516/2017/10/049} {\bibfield  {journal} {\bibinfo  {journal} {Journal of Cosmology and Astroparticle Physics}\ }\textbf {\bibinfo {volume} {2017}}\bibinfo  {number} { (10)},\ \bibinfo {pages} {049–049}}\BibitemShut {NoStop}%
\bibitem [{\citenamefont {An}\ \emph {et~al.}(2018)\citenamefont {An}, \citenamefont {Feng},\ and\ \citenamefont {Wang}}]{An_2018}%
  \BibitemOpen
\bibfield  {number} {  }\bibfield  {author} {\bibinfo {author} {\bibfnamefont {R.}~\bibnamefont {An}}, \bibinfo {author} {\bibfnamefont {C.}~\bibnamefont {Feng}},\ and\ \bibinfo {author} {\bibfnamefont {B.}~\bibnamefont {Wang}},\ }\href {https://doi.org/10.1088/1475-7516/2018/02/038} {\bibfield  {journal} {\bibinfo  {journal} {Journal of Cosmology and Astroparticle Physics}\ }\textbf {\bibinfo {volume} {2018}}\bibinfo  {number} { (02)},\ \bibinfo {pages} {038–038}}\BibitemShut {NoStop}%
\bibitem [{\citenamefont {Bachega}\ \emph {et~al.}(2020)\citenamefont {Bachega}, \citenamefont {Costa}, \citenamefont {Abdalla},\ and\ \citenamefont {Fornazier}}]{Bachega_2020}%
  \BibitemOpen
\bibfield  {number} {  }\bibfield  {author} {\bibinfo {author} {\bibfnamefont {R.~R.}\ \bibnamefont {Bachega}}, \bibinfo {author} {\bibfnamefont {A.~A.}\ \bibnamefont {Costa}}, \bibinfo {author} {\bibfnamefont {E.}~\bibnamefont {Abdalla}},\ and\ \bibinfo {author} {\bibfnamefont {K.}~\bibnamefont {Fornazier}},\ }\href {https://doi.org/10.1088/1475-7516/2020/05/021} {\bibfield  {journal} {\bibinfo  {journal} {Journal of Cosmology and Astroparticle Physics}\ }\textbf {\bibinfo {volume} {2020}}\bibinfo  {number} { (05)},\ \bibinfo {pages} {021–021}}\BibitemShut {NoStop}%
\bibitem [{\citenamefont {Li}\ \emph {et~al.}(2020{\natexlab{a}})\citenamefont {Li}, \citenamefont {Ren}, \citenamefont {Khurshudyan},\ and\ \citenamefont {Cai}}]{Li_2020}%
  \BibitemOpen
\bibfield  {number} {  }\bibfield  {author} {\bibinfo {author} {\bibfnamefont {C.}~\bibnamefont {Li}}, \bibinfo {author} {\bibfnamefont {X.}~\bibnamefont {Ren}}, \bibinfo {author} {\bibfnamefont {M.}~\bibnamefont {Khurshudyan}},\ and\ \bibinfo {author} {\bibfnamefont {Y.-F.}\ \bibnamefont {Cai}},\ }\href {https://doi.org/10.1016/j.physletb.2019.135141} {\bibfield  {journal} {\bibinfo  {journal} {Physics Letters B}\ }\textbf {\bibinfo {volume} {801}},\ \bibinfo {pages} {135141} (\bibinfo {year} {2020}{\natexlab{a}})}\BibitemShut {NoStop}%
\bibitem [{\citenamefont {Halder}\ and\ \citenamefont {Pandey}(2021)}]{Halder_2021}%
  \BibitemOpen
  \bibfield  {author} {\bibinfo {author} {\bibfnamefont {A.}~\bibnamefont {Halder}}\ and\ \bibinfo {author} {\bibfnamefont {M.}~\bibnamefont {Pandey}},\ }\href {https://doi.org/10.1093/mnras/stab2795} {\bibfield  {journal} {\bibinfo  {journal} {Monthly Notices of the Royal Astronomical Society}\ }\textbf {\bibinfo {volume} {508}},\ \bibinfo {pages} {3446–3454} (\bibinfo {year} {2021})}\BibitemShut {NoStop}%
\bibitem [{\citenamefont {Mukhopadhyay}\ \emph {et~al.}(2021)\citenamefont {Mukhopadhyay}, \citenamefont {Majumdar},\ and\ \citenamefont {Datta}}]{Mukhopadhyay_2021}%
  \BibitemOpen
  \bibfield  {author} {\bibinfo {author} {\bibfnamefont {U.}~\bibnamefont {Mukhopadhyay}}, \bibinfo {author} {\bibfnamefont {D.}~\bibnamefont {Majumdar}},\ and\ \bibinfo {author} {\bibfnamefont {K.~K.}\ \bibnamefont {Datta}},\ }\bibfield  {journal} {\bibinfo  {journal} {Physical Review D}\ }\textbf {\bibinfo {volume} {103}},\ \href {https://doi.org/10.1103/physrevd.103.063510} {10.1103/physrevd.103.063510} (\bibinfo {year} {2021})\BibitemShut {NoStop}%
\bibitem [{\citenamefont {Xiao}\ \emph {et~al.}(2021)\citenamefont {Xiao}, \citenamefont {Costa},\ and\ \citenamefont {Wang}}]{Xiao_2021}%
  \BibitemOpen
  \bibfield  {author} {\bibinfo {author} {\bibfnamefont {L.}~\bibnamefont {Xiao}}, \bibinfo {author} {\bibfnamefont {A.~A.}\ \bibnamefont {Costa}},\ and\ \bibinfo {author} {\bibfnamefont {B.}~\bibnamefont {Wang}},\ }\href {https://doi.org/10.1093/mnras/stab3256} {\bibfield  {journal} {\bibinfo  {journal} {Monthly Notices of the Royal Astronomical Society}\ }\textbf {\bibinfo {volume} {510}},\ \bibinfo {pages} {1495–1514} (\bibinfo {year} {2021})}\BibitemShut {NoStop}%
\bibitem [{\citenamefont {Aljaf}\ \emph {et~al.}(2021)\citenamefont {Aljaf}, \citenamefont {Gregoris},\ and\ \citenamefont {Khurshudyan}}]{Aljaf_2021}%
  \BibitemOpen
  \bibfield  {author} {\bibinfo {author} {\bibfnamefont {M.}~\bibnamefont {Aljaf}}, \bibinfo {author} {\bibfnamefont {D.}~\bibnamefont {Gregoris}},\ and\ \bibinfo {author} {\bibfnamefont {M.}~\bibnamefont {Khurshudyan}},\ }\bibfield  {journal} {\bibinfo  {journal} {The European Physical Journal C}\ }\textbf {\bibinfo {volume} {81}},\ \href {https://doi.org/10.1140/epjc/s10052-021-09306-2} {10.1140/epjc/s10052-021-09306-2} (\bibinfo {year} {2021})\BibitemShut {NoStop}%
\bibitem [{\citenamefont {Zhang}\ \emph {et~al.}(2018)\citenamefont {Zhang}, \citenamefont {An}, \citenamefont {Liao}, \citenamefont {Luo}, \citenamefont {Li},\ and\ \citenamefont {Wang}}]{Zhang_2018}%
  \BibitemOpen
  \bibfield  {author} {\bibinfo {author} {\bibfnamefont {J.}~\bibnamefont {Zhang}}, \bibinfo {author} {\bibfnamefont {R.}~\bibnamefont {An}}, \bibinfo {author} {\bibfnamefont {S.}~\bibnamefont {Liao}}, \bibinfo {author} {\bibfnamefont {W.}~\bibnamefont {Luo}}, \bibinfo {author} {\bibfnamefont {Z.}~\bibnamefont {Li}},\ and\ \bibinfo {author} {\bibfnamefont {B.}~\bibnamefont {Wang}},\ }\bibfield  {journal} {\bibinfo  {journal} {Physical Review D}\ }\textbf {\bibinfo {volume} {98}},\ \href {https://doi.org/10.1103/physrevd.98.103530} {10.1103/physrevd.98.103530} (\bibinfo {year} {2018})\BibitemShut {NoStop}%
\bibitem [{\citenamefont {Zhang}\ \emph {et~al.}(2019)\citenamefont {Zhang}, \citenamefont {An}, \citenamefont {Luo}, \citenamefont {Li}, \citenamefont {Liao},\ and\ \citenamefont {Wang}}]{Zhang_2019}%
  \BibitemOpen
  \bibfield  {author} {\bibinfo {author} {\bibfnamefont {J.}~\bibnamefont {Zhang}}, \bibinfo {author} {\bibfnamefont {R.}~\bibnamefont {An}}, \bibinfo {author} {\bibfnamefont {W.}~\bibnamefont {Luo}}, \bibinfo {author} {\bibfnamefont {Z.}~\bibnamefont {Li}}, \bibinfo {author} {\bibfnamefont {S.}~\bibnamefont {Liao}},\ and\ \bibinfo {author} {\bibfnamefont {B.}~\bibnamefont {Wang}},\ }\href {https://doi.org/10.3847/2041-8213/ab133f} {\bibfield  {journal} {\bibinfo  {journal} {The Astrophysical Journal Letters}\ }\textbf {\bibinfo {volume} {875}},\ \bibinfo {pages} {L11} (\bibinfo {year} {2019})}\BibitemShut {NoStop}%
\bibitem [{\citenamefont {Liu}\ \emph {et~al.}(2022)\citenamefont {Liu}, \citenamefont {Liao}, \citenamefont {Liu}, \citenamefont {Zhang}, \citenamefont {An},\ and\ \citenamefont {Fan}}]{Liu_2022}%
  \BibitemOpen
  \bibfield  {author} {\bibinfo {author} {\bibfnamefont {Y.}~\bibnamefont {Liu}}, \bibinfo {author} {\bibfnamefont {S.}~\bibnamefont {Liao}}, \bibinfo {author} {\bibfnamefont {X.}~\bibnamefont {Liu}}, \bibinfo {author} {\bibfnamefont {J.}~\bibnamefont {Zhang}}, \bibinfo {author} {\bibfnamefont {R.}~\bibnamefont {An}},\ and\ \bibinfo {author} {\bibfnamefont {Z.}~\bibnamefont {Fan}},\ }\href {https://doi.org/10.1093/mnras/stac229} {\bibfield  {journal} {\bibinfo  {journal} {Monthly Notices of the Royal Astronomical Society}\ }\textbf {\bibinfo {volume} {511}},\ \bibinfo {pages} {3076–3088} (\bibinfo {year} {2022})}\BibitemShut {NoStop}%
\bibitem [{\citenamefont {Sharma}\ \emph {et~al.}(2021)\citenamefont {Sharma}, \citenamefont {Varshney},\ and\ \citenamefont {Dubey}}]{Sharma_2021}%
  \BibitemOpen
  \bibfield  {author} {\bibinfo {author} {\bibfnamefont {U.~K.}\ \bibnamefont {Sharma}}, \bibinfo {author} {\bibfnamefont {G.}~\bibnamefont {Varshney}},\ and\ \bibinfo {author} {\bibfnamefont {V.~C.}\ \bibnamefont {Dubey}},\ }\href {https://doi.org/10.1142/s0218271821500218} {\bibfield  {journal} {\bibinfo  {journal} {International Journal of Modern Physics D}\ }\textbf {\bibinfo {volume} {30}},\ \bibinfo {pages} {2150021} (\bibinfo {year} {2021})}\BibitemShut {NoStop}%
\bibitem [{\citenamefont {Yang}\ \emph {et~al.}(2019)\citenamefont {Yang}, \citenamefont {Vagnozzi}, \citenamefont {Di~Valentino}, \citenamefont {Nunes}, \citenamefont {Pan},\ and\ \citenamefont {Mota}}]{Yang_2019}%
  \BibitemOpen
  \bibfield  {author} {\bibinfo {author} {\bibfnamefont {W.}~\bibnamefont {Yang}}, \bibinfo {author} {\bibfnamefont {S.}~\bibnamefont {Vagnozzi}}, \bibinfo {author} {\bibfnamefont {E.}~\bibnamefont {Di~Valentino}}, \bibinfo {author} {\bibfnamefont {R.~C.}\ \bibnamefont {Nunes}}, \bibinfo {author} {\bibfnamefont {S.}~\bibnamefont {Pan}},\ and\ \bibinfo {author} {\bibfnamefont {D.~F.}\ \bibnamefont {Mota}},\ }\href {https://doi.org/10.1088/1475-7516/2019/07/037} {\bibfield  {journal} {\bibinfo  {journal} {JCAP}\ }\textbf {\bibinfo {volume} {07}},\ \bibinfo {pages} {037}},\ \Eprint {https://arxiv.org/abs/1905.08286} {arXiv:1905.08286 [astro-ph.CO]} \BibitemShut {NoStop}%
\bibitem [{\citenamefont {Sun}\ and\ \citenamefont {Yue}(2012)}]{Sun_2012}%
  \BibitemOpen
  \bibfield  {author} {\bibinfo {author} {\bibfnamefont {C.-Y.}\ \bibnamefont {Sun}}\ and\ \bibinfo {author} {\bibfnamefont {R.-H.}\ \bibnamefont {Yue}},\ }\bibfield  {journal} {\bibinfo  {journal} {Physical Review D}\ }\textbf {\bibinfo {volume} {85}},\ \href {https://doi.org/10.1103/physrevd.85.043010} {10.1103/physrevd.85.043010} (\bibinfo {year} {2012})\BibitemShut {NoStop}%
\bibitem [{\citenamefont {Zhang}\ \emph {et~al.}(2014)\citenamefont {Zhang}, \citenamefont {Sun}, \citenamefont {Yang},\ and\ \citenamefont {Yue}}]{Zhang_2014}%
  \BibitemOpen
  \bibfield  {author} {\bibinfo {author} {\bibfnamefont {M.}~\bibnamefont {Zhang}}, \bibinfo {author} {\bibfnamefont {C.}~\bibnamefont {Sun}}, \bibinfo {author} {\bibfnamefont {Z.}~\bibnamefont {Yang}},\ and\ \bibinfo {author} {\bibfnamefont {R.}~\bibnamefont {Yue}},\ }\href {https://doi.org/10.1007/s11433-014-5550-x} {\bibfield  {journal} {\bibinfo  {journal} {Sci. China Phys. Mech. Astron.}\ }\textbf {\bibinfo {volume} {57}},\ \bibinfo {pages} {1805} (\bibinfo {year} {2014})},\ \Eprint {https://arxiv.org/abs/1406.6216} {arXiv:1406.6216 [hep-th]} \BibitemShut {NoStop}%
\bibitem [{\citenamefont {Halder}\ \emph {et~al.}(2024{\natexlab{b}})\citenamefont {Halder}, \citenamefont {de~Haro}, \citenamefont {Saha},\ and\ \citenamefont {Pan}}]{halder2024phasespaceanalysissignshifting}%
  \BibitemOpen
  \bibfield  {author} {\bibinfo {author} {\bibfnamefont {S.}~\bibnamefont {Halder}}, \bibinfo {author} {\bibfnamefont {J.}~\bibnamefont {de~Haro}}, \bibinfo {author} {\bibfnamefont {T.}~\bibnamefont {Saha}},\ and\ \bibinfo {author} {\bibfnamefont {S.}~\bibnamefont {Pan}},\ }\href {https://arxiv.org/abs/2403.01397} {\bibinfo {title} {Phase space analysis of sign-shifting interacting dark energy models}} (\bibinfo {year} {2024}{\natexlab{b}}),\ \Eprint {https://arxiv.org/abs/2403.01397} {arXiv:2403.01397 [gr-qc]} \BibitemShut {NoStop}%
\bibitem [{\citenamefont {Khurshudyan}\ and\ \citenamefont {Pourhassan}(2015)}]{Khurshudyan_2015}%
  \BibitemOpen
  \bibfield  {author} {\bibinfo {author} {\bibfnamefont {M.}~\bibnamefont {Khurshudyan}}\ and\ \bibinfo {author} {\bibfnamefont {B.}~\bibnamefont {Pourhassan}},\ }\href {https://doi.org/10.1007/s10773-015-2564-8} {\bibfield  {journal} {\bibinfo  {journal} {International Journal of Theoretical Physics}\ }\textbf {\bibinfo {volume} {54}},\ \bibinfo {pages} {3251–3267} (\bibinfo {year} {2015})}\BibitemShut {NoStop}%
\bibitem [{\citenamefont {Väliviita}\ \emph {et~al.}(2010)\citenamefont {Väliviita}, \citenamefont {Maartens},\ and\ \citenamefont {Majerotto}}]{V_liviita_2010}%
  \BibitemOpen
  \bibfield  {author} {\bibinfo {author} {\bibfnamefont {J.}~\bibnamefont {Väliviita}}, \bibinfo {author} {\bibfnamefont {R.}~\bibnamefont {Maartens}},\ and\ \bibinfo {author} {\bibfnamefont {E.}~\bibnamefont {Majerotto}},\ }\href {https://doi.org/10.1111/j.1365-2966.2009.16115.x} {\bibfield  {journal} {\bibinfo  {journal} {Monthly Notices of the Royal Astronomical Society}\ }\textbf {\bibinfo {volume} {402}},\ \bibinfo {pages} {2355–2368} (\bibinfo {year} {2010})}\BibitemShut {NoStop}%
\bibitem [{\citenamefont {Izquierdo}\ \emph {et~al.}(2017)\citenamefont {Izquierdo}, \citenamefont {Blanquet-Jaramillo},\ and\ \citenamefont {Sussman}}]{Izquierdo_2017}%
  \BibitemOpen
  \bibfield  {author} {\bibinfo {author} {\bibfnamefont {G.}~\bibnamefont {Izquierdo}}, \bibinfo {author} {\bibfnamefont {R.~C.}\ \bibnamefont {Blanquet-Jaramillo}},\ and\ \bibinfo {author} {\bibfnamefont {R.~A.}\ \bibnamefont {Sussman}},\ }\bibfield  {journal} {\bibinfo  {journal} {General Relativity and Gravitation}\ }\textbf {\bibinfo {volume} {50}},\ \href {https://doi.org/10.1007/s10714-017-2321-9} {10.1007/s10714-017-2321-9} (\bibinfo {year} {2017})\BibitemShut {NoStop}%
\bibitem [{\citenamefont {Rodriguez-Benites}\ \emph {et~al.}(2024)\citenamefont {Rodriguez-Benites}, \citenamefont {Gonzalez-Espinoza}, \citenamefont {Otalora},\ and\ \citenamefont {Alva-Morales}}]{Rodriguez_Benites_2024}%
  \BibitemOpen
  \bibfield  {author} {\bibinfo {author} {\bibfnamefont {C.}~\bibnamefont {Rodriguez-Benites}}, \bibinfo {author} {\bibfnamefont {M.}~\bibnamefont {Gonzalez-Espinoza}}, \bibinfo {author} {\bibfnamefont {G.}~\bibnamefont {Otalora}},\ and\ \bibinfo {author} {\bibfnamefont {M.}~\bibnamefont {Alva-Morales}},\ }\bibfield  {journal} {\bibinfo  {journal} {The European Physical Journal C}\ }\textbf {\bibinfo {volume} {84}},\ \href {https://doi.org/10.1140/epjc/s10052-024-12613-z} {10.1140/epjc/s10052-024-12613-z} (\bibinfo {year} {2024})\BibitemShut {NoStop}%
\bibitem [{\citenamefont {Gavela}\ \emph {et~al.}(2009)\citenamefont {Gavela}, \citenamefont {Hernández}, \citenamefont {Honorez}, \citenamefont {Mena},\ and\ \citenamefont {Rigolin}}]{M.B.Gavela_2009}%
  \BibitemOpen
  \bibfield  {author} {\bibinfo {author} {\bibfnamefont {M.}~\bibnamefont {Gavela}}, \bibinfo {author} {\bibfnamefont {D.}~\bibnamefont {Hernández}}, \bibinfo {author} {\bibfnamefont {L.~L.}\ \bibnamefont {Honorez}}, \bibinfo {author} {\bibfnamefont {O.}~\bibnamefont {Mena}},\ and\ \bibinfo {author} {\bibfnamefont {S.}~\bibnamefont {Rigolin}},\ }\href {https://doi.org/10.1088/1475-7516/2009/07/034} {\bibfield  {journal} {\bibinfo  {journal} {Journal of Cosmology and Astroparticle Physics}\ }\textbf {\bibinfo {volume} {2009}}\bibinfo  {number} { (07)},\ \bibinfo {pages} {034}}\BibitemShut {NoStop}%
\bibitem [{\citenamefont {Honorez}\ \emph {et~al.}(2010)\citenamefont {Honorez}, \citenamefont {Reid}, \citenamefont {Mena}, \citenamefont {Verde},\ and\ \citenamefont {Jimenez}}]{Honorez_2010}%
  \BibitemOpen
\bibfield  {number} {  }\bibfield  {author} {\bibinfo {author} {\bibfnamefont {L.~L.}\ \bibnamefont {Honorez}}, \bibinfo {author} {\bibfnamefont {B.~A.}\ \bibnamefont {Reid}}, \bibinfo {author} {\bibfnamefont {O.}~\bibnamefont {Mena}}, \bibinfo {author} {\bibfnamefont {L.}~\bibnamefont {Verde}},\ and\ \bibinfo {author} {\bibfnamefont {R.}~\bibnamefont {Jimenez}},\ }\href {https://doi.org/10.1088/1475-7516/2010/09/029} {\bibfield  {journal} {\bibinfo  {journal} {Journal of Cosmology and Astroparticle Physics}\ }\textbf {\bibinfo {volume} {2010}}\bibinfo  {number} { (09)},\ \bibinfo {pages} {029–029}}\BibitemShut {NoStop}%
\bibitem [{\citenamefont {Khyllep}\ \emph {et~al.}(2022)\citenamefont {Khyllep}, \citenamefont {Dutta}, \citenamefont {Basilakos},\ and\ \citenamefont {Saridakis}}]{Khyllep_2022}%
  \BibitemOpen
\bibfield  {number} {  }\bibfield  {author} {\bibinfo {author} {\bibfnamefont {W.}~\bibnamefont {Khyllep}}, \bibinfo {author} {\bibfnamefont {J.}~\bibnamefont {Dutta}}, \bibinfo {author} {\bibfnamefont {S.}~\bibnamefont {Basilakos}},\ and\ \bibinfo {author} {\bibfnamefont {E.~N.}\ \bibnamefont {Saridakis}},\ }\bibfield  {journal} {\bibinfo  {journal} {Physical Review D}\ }\textbf {\bibinfo {volume} {105}},\ \href {https://doi.org/10.1103/physrevd.105.043511} {10.1103/physrevd.105.043511} (\bibinfo {year} {2022})\BibitemShut {NoStop}%
\bibitem [{\citenamefont {Pooya}(2024{\natexlab{b}})}]{pooya2024growthmatterperturbationsinteracting}%
  \BibitemOpen
  \bibfield  {author} {\bibinfo {author} {\bibfnamefont {N.~N.}\ \bibnamefont {Pooya}},\ }\href {https://arxiv.org/abs/2407.03766} {\bibinfo {title} {Growth of matter perturbations in the interacting dark energy/dark matter scenarios}} (\bibinfo {year} {2024}{\natexlab{b}}),\ \Eprint {https://arxiv.org/abs/2407.03766} {arXiv:2407.03766 [astro-ph.CO]} \BibitemShut {NoStop}%
\bibitem [{\citenamefont {Li}\ \emph {et~al.}(2020{\natexlab{b}})\citenamefont {Li}, \citenamefont {Zhang},\ and\ \citenamefont {Zhang}}]{Li_2020_2}%
  \BibitemOpen
  \bibfield  {author} {\bibinfo {author} {\bibfnamefont {H.-L.}\ \bibnamefont {Li}}, \bibinfo {author} {\bibfnamefont {J.-F.}\ \bibnamefont {Zhang}},\ and\ \bibinfo {author} {\bibfnamefont {X.}~\bibnamefont {Zhang}},\ }\href {https://doi.org/10.1088/1572-9494/abb7c9} {\bibfield  {journal} {\bibinfo  {journal} {Communications in Theoretical Physics}\ }\textbf {\bibinfo {volume} {72}},\ \bibinfo {pages} {125401} (\bibinfo {year} {2020}{\natexlab{b}})}\BibitemShut {NoStop}%
\bibitem [{\citenamefont {Santos}\ \emph {et~al.}(2017)\citenamefont {Santos}, \citenamefont {Zhao}, \citenamefont {Ferreira},\ and\ \citenamefont {Quintin}}]{Santos_2017}%
  \BibitemOpen
  \bibfield  {author} {\bibinfo {author} {\bibfnamefont {L.}~\bibnamefont {Santos}}, \bibinfo {author} {\bibfnamefont {W.}~\bibnamefont {Zhao}}, \bibinfo {author} {\bibfnamefont {E.~G.}\ \bibnamefont {Ferreira}},\ and\ \bibinfo {author} {\bibfnamefont {J.}~\bibnamefont {Quintin}},\ }\bibfield  {journal} {\bibinfo  {journal} {Physical Review D}\ }\textbf {\bibinfo {volume} {96}},\ \href {https://doi.org/10.1103/physrevd.96.103529} {10.1103/physrevd.96.103529} (\bibinfo {year} {2017})\BibitemShut {NoStop}%
\bibitem [{\citenamefont {Grandón}\ and\ \citenamefont {Cárdenas}(2019)}]{Grand_n_2019}%
  \BibitemOpen
  \bibfield  {author} {\bibinfo {author} {\bibfnamefont {D.}~\bibnamefont {Grandón}}\ and\ \bibinfo {author} {\bibfnamefont {V.~H.}\ \bibnamefont {Cárdenas}},\ }\bibfield  {journal} {\bibinfo  {journal} {General Relativity and Gravitation}\ }\textbf {\bibinfo {volume} {51}},\ \href {https://doi.org/10.1007/s10714-019-2526-1} {10.1007/s10714-019-2526-1} (\bibinfo {year} {2019})\BibitemShut {NoStop}%
\bibitem [{\citenamefont {Li}\ \emph {et~al.}(2024{\natexlab{c}})\citenamefont {Li}, \citenamefont {Wu}, \citenamefont {Du}, \citenamefont {Jin}, \citenamefont {Li}, \citenamefont {Zhang},\ and\ \citenamefont {Zhang}}]{Li_2024}%
  \BibitemOpen
  \bibfield  {author} {\bibinfo {author} {\bibfnamefont {T.-N.}\ \bibnamefont {Li}}, \bibinfo {author} {\bibfnamefont {P.-J.}\ \bibnamefont {Wu}}, \bibinfo {author} {\bibfnamefont {G.-H.}\ \bibnamefont {Du}}, \bibinfo {author} {\bibfnamefont {S.-J.}\ \bibnamefont {Jin}}, \bibinfo {author} {\bibfnamefont {H.-L.}\ \bibnamefont {Li}}, \bibinfo {author} {\bibfnamefont {J.-F.}\ \bibnamefont {Zhang}},\ and\ \bibinfo {author} {\bibfnamefont {X.}~\bibnamefont {Zhang}},\ }\href {https://doi.org/10.3847/1538-4357/ad87f0} {\bibfield  {journal} {\bibinfo  {journal} {The Astrophysical Journal}\ }\textbf {\bibinfo {volume} {976}},\ \bibinfo {pages} {1} (\bibinfo {year} {2024}{\natexlab{c}})}\BibitemShut {NoStop}%
\bibitem [{\citenamefont {Yan}\ \emph {et~al.}(2025)\citenamefont {Yan}, \citenamefont {Pan}, \citenamefont {Wang}, \citenamefont {Xu},\ and\ \citenamefont {Peng}}]{yan2025investigatinginteractingdarkenergy}%
  \BibitemOpen
  \bibfield  {author} {\bibinfo {author} {\bibfnamefont {H.}~\bibnamefont {Yan}}, \bibinfo {author} {\bibfnamefont {Y.}~\bibnamefont {Pan}}, \bibinfo {author} {\bibfnamefont {J.-X.}\ \bibnamefont {Wang}}, \bibinfo {author} {\bibfnamefont {W.-X.}\ \bibnamefont {Xu}},\ and\ \bibinfo {author} {\bibfnamefont {Z.-H.}\ \bibnamefont {Peng}},\ }\href {https://arxiv.org/abs/2507.16308} {\bibinfo {title} {Investigating interacting dark energy models using fast radio burst observations}} (\bibinfo {year} {2025}),\ \Eprint {https://arxiv.org/abs/2507.16308} {arXiv:2507.16308 [astro-ph.CO]} \BibitemShut {NoStop}%
\bibitem [{\citenamefont {Caprini}\ and\ \citenamefont {Tamanini}(2016)}]{Caprini_2016}%
  \BibitemOpen
  \bibfield  {author} {\bibinfo {author} {\bibfnamefont {C.}~\bibnamefont {Caprini}}\ and\ \bibinfo {author} {\bibfnamefont {N.}~\bibnamefont {Tamanini}},\ }\href {https://doi.org/10.1088/1475-7516/2016/10/006} {\bibfield  {journal} {\bibinfo  {journal} {Journal of Cosmology and Astroparticle Physics}\ }\textbf {\bibinfo {volume} {2016}}\bibinfo  {number} { (10)},\ \bibinfo {pages} {006–006}}\BibitemShut {NoStop}%
\bibitem [{\citenamefont {Wang}\ \emph {et~al.}(2025)\citenamefont {Wang}, \citenamefont {He}, \citenamefont {Wang}, \citenamefont {Li},\ and\ \citenamefont {Zhang}}]{wang2025prospectsconstraininginteractingdark}%
  \BibitemOpen
\bibfield  {number} {  }\bibfield  {author} {\bibinfo {author} {\bibfnamefont {B.}~\bibnamefont {Wang}}, \bibinfo {author} {\bibfnamefont {D.-Z.}\ \bibnamefont {He}}, \bibinfo {author} {\bibfnamefont {L.-F.}\ \bibnamefont {Wang}}, \bibinfo {author} {\bibfnamefont {H.-L.}\ \bibnamefont {Li}},\ and\ \bibinfo {author} {\bibfnamefont {Y.}~\bibnamefont {Zhang}},\ }\href {https://arxiv.org/abs/2210.04000} {\bibinfo {title} {Prospects for constraining interacting dark energy cosmology with gravitational-wave bright sirens detected by future ska-era pulsar timing arrays}} (\bibinfo {year} {2025}),\ \Eprint {https://arxiv.org/abs/2210.04000} {arXiv:2210.04000 [astro-ph.CO]} \BibitemShut {NoStop}%
\bibitem [{\citenamefont {Costa}\ \emph {et~al.}(2018)\citenamefont {Costa}, \citenamefont {Landim}, \citenamefont {Wang},\ and\ \citenamefont {Abdalla}}]{Costa_2018}%
  \BibitemOpen
  \bibfield  {author} {\bibinfo {author} {\bibfnamefont {A.~A.}\ \bibnamefont {Costa}}, \bibinfo {author} {\bibfnamefont {R.~C.~G.}\ \bibnamefont {Landim}}, \bibinfo {author} {\bibfnamefont {B.}~\bibnamefont {Wang}},\ and\ \bibinfo {author} {\bibfnamefont {E.}~\bibnamefont {Abdalla}},\ }\bibfield  {journal} {\bibinfo  {journal} {The European Physical Journal C}\ }\textbf {\bibinfo {volume} {78}},\ \href {https://doi.org/10.1140/epjc/s10052-018-6237-7} {10.1140/epjc/s10052-018-6237-7} (\bibinfo {year} {2018})\BibitemShut {NoStop}%
\bibitem [{\citenamefont {Guo}\ \emph {et~al.}(2017)\citenamefont {Guo}, \citenamefont {Li}, \citenamefont {Zhang},\ and\ \citenamefont {Zhang}}]{Guo_2017}%
  \BibitemOpen
  \bibfield  {author} {\bibinfo {author} {\bibfnamefont {R.-Y.}\ \bibnamefont {Guo}}, \bibinfo {author} {\bibfnamefont {Y.-H.}\ \bibnamefont {Li}}, \bibinfo {author} {\bibfnamefont {J.-F.}\ \bibnamefont {Zhang}},\ and\ \bibinfo {author} {\bibfnamefont {X.}~\bibnamefont {Zhang}},\ }\href {https://doi.org/10.1088/1475-7516/2017/05/040} {\bibfield  {journal} {\bibinfo  {journal} {Journal of Cosmology and Astroparticle Physics}\ }\textbf {\bibinfo {volume} {2017}}\bibinfo  {number} { (05)},\ \bibinfo {pages} {040–040}}\BibitemShut {NoStop}%
\bibitem [{\citenamefont {Feng}\ \emph {et~al.}(2019)\citenamefont {Feng}, \citenamefont {Zhang},\ and\ \citenamefont {Zhang}}]{Feng_2019_2}%
  \BibitemOpen
\bibfield  {number} {  }\bibfield  {author} {\bibinfo {author} {\bibfnamefont {L.}~\bibnamefont {Feng}}, \bibinfo {author} {\bibfnamefont {J.-F.}\ \bibnamefont {Zhang}},\ and\ \bibinfo {author} {\bibfnamefont {X.}~\bibnamefont {Zhang}},\ }\href {https://doi.org/10.1016/j.dark.2018.100261} {\bibfield  {journal} {\bibinfo  {journal} {Physics of the Dark Universe}\ }\textbf {\bibinfo {volume} {23}},\ \bibinfo {pages} {100261} (\bibinfo {year} {2019})}\BibitemShut {NoStop}%
\bibitem [{\citenamefont {Feng}\ \emph {et~al.}(2020{\natexlab{a}})\citenamefont {Feng}, \citenamefont {He}, \citenamefont {Li}, \citenamefont {Zhang},\ and\ \citenamefont {Zhang}}]{Feng_2020}%
  \BibitemOpen
  \bibfield  {author} {\bibinfo {author} {\bibfnamefont {L.}~\bibnamefont {Feng}}, \bibinfo {author} {\bibfnamefont {D.-Z.}\ \bibnamefont {He}}, \bibinfo {author} {\bibfnamefont {H.-L.}\ \bibnamefont {Li}}, \bibinfo {author} {\bibfnamefont {J.-F.}\ \bibnamefont {Zhang}},\ and\ \bibinfo {author} {\bibfnamefont {X.}~\bibnamefont {Zhang}},\ }\href {https://doi.org/10.1007/s11433-019-1511-8} {\bibfield  {journal} {\bibinfo  {journal} {Sci. China Phys. Mech. Astron.}\ }\textbf {\bibinfo {volume} {63}},\ \bibinfo {pages} {290404} (\bibinfo {year} {2020}{\natexlab{a}})},\ \Eprint {https://arxiv.org/abs/1910.03872} {arXiv:1910.03872 [astro-ph.CO]} \BibitemShut {NoStop}%
\bibitem [{\citenamefont {Zhao}\ \emph {et~al.}(2020)\citenamefont {Zhao}, \citenamefont {Guo}, \citenamefont {He}, \citenamefont {Zhang},\ and\ \citenamefont {Zhang}}]{Zhao_2020}%
  \BibitemOpen
  \bibfield  {author} {\bibinfo {author} {\bibfnamefont {M.}~\bibnamefont {Zhao}}, \bibinfo {author} {\bibfnamefont {R.}~\bibnamefont {Guo}}, \bibinfo {author} {\bibfnamefont {D.}~\bibnamefont {He}}, \bibinfo {author} {\bibfnamefont {J.}~\bibnamefont {Zhang}},\ and\ \bibinfo {author} {\bibfnamefont {X.}~\bibnamefont {Zhang}},\ }\href {https://doi.org/10.1007/s11433-019-1474-8} {\bibfield  {journal} {\bibinfo  {journal} {Sci. China Phys. Mech. Astron.}\ }\textbf {\bibinfo {volume} {63}},\ \bibinfo {pages} {230412} (\bibinfo {year} {2020})},\ \Eprint {https://arxiv.org/abs/1810.11658} {arXiv:1810.11658 [astro-ph.CO]} \BibitemShut {NoStop}%
\bibitem [{\citenamefont {Guo}\ \emph {et~al.}(2018)\citenamefont {Guo}, \citenamefont {Zhang}, \citenamefont {Li}, \citenamefont {He},\ and\ \citenamefont {Zhang}}]{Guo_2018}%
  \BibitemOpen
  \bibfield  {author} {\bibinfo {author} {\bibfnamefont {J.-J.}\ \bibnamefont {Guo}}, \bibinfo {author} {\bibfnamefont {J.-F.}\ \bibnamefont {Zhang}}, \bibinfo {author} {\bibfnamefont {Y.-H.}\ \bibnamefont {Li}}, \bibinfo {author} {\bibfnamefont {D.-Z.}\ \bibnamefont {He}},\ and\ \bibinfo {author} {\bibfnamefont {X.}~\bibnamefont {Zhang}},\ }\href {https://doi.org/10.1007/s11433-017-9131-9} {\bibfield  {journal} {\bibinfo  {journal} {Sci. China Phys. Mech. Astron.}\ }\textbf {\bibinfo {volume} {61}},\ \bibinfo {pages} {030011} (\bibinfo {year} {2018})},\ \Eprint {https://arxiv.org/abs/1710.03068} {arXiv:1710.03068 [astro-ph.CO]} \BibitemShut {NoStop}%
\bibitem [{\citenamefont {Li}\ \emph {et~al.}(2025)\citenamefont {Li}, \citenamefont {Du}, \citenamefont {Li}, \citenamefont {Wu}, \citenamefont {Jin}, \citenamefont {Zhang},\ and\ \citenamefont {Zhang}}]{li2025probingsignchangeableinteractiondark}%
  \BibitemOpen
  \bibfield  {author} {\bibinfo {author} {\bibfnamefont {T.-N.}\ \bibnamefont {Li}}, \bibinfo {author} {\bibfnamefont {G.-H.}\ \bibnamefont {Du}}, \bibinfo {author} {\bibfnamefont {Y.-H.}\ \bibnamefont {Li}}, \bibinfo {author} {\bibfnamefont {P.-J.}\ \bibnamefont {Wu}}, \bibinfo {author} {\bibfnamefont {S.-J.}\ \bibnamefont {Jin}}, \bibinfo {author} {\bibfnamefont {J.-F.}\ \bibnamefont {Zhang}},\ and\ \bibinfo {author} {\bibfnamefont {X.}~\bibnamefont {Zhang}},\ }\href@noop {} {\bibinfo {title} {{Probing the sign-changeable interaction between dark energy and dark matter with DESI baryon acoustic oscillations and DES supernovae data}}} (\bibinfo {year} {2025}),\ \Eprint {https://arxiv.org/abs/2501.07361} {arXiv:2501.07361 [astro-ph.CO]} \BibitemShut {NoStop}%
\bibitem [{\citenamefont {Feng}\ \emph {et~al.}(2018)\citenamefont {Feng}, \citenamefont {Li}, \citenamefont {Yu}, \citenamefont {Zhang},\ and\ \citenamefont {Zhang}}]{Feng_2018}%
  \BibitemOpen
  \bibfield  {author} {\bibinfo {author} {\bibfnamefont {L.}~\bibnamefont {Feng}}, \bibinfo {author} {\bibfnamefont {Y.-H.}\ \bibnamefont {Li}}, \bibinfo {author} {\bibfnamefont {F.}~\bibnamefont {Yu}}, \bibinfo {author} {\bibfnamefont {J.-F.}\ \bibnamefont {Zhang}},\ and\ \bibinfo {author} {\bibfnamefont {X.}~\bibnamefont {Zhang}},\ }\bibfield  {journal} {\bibinfo  {journal} {The European Physical Journal C}\ }\textbf {\bibinfo {volume} {78}},\ \href {https://doi.org/10.1140/epjc/s10052-018-6338-3} {10.1140/epjc/s10052-018-6338-3} (\bibinfo {year} {2018})\BibitemShut {NoStop}%
\bibitem [{\citenamefont {Sadri}(2019)}]{Sadri_2019}%
  \BibitemOpen
  \bibfield  {author} {\bibinfo {author} {\bibfnamefont {E.}~\bibnamefont {Sadri}},\ }\bibfield  {journal} {\bibinfo  {journal} {The European Physical Journal C}\ }\textbf {\bibinfo {volume} {79}},\ \href {https://doi.org/10.1140/epjc/s10052-019-7263-9} {10.1140/epjc/s10052-019-7263-9} (\bibinfo {year} {2019})\BibitemShut {NoStop}%
\bibitem [{\citenamefont {Bégué}\ \emph {et~al.}(2019)\citenamefont {Bégué}, \citenamefont {Stahl},\ and\ \citenamefont {Xue}}]{B_gu__2019}%
  \BibitemOpen
  \bibfield  {author} {\bibinfo {author} {\bibfnamefont {D.}~\bibnamefont {Bégué}}, \bibinfo {author} {\bibfnamefont {C.}~\bibnamefont {Stahl}},\ and\ \bibinfo {author} {\bibfnamefont {S.-S.}\ \bibnamefont {Xue}},\ }\href {https://doi.org/10.1016/j.nuclphysb.2019.01.001} {\bibfield  {journal} {\bibinfo  {journal} {Nuclear Physics B}\ }\textbf {\bibinfo {volume} {940}},\ \bibinfo {pages} {312–320} (\bibinfo {year} {2019})}\BibitemShut {NoStop}%
\bibitem [{\citenamefont {Banerjee}\ \emph {et~al.}(2024)\citenamefont {Banerjee}, \citenamefont {Mandal}, \citenamefont {Biswas},\ and\ \citenamefont {Biswas}}]{Banerjee_2024}%
  \BibitemOpen
  \bibfield  {author} {\bibinfo {author} {\bibfnamefont {T.}~\bibnamefont {Banerjee}}, \bibinfo {author} {\bibfnamefont {G.}~\bibnamefont {Mandal}}, \bibinfo {author} {\bibfnamefont {A.}~\bibnamefont {Biswas}},\ and\ \bibinfo {author} {\bibfnamefont {S.~K.}\ \bibnamefont {Biswas}},\ }\href {https://doi.org/10.1093/mnras/stae1047} {\bibfield  {journal} {\bibinfo  {journal} {Monthly Notices of the Royal Astronomical Society}\ }\textbf {\bibinfo {volume} {531}},\ \bibinfo {pages} {1–23} (\bibinfo {year} {2024})}\BibitemShut {NoStop}%
\bibitem [{\citenamefont {Guin}\ \emph {et~al.}(2025)\citenamefont {Guin}, \citenamefont {Paul},\ and\ \citenamefont {Gangopadhyay}}]{Guin:2025xki}%
  \BibitemOpen
  \bibfield  {author} {\bibinfo {author} {\bibfnamefont {G.}~\bibnamefont {Guin}}, \bibinfo {author} {\bibfnamefont {S.}~\bibnamefont {Paul}},\ and\ \bibinfo {author} {\bibfnamefont {S.}~\bibnamefont {Gangopadhyay}},\ }\href@noop {} {\bibinfo {title} {{Barrow holographic dark energy interacting model in the presence of radiation and matter}}} (\bibinfo {year} {2025}),\ \Eprint {https://arxiv.org/abs/2507.01070} {arXiv:2507.01070 [gr-qc]} \BibitemShut {NoStop}%
\bibitem [{\citenamefont {Zhang}(2009)}]{zhang2009crossingphantomdivide}%
  \BibitemOpen
  \bibfield  {author} {\bibinfo {author} {\bibfnamefont {H.}~\bibnamefont {Zhang}},\ }\href {https://arxiv.org/abs/0909.3013} {\bibinfo {title} {Crossing the phantom divide}} (\bibinfo {year} {2009}),\ \Eprint {https://arxiv.org/abs/0909.3013} {arXiv:0909.3013 [astro-ph.CO]} \BibitemShut {NoStop}%
\bibitem [{\citenamefont {Rezaei}(2020)}]{Rezaei_2020}%
  \BibitemOpen
  \bibfield  {author} {\bibinfo {author} {\bibfnamefont {Z.}~\bibnamefont {Rezaei}},\ }\href {https://doi.org/10.3847/1538-4357/abb59d} {\bibfield  {journal} {\bibinfo  {journal} {The Astrophysical Journal}\ }\textbf {\bibinfo {volume} {902}},\ \bibinfo {pages} {102} (\bibinfo {year} {2020})}\BibitemShut {NoStop}%
\bibitem [{\citenamefont {Zhao}\ \emph {et~al.}(2023)\citenamefont {Zhao}, \citenamefont {Liu}, \citenamefont {Liao}, \citenamefont {Zhang}, \citenamefont {Liu},\ and\ \citenamefont {Du}}]{Zhao_2023}%
  \BibitemOpen
  \bibfield  {author} {\bibinfo {author} {\bibfnamefont {Y.}~\bibnamefont {Zhao}}, \bibinfo {author} {\bibfnamefont {Y.}~\bibnamefont {Liu}}, \bibinfo {author} {\bibfnamefont {S.}~\bibnamefont {Liao}}, \bibinfo {author} {\bibfnamefont {J.}~\bibnamefont {Zhang}}, \bibinfo {author} {\bibfnamefont {X.}~\bibnamefont {Liu}},\ and\ \bibinfo {author} {\bibfnamefont {W.}~\bibnamefont {Du}},\ }\href {https://doi.org/10.1093/mnras/stad1814} {\bibfield  {journal} {\bibinfo  {journal} {Monthly Notices of the Royal Astronomical Society}\ }\textbf {\bibinfo {volume} {523}},\ \bibinfo {pages} {5962–5971} (\bibinfo {year} {2023})}\BibitemShut {NoStop}%
\bibitem [{\citenamefont {Nagpal}\ \emph {et~al.}(2025)\citenamefont {Nagpal}, \citenamefont {Pacif}, \citenamefont {Atamurotov},\ and\ \citenamefont {Pati}}]{nagpal2025darksectorinteractionsprobing}%
  \BibitemOpen
  \bibfield  {author} {\bibinfo {author} {\bibfnamefont {R.}~\bibnamefont {Nagpal}}, \bibinfo {author} {\bibfnamefont {S.~K.~J.}\ \bibnamefont {Pacif}}, \bibinfo {author} {\bibfnamefont {F.}~\bibnamefont {Atamurotov}},\ and\ \bibinfo {author} {\bibfnamefont {R.}~\bibnamefont {Pati}},\ }\href {https://arxiv.org/abs/2503.06319} {\bibinfo {title} {Dark sector interactions: Probing the hubble parameter and the sound horizon}} (\bibinfo {year} {2025}),\ \Eprint {https://arxiv.org/abs/2503.06319} {arXiv:2503.06319 [gr-qc]} \BibitemShut {NoStop}%
\bibitem [{\citenamefont {Bahamonde}\ \emph {et~al.}(2018)\citenamefont {Bahamonde}, \citenamefont {Böhmer}, \citenamefont {Carloni}, \citenamefont {Copeland}, \citenamefont {Fang},\ and\ \citenamefont {Tamanini}}]{Bahamonde_2018}%
  \BibitemOpen
  \bibfield  {author} {\bibinfo {author} {\bibfnamefont {S.}~\bibnamefont {Bahamonde}}, \bibinfo {author} {\bibfnamefont {C.~G.}\ \bibnamefont {Böhmer}}, \bibinfo {author} {\bibfnamefont {S.}~\bibnamefont {Carloni}}, \bibinfo {author} {\bibfnamefont {E.~J.}\ \bibnamefont {Copeland}}, \bibinfo {author} {\bibfnamefont {W.}~\bibnamefont {Fang}},\ and\ \bibinfo {author} {\bibfnamefont {N.}~\bibnamefont {Tamanini}},\ }\href {https://doi.org/10.1016/j.physrep.2018.09.001} {\bibfield  {journal} {\bibinfo  {journal} {Physics Reports}\ }\textbf {\bibinfo {volume} {775–777}},\ \bibinfo {pages} {1–122} (\bibinfo {year} {2018})}\BibitemShut {NoStop}%
\bibitem [{\citenamefont {Izquierdo}\ \emph {et~al.}(2018)\citenamefont {Izquierdo}, \citenamefont {Blanquet-Jaramillo},\ and\ \citenamefont {Sussman}}]{Izquierdo_2018}%
  \BibitemOpen
  \bibfield  {author} {\bibinfo {author} {\bibfnamefont {G.}~\bibnamefont {Izquierdo}}, \bibinfo {author} {\bibfnamefont {R.~C.}\ \bibnamefont {Blanquet-Jaramillo}},\ and\ \bibinfo {author} {\bibfnamefont {R.~A.}\ \bibnamefont {Sussman}},\ }\bibfield  {journal} {\bibinfo  {journal} {The European Physical Journal C}\ }\textbf {\bibinfo {volume} {78}},\ \href {https://doi.org/10.1140/epjc/s10052-018-5699-y} {10.1140/epjc/s10052-018-5699-y} (\bibinfo {year} {2018})\BibitemShut {NoStop}%
\bibitem [{\citenamefont {Panotopoulos}\ \emph {et~al.}(2020)\citenamefont {Panotopoulos}, \citenamefont {Rinc{\'o}n}, \citenamefont {Otalora},\ and\ \citenamefont {Videla}}]{Panotopoulos_2020}%
  \BibitemOpen
  \bibfield  {author} {\bibinfo {author} {\bibfnamefont {G.}~\bibnamefont {Panotopoulos}}, \bibinfo {author} {\bibfnamefont {{\'A}.}~\bibnamefont {Rinc{\'o}n}}, \bibinfo {author} {\bibfnamefont {G.}~\bibnamefont {Otalora}},\ and\ \bibinfo {author} {\bibfnamefont {N.}~\bibnamefont {Videla}},\ }\href {https://doi.org/10.1140/epjc/s10052-020-7828-7} {\bibfield  {journal} {\bibinfo  {journal} {Eur. Phys. J. C}\ }\textbf {\bibinfo {volume} {80}},\ \bibinfo {pages} {286} (\bibinfo {year} {2020})},\ \Eprint {https://arxiv.org/abs/1912.01723} {arXiv:1912.01723 [gr-qc]} \BibitemShut {NoStop}%
\bibitem [{\citenamefont {Deogharia}\ \emph {et~al.}(2021)\citenamefont {Deogharia}, \citenamefont {Bandyopadhyay},\ and\ \citenamefont {Biswas}}]{Deogharia_2021}%
  \BibitemOpen
  \bibfield  {author} {\bibinfo {author} {\bibfnamefont {G.}~\bibnamefont {Deogharia}}, \bibinfo {author} {\bibfnamefont {M.}~\bibnamefont {Bandyopadhyay}},\ and\ \bibinfo {author} {\bibfnamefont {R.}~\bibnamefont {Biswas}},\ }\bibfield  {journal} {\bibinfo  {journal} {Modern Physics Letters A}\ }\textbf {\bibinfo {volume} {36}},\ \href {https://doi.org/10.1142/s0217732321502758} {10.1142/s0217732321502758} (\bibinfo {year} {2021})\BibitemShut {NoStop}%
\bibitem [{\citenamefont {von Marttens}\ \emph {et~al.}(2020{\natexlab{b}})\citenamefont {von Marttens}, \citenamefont {Borges}, \citenamefont {Carneiro}, \citenamefont {Alcaniz},\ and\ \citenamefont {Zimdahl}}]{von_Marttens_2020}%
  \BibitemOpen
  \bibfield  {author} {\bibinfo {author} {\bibfnamefont {R.}~\bibnamefont {von Marttens}}, \bibinfo {author} {\bibfnamefont {H.~A.}\ \bibnamefont {Borges}}, \bibinfo {author} {\bibfnamefont {S.}~\bibnamefont {Carneiro}}, \bibinfo {author} {\bibfnamefont {J.~S.}\ \bibnamefont {Alcaniz}},\ and\ \bibinfo {author} {\bibfnamefont {W.}~\bibnamefont {Zimdahl}},\ }\bibfield  {journal} {\bibinfo  {journal} {The European Physical Journal C}\ }\textbf {\bibinfo {volume} {80}},\ \href {https://doi.org/10.1140/epjc/s10052-020-08682-5} {10.1140/epjc/s10052-020-08682-5} (\bibinfo {year} {2020}{\natexlab{b}})\BibitemShut {NoStop}%
\bibitem [{\citenamefont {Marcondes}\ \emph {et~al.}(2016)\citenamefont {Marcondes}, \citenamefont {Landim}, \citenamefont {Costa}, \citenamefont {Wang},\ and\ \citenamefont {Abdalla}}]{Marcondes_2016}%
  \BibitemOpen
  \bibfield  {author} {\bibinfo {author} {\bibfnamefont {R.~J.}\ \bibnamefont {Marcondes}}, \bibinfo {author} {\bibfnamefont {R.~C.}\ \bibnamefont {Landim}}, \bibinfo {author} {\bibfnamefont {A.~A.}\ \bibnamefont {Costa}}, \bibinfo {author} {\bibfnamefont {B.}~\bibnamefont {Wang}},\ and\ \bibinfo {author} {\bibfnamefont {E.}~\bibnamefont {Abdalla}},\ }\href {https://doi.org/10.1088/1475-7516/2016/12/009} {\bibfield  {journal} {\bibinfo  {journal} {Journal of Cosmology and Astroparticle Physics}\ }\textbf {\bibinfo {volume} {2016}}\bibinfo  {number} { (12)},\ \bibinfo {pages} {009–009}}\BibitemShut {NoStop}%
\bibitem [{\citenamefont {Yang}\ \emph {et~al.}(2016)\citenamefont {Yang}, \citenamefont {Li}, \citenamefont {Wu},\ and\ \citenamefont {Lu}}]{Yang_2016}%
  \BibitemOpen
\bibfield  {number} {  }\bibfield  {author} {\bibinfo {author} {\bibfnamefont {W.}~\bibnamefont {Yang}}, \bibinfo {author} {\bibfnamefont {H.}~\bibnamefont {Li}}, \bibinfo {author} {\bibfnamefont {Y.}~\bibnamefont {Wu}},\ and\ \bibinfo {author} {\bibfnamefont {J.}~\bibnamefont {Lu}},\ }\href {https://doi.org/10.1088/1475-7516/2016/10/007} {\bibfield  {journal} {\bibinfo  {journal} {Journal of Cosmology and Astroparticle Physics}\ }\textbf {\bibinfo {volume} {2016}}\bibinfo  {number} { (10)},\ \bibinfo {pages} {007–007}}\BibitemShut {NoStop}%
\bibitem [{\citenamefont {Silva}\ \emph {et~al.}(2024{\natexlab{b}})\citenamefont {Silva}, \citenamefont {Zúñiga-Bolaño}, \citenamefont {Nunes},\ and\ \citenamefont {Di~Valentino}}]{Silva_2024}%
  \BibitemOpen
\bibfield  {number} {  }\bibfield  {author} {\bibinfo {author} {\bibfnamefont {E.}~\bibnamefont {Silva}}, \bibinfo {author} {\bibfnamefont {U.}~\bibnamefont {Zúñiga-Bolaño}}, \bibinfo {author} {\bibfnamefont {R.~C.}\ \bibnamefont {Nunes}},\ and\ \bibinfo {author} {\bibfnamefont {E.}~\bibnamefont {Di~Valentino}},\ }\bibfield  {journal} {\bibinfo  {journal} {The European Physical Journal C}\ }\textbf {\bibinfo {volume} {84}},\ \href {https://doi.org/10.1140/epjc/s10052-024-13487-x} {10.1140/epjc/s10052-024-13487-x} (\bibinfo {year} {2024}{\natexlab{b}})\BibitemShut {NoStop}%
\bibitem [{\citenamefont {Salvatelli}\ \emph {et~al.}(2013)\citenamefont {Salvatelli}, \citenamefont {Marchini}, \citenamefont {Lopez-Honorez},\ and\ \citenamefont {Mena}}]{Salvatelli_2013}%
  \BibitemOpen
  \bibfield  {author} {\bibinfo {author} {\bibfnamefont {V.}~\bibnamefont {Salvatelli}}, \bibinfo {author} {\bibfnamefont {A.}~\bibnamefont {Marchini}}, \bibinfo {author} {\bibfnamefont {L.}~\bibnamefont {Lopez-Honorez}},\ and\ \bibinfo {author} {\bibfnamefont {O.}~\bibnamefont {Mena}},\ }\bibfield  {journal} {\bibinfo  {journal} {Physical Review D}\ }\textbf {\bibinfo {volume} {88}},\ \href {https://doi.org/10.1103/physrevd.88.023531} {10.1103/physrevd.88.023531} (\bibinfo {year} {2013})\BibitemShut {NoStop}%
\bibitem [{\citenamefont {Di~Valentino}\ \emph {et~al.}(2017{\natexlab{b}})\citenamefont {Di~Valentino}, \citenamefont {Melchiorri},\ and\ \citenamefont {Mena}}]{Di_Valentino_2017}%
  \BibitemOpen
  \bibfield  {author} {\bibinfo {author} {\bibfnamefont {E.}~\bibnamefont {Di~Valentino}}, \bibinfo {author} {\bibfnamefont {A.}~\bibnamefont {Melchiorri}},\ and\ \bibinfo {author} {\bibfnamefont {O.}~\bibnamefont {Mena}},\ }\bibfield  {journal} {\bibinfo  {journal} {Physical Review D}\ }\textbf {\bibinfo {volume} {96}},\ \href {https://doi.org/10.1103/physrevd.96.043503} {10.1103/physrevd.96.043503} (\bibinfo {year} {2017}{\natexlab{b}})\BibitemShut {NoStop}%
\bibitem [{\citenamefont {Yang}\ \emph {et~al.}(2020{\natexlab{b}})\citenamefont {Yang}, \citenamefont {Pan}, \citenamefont {Nunes},\ and\ \citenamefont {Mota}}]{Yang_2020_2}%
  \BibitemOpen
  \bibfield  {author} {\bibinfo {author} {\bibfnamefont {W.}~\bibnamefont {Yang}}, \bibinfo {author} {\bibfnamefont {S.}~\bibnamefont {Pan}}, \bibinfo {author} {\bibfnamefont {R.~C.}\ \bibnamefont {Nunes}},\ and\ \bibinfo {author} {\bibfnamefont {D.~F.}\ \bibnamefont {Mota}},\ }\href {https://doi.org/10.1088/1475-7516/2020/04/008} {\bibfield  {journal} {\bibinfo  {journal} {Journal of Cosmology and Astroparticle Physics}\ }\textbf {\bibinfo {volume} {2020}}\bibinfo  {number} { (04)},\ \bibinfo {pages} {008–008}}\BibitemShut {NoStop}%
\bibitem [{\citenamefont {Lucca}\ and\ \citenamefont {Hooper}(2020{\natexlab{b}})}]{Lucca_2020}%
  \BibitemOpen
\bibfield  {number} {  }\bibfield  {author} {\bibinfo {author} {\bibfnamefont {M.}~\bibnamefont {Lucca}}\ and\ \bibinfo {author} {\bibfnamefont {D.~C.}\ \bibnamefont {Hooper}},\ }\bibfield  {journal} {\bibinfo  {journal} {Physical Review D}\ }\textbf {\bibinfo {volume} {102}},\ \href {https://doi.org/10.1103/physrevd.102.123502} {10.1103/physrevd.102.123502} (\bibinfo {year} {2020}{\natexlab{b}})\BibitemShut {NoStop}%
\bibitem [{\citenamefont {Di~Valentino}\ \emph {et~al.}(2020{\natexlab{c}})\citenamefont {Di~Valentino}, \citenamefont {Melchiorri}, \citenamefont {Mena},\ and\ \citenamefont {Vagnozzi}}]{Di_Valentino_2020}%
  \BibitemOpen
  \bibfield  {author} {\bibinfo {author} {\bibfnamefont {E.}~\bibnamefont {Di~Valentino}}, \bibinfo {author} {\bibfnamefont {A.}~\bibnamefont {Melchiorri}}, \bibinfo {author} {\bibfnamefont {O.}~\bibnamefont {Mena}},\ and\ \bibinfo {author} {\bibfnamefont {S.}~\bibnamefont {Vagnozzi}},\ }\bibfield  {journal} {\bibinfo  {journal} {Physical Review D}\ }\textbf {\bibinfo {volume} {101}},\ \href {https://doi.org/10.1103/physrevd.101.063502} {10.1103/physrevd.101.063502} (\bibinfo {year} {2020}{\natexlab{c}})\BibitemShut {NoStop}%
\bibitem [{\citenamefont {Yang}\ \emph {et~al.}(2023)\citenamefont {Yang}, \citenamefont {Pan}, \citenamefont {Mena},\ and\ \citenamefont {Di~Valentino}}]{Yang_2023}%
  \BibitemOpen
  \bibfield  {author} {\bibinfo {author} {\bibfnamefont {W.}~\bibnamefont {Yang}}, \bibinfo {author} {\bibfnamefont {S.}~\bibnamefont {Pan}}, \bibinfo {author} {\bibfnamefont {O.}~\bibnamefont {Mena}},\ and\ \bibinfo {author} {\bibfnamefont {E.}~\bibnamefont {Di~Valentino}},\ }\href {https://doi.org/10.1016/j.jheap.2023.09.001} {\bibfield  {journal} {\bibinfo  {journal} {Journal of High Energy Astrophysics}\ }\textbf {\bibinfo {volume} {40}},\ \bibinfo {pages} {19–40} (\bibinfo {year} {2023})}\BibitemShut {NoStop}%
\bibitem [{\citenamefont {Bernui}\ \emph {et~al.}(2023{\natexlab{b}})\citenamefont {Bernui}, \citenamefont {Di~Valentino}, \citenamefont {Giarè}, \citenamefont {Kumar},\ and\ \citenamefont {Nunes}}]{Bernui_2023}%
  \BibitemOpen
  \bibfield  {author} {\bibinfo {author} {\bibfnamefont {A.}~\bibnamefont {Bernui}}, \bibinfo {author} {\bibfnamefont {E.}~\bibnamefont {Di~Valentino}}, \bibinfo {author} {\bibfnamefont {W.}~\bibnamefont {Giarè}}, \bibinfo {author} {\bibfnamefont {S.}~\bibnamefont {Kumar}},\ and\ \bibinfo {author} {\bibfnamefont {R.~C.}\ \bibnamefont {Nunes}},\ }\bibfield  {journal} {\bibinfo  {journal} {Physical Review D}\ }\textbf {\bibinfo {volume} {107}},\ \href {https://doi.org/10.1103/physrevd.107.103531} {10.1103/physrevd.107.103531} (\bibinfo {year} {2023}{\natexlab{b}})\BibitemShut {NoStop}%
\bibitem [{\citenamefont {Giar\`e}\ \emph {et~al.}(2024{\natexlab{c}})\citenamefont {Giar\`e}, \citenamefont {Sabogal}, \citenamefont {Nunes},\ and\ \citenamefont {Di~Valentino}}]{Giare:2024smz}%
  \BibitemOpen
  \bibfield  {author} {\bibinfo {author} {\bibfnamefont {W.}~\bibnamefont {Giar\`e}}, \bibinfo {author} {\bibfnamefont {M.~A.}\ \bibnamefont {Sabogal}}, \bibinfo {author} {\bibfnamefont {R.~C.}\ \bibnamefont {Nunes}},\ and\ \bibinfo {author} {\bibfnamefont {E.}~\bibnamefont {Di~Valentino}},\ }\href@noop {} {\bibinfo {title} {{Interacting Dark Energy after DESI Baryon Acoustic Oscillation measurements}}} (\bibinfo {year} {2024}{\natexlab{c}}),\ \Eprint {https://arxiv.org/abs/2404.15232} {arXiv:2404.15232 [astro-ph.CO]} \BibitemShut {NoStop}%
\bibitem [{\citenamefont {Clemson}\ \emph {et~al.}(2012)\citenamefont {Clemson}, \citenamefont {Koyama}, \citenamefont {Zhao}, \citenamefont {Maartens},\ and\ \citenamefont {Väliviita}}]{Clemson_2012}%
  \BibitemOpen
  \bibfield  {author} {\bibinfo {author} {\bibfnamefont {T.}~\bibnamefont {Clemson}}, \bibinfo {author} {\bibfnamefont {K.}~\bibnamefont {Koyama}}, \bibinfo {author} {\bibfnamefont {G.-B.}\ \bibnamefont {Zhao}}, \bibinfo {author} {\bibfnamefont {R.}~\bibnamefont {Maartens}},\ and\ \bibinfo {author} {\bibfnamefont {J.}~\bibnamefont {Väliviita}},\ }\bibfield  {journal} {\bibinfo  {journal} {Physical Review D}\ }\textbf {\bibinfo {volume} {85}},\ \href {https://doi.org/10.1103/physrevd.85.043007} {10.1103/physrevd.85.043007} (\bibinfo {year} {2012})\BibitemShut {NoStop}%
\bibitem [{\citenamefont {Nunes}\ \emph {et~al.}(2022)\citenamefont {Nunes}, \citenamefont {Vagnozzi}, \citenamefont {Kumar}, \citenamefont {Di~Valentino},\ and\ \citenamefont {Mena}}]{Nunes_2022}%
  \BibitemOpen
  \bibfield  {author} {\bibinfo {author} {\bibfnamefont {R.~C.}\ \bibnamefont {Nunes}}, \bibinfo {author} {\bibfnamefont {S.}~\bibnamefont {Vagnozzi}}, \bibinfo {author} {\bibfnamefont {S.}~\bibnamefont {Kumar}}, \bibinfo {author} {\bibfnamefont {E.}~\bibnamefont {Di~Valentino}},\ and\ \bibinfo {author} {\bibfnamefont {O.}~\bibnamefont {Mena}},\ }\bibfield  {journal} {\bibinfo  {journal} {Physical Review D}\ }\textbf {\bibinfo {volume} {105}},\ \href {https://doi.org/10.1103/physrevd.105.123506} {10.1103/physrevd.105.123506} (\bibinfo {year} {2022})\BibitemShut {NoStop}%
\bibitem [{\citenamefont {Zhai}\ \emph {et~al.}(2023{\natexlab{b}})\citenamefont {Zhai}, \citenamefont {Giarè}, \citenamefont {van~de Bruck}, \citenamefont {Valentino}, \citenamefont {Mena},\ and\ \citenamefont {Nunes}}]{Zhai_2023}%
  \BibitemOpen
  \bibfield  {author} {\bibinfo {author} {\bibfnamefont {Y.}~\bibnamefont {Zhai}}, \bibinfo {author} {\bibfnamefont {W.}~\bibnamefont {Giarè}}, \bibinfo {author} {\bibfnamefont {C.}~\bibnamefont {van~de Bruck}}, \bibinfo {author} {\bibfnamefont {E.~D.}\ \bibnamefont {Valentino}}, \bibinfo {author} {\bibfnamefont {O.}~\bibnamefont {Mena}},\ and\ \bibinfo {author} {\bibfnamefont {R.~C.}\ \bibnamefont {Nunes}},\ }\href {https://doi.org/10.1088/1475-7516/2023/07/032} {\bibfield  {journal} {\bibinfo  {journal} {Journal of Cosmology and Astroparticle Physics}\ }\textbf {\bibinfo {volume} {2023}}\bibinfo  {number} { (07)},\ \bibinfo {pages} {032}}\BibitemShut {NoStop}%
\bibitem [{\citenamefont {Giarè}\ \emph {et~al.}(2024)\citenamefont {Giarè}, \citenamefont {Zhai}, \citenamefont {Pan}, \citenamefont {Di~Valentino}, \citenamefont {Nunes},\ and\ \citenamefont {van~de Bruck}}]{Giar__2024}%
  \BibitemOpen
\bibfield  {number} {  }\bibfield  {author} {\bibinfo {author} {\bibfnamefont {W.}~\bibnamefont {Giarè}}, \bibinfo {author} {\bibfnamefont {Y.}~\bibnamefont {Zhai}}, \bibinfo {author} {\bibfnamefont {S.}~\bibnamefont {Pan}}, \bibinfo {author} {\bibfnamefont {E.}~\bibnamefont {Di~Valentino}}, \bibinfo {author} {\bibfnamefont {R.~C.}\ \bibnamefont {Nunes}},\ and\ \bibinfo {author} {\bibfnamefont {C.}~\bibnamefont {van~de Bruck}},\ }\bibfield  {journal} {\bibinfo  {journal} {Physical Review D}\ }\textbf {\bibinfo {volume} {110}},\ \href {https://doi.org/10.1103/physrevd.110.063527} {10.1103/physrevd.110.063527} (\bibinfo {year} {2024})\BibitemShut {NoStop}%
\bibitem [{\citenamefont {Yang}\ \emph {et~al.}(2020{\natexlab{c}})\citenamefont {Yang}, \citenamefont {Pan}, \citenamefont {Valentino}, \citenamefont {Wang},\ and\ \citenamefont {Wang}}]{Yang_2020_3}%
  \BibitemOpen
  \bibfield  {author} {\bibinfo {author} {\bibfnamefont {W.}~\bibnamefont {Yang}}, \bibinfo {author} {\bibfnamefont {S.}~\bibnamefont {Pan}}, \bibinfo {author} {\bibfnamefont {E.~D.}\ \bibnamefont {Valentino}}, \bibinfo {author} {\bibfnamefont {B.}~\bibnamefont {Wang}},\ and\ \bibinfo {author} {\bibfnamefont {A.}~\bibnamefont {Wang}},\ }\href {https://doi.org/10.1088/1475-7516/2020/05/050} {\bibfield  {journal} {\bibinfo  {journal} {Journal of Cosmology and Astroparticle Physics}\ }\textbf {\bibinfo {volume} {2020}}\bibinfo  {number} { (05)},\ \bibinfo {pages} {050–050}}\BibitemShut {NoStop}%
\bibitem [{\citenamefont {Feng}\ \emph {et~al.}(2020{\natexlab{b}})\citenamefont {Feng}, \citenamefont {Li}, \citenamefont {Zhang},\ and\ \citenamefont {Zhang}}]{Feng_2019}%
  \BibitemOpen
\bibfield  {number} {  }\bibfield  {author} {\bibinfo {author} {\bibfnamefont {L.}~\bibnamefont {Feng}}, \bibinfo {author} {\bibfnamefont {H.-L.}\ \bibnamefont {Li}}, \bibinfo {author} {\bibfnamefont {J.-F.}\ \bibnamefont {Zhang}},\ and\ \bibinfo {author} {\bibfnamefont {X.}~\bibnamefont {Zhang}},\ }\href {https://doi.org/10.1007/s11433-019-9431-9} {\bibfield  {journal} {\bibinfo  {journal} {Sci. China Phys. Mech. Astron.}\ }\textbf {\bibinfo {volume} {63}},\ \bibinfo {pages} {220401} (\bibinfo {year} {2020}{\natexlab{b}})},\ \Eprint {https://arxiv.org/abs/1903.08848} {arXiv:1903.08848 [astro-ph.CO]} \BibitemShut {NoStop}%
\bibitem [{\citenamefont {Yang}\ \emph {et~al.}(2020{\natexlab{d}})\citenamefont {Yang}, \citenamefont {Di~Valentino}, \citenamefont {Mena}, \citenamefont {Pan},\ and\ \citenamefont {Nunes}}]{Yang_2020}%
  \BibitemOpen
  \bibfield  {author} {\bibinfo {author} {\bibfnamefont {W.}~\bibnamefont {Yang}}, \bibinfo {author} {\bibfnamefont {E.}~\bibnamefont {Di~Valentino}}, \bibinfo {author} {\bibfnamefont {O.}~\bibnamefont {Mena}}, \bibinfo {author} {\bibfnamefont {S.}~\bibnamefont {Pan}},\ and\ \bibinfo {author} {\bibfnamefont {R.~C.}\ \bibnamefont {Nunes}},\ }\bibfield  {journal} {\bibinfo  {journal} {Physical Review D}\ }\textbf {\bibinfo {volume} {101}},\ \href {https://doi.org/10.1103/physrevd.101.083509} {10.1103/physrevd.101.083509} (\bibinfo {year} {2020}{\natexlab{d}})\BibitemShut {NoStop}%
\bibitem [{\citenamefont {Silva}\ \emph {et~al.}(2025{\natexlab{b}})\citenamefont {Silva}, \citenamefont {Sabogal}, \citenamefont {Souza}, \citenamefont {Nunes}, \citenamefont {Valentino},\ and\ \citenamefont {Kumar}}]{silva2025newconstraintsinteractingdark}%
  \BibitemOpen
  \bibfield  {author} {\bibinfo {author} {\bibfnamefont {E.}~\bibnamefont {Silva}}, \bibinfo {author} {\bibfnamefont {M.~A.}\ \bibnamefont {Sabogal}}, \bibinfo {author} {\bibfnamefont {M.~S.}\ \bibnamefont {Souza}}, \bibinfo {author} {\bibfnamefont {R.~C.}\ \bibnamefont {Nunes}}, \bibinfo {author} {\bibfnamefont {E.~D.}\ \bibnamefont {Valentino}},\ and\ \bibinfo {author} {\bibfnamefont {S.}~\bibnamefont {Kumar}},\ }\href {https://arxiv.org/abs/2503.23225} {\bibinfo {title} {New constraints on interacting dark energy from desi dr2 bao observations}} (\bibinfo {year} {2025}{\natexlab{b}}),\ \Eprint {https://arxiv.org/abs/2503.23225} {arXiv:2503.23225 [astro-ph.CO]} \BibitemShut {NoStop}%
\bibitem [{\citenamefont {Nayak}(2020)}]{Nayak_2020}%
  \BibitemOpen
  \bibfield  {author} {\bibinfo {author} {\bibfnamefont {B.}~\bibnamefont {Nayak}},\ }\href {https://doi.org/10.1134/s020228932003010x} {\bibfield  {journal} {\bibinfo  {journal} {Gravitation and Cosmology}\ }\textbf {\bibinfo {volume} {26}},\ \bibinfo {pages} {273–280} (\bibinfo {year} {2020})}\BibitemShut {NoStop}%
\bibitem [{\citenamefont {Sinha}\ and\ \citenamefont {Banerjee}(2020)}]{Sinha_2020}%
  \BibitemOpen
  \bibfield  {author} {\bibinfo {author} {\bibfnamefont {S.}~\bibnamefont {Sinha}}\ and\ \bibinfo {author} {\bibfnamefont {N.}~\bibnamefont {Banerjee}},\ }\bibfield  {journal} {\bibinfo  {journal} {The European Physical Journal Plus}\ }\textbf {\bibinfo {volume} {135}},\ \href {https://doi.org/10.1140/epjp/s13360-020-00803-z} {10.1140/epjp/s13360-020-00803-z} (\bibinfo {year} {2020})\BibitemShut {NoStop}%
\bibitem [{\citenamefont {Ghodsi~Yengejeh}\ \emph {et~al.}(2023)\citenamefont {Ghodsi~Yengejeh}, \citenamefont {Fakhry}, \citenamefont {T.~Firouzjaee},\ and\ \citenamefont {Fathi}}]{Ghodsi_Yengejeh_2023}%
  \BibitemOpen
  \bibfield  {author} {\bibinfo {author} {\bibfnamefont {M.}~\bibnamefont {Ghodsi~Yengejeh}}, \bibinfo {author} {\bibfnamefont {S.}~\bibnamefont {Fakhry}}, \bibinfo {author} {\bibfnamefont {J.}~\bibnamefont {T.~Firouzjaee}},\ and\ \bibinfo {author} {\bibfnamefont {H.}~\bibnamefont {Fathi}},\ }\href {https://doi.org/10.1016/j.dark.2022.101144} {\bibfield  {journal} {\bibinfo  {journal} {Physics of the Dark Universe}\ }\textbf {\bibinfo {volume} {39}},\ \bibinfo {pages} {101144} (\bibinfo {year} {2023})}\BibitemShut {NoStop}%
\bibitem [{\citenamefont {Di~Valentino}\ \emph {et~al.}(2021{\natexlab{b}})\citenamefont {Di~Valentino}, \citenamefont {Melchiorri}, \citenamefont {Mena}, \citenamefont {Pan},\ and\ \citenamefont {Yang}}]{Di_Valentino_2021_closed}%
  \BibitemOpen
  \bibfield  {author} {\bibinfo {author} {\bibfnamefont {E.}~\bibnamefont {Di~Valentino}}, \bibinfo {author} {\bibfnamefont {A.}~\bibnamefont {Melchiorri}}, \bibinfo {author} {\bibfnamefont {O.}~\bibnamefont {Mena}}, \bibinfo {author} {\bibfnamefont {S.}~\bibnamefont {Pan}},\ and\ \bibinfo {author} {\bibfnamefont {W.}~\bibnamefont {Yang}},\ }\href {https://doi.org/10.1093/mnrasl/slaa207} {\bibfield  {journal} {\bibinfo  {journal} {Monthly Notices of the Royal Astronomical Society: Letters}\ }\textbf {\bibinfo {volume} {502}},\ \bibinfo {pages} {L23–L28} (\bibinfo {year} {2021}{\natexlab{b}})}\BibitemShut {NoStop}%
\bibitem [{\citenamefont {Yang}\ \emph {et~al.}(2021)\citenamefont {Yang}, \citenamefont {Pan}, \citenamefont {Di~Valentino}, \citenamefont {Mena},\ and\ \citenamefont {Melchiorri}}]{Yang_2021}%
  \BibitemOpen
  \bibfield  {author} {\bibinfo {author} {\bibfnamefont {W.}~\bibnamefont {Yang}}, \bibinfo {author} {\bibfnamefont {S.}~\bibnamefont {Pan}}, \bibinfo {author} {\bibfnamefont {E.}~\bibnamefont {Di~Valentino}}, \bibinfo {author} {\bibfnamefont {O.}~\bibnamefont {Mena}},\ and\ \bibinfo {author} {\bibfnamefont {A.}~\bibnamefont {Melchiorri}},\ }\href {https://doi.org/10.1088/1475-7516/2021/10/008} {\bibfield  {journal} {\bibinfo  {journal} {Journal of Cosmology and Astroparticle Physics}\ }\textbf {\bibinfo {volume} {2021}}\bibinfo  {number} { (10)},\ \bibinfo {pages} {008}}\BibitemShut {NoStop}%
\bibitem [{\citenamefont {Joseph}\ and\ \citenamefont {Saha}(2022)}]{Joseph_2022}%
  \BibitemOpen
\bibfield  {number} {  }\bibfield  {author} {\bibinfo {author} {\bibfnamefont {A.}~\bibnamefont {Joseph}}\ and\ \bibinfo {author} {\bibfnamefont {R.}~\bibnamefont {Saha}},\ }\href {https://doi.org/10.1093/mnras/stac3586} {\bibfield  {journal} {\bibinfo  {journal} {Monthly Notices of the Royal Astronomical Society}\ }\textbf {\bibinfo {volume} {519}},\ \bibinfo {pages} {1809–1822} (\bibinfo {year} {2022})}\BibitemShut {NoStop}%
\bibitem [{\citenamefont {Forconi}\ \emph {et~al.}(2024{\natexlab{b}})\citenamefont {Forconi}, \citenamefont {Giarè}, \citenamefont {Mena}, \citenamefont {Ruchika}, \citenamefont {Valentino}, \citenamefont {Melchiorri},\ and\ \citenamefont {Nunes}}]{Forconi_2024}%
  \BibitemOpen
  \bibfield  {author} {\bibinfo {author} {\bibfnamefont {M.}~\bibnamefont {Forconi}}, \bibinfo {author} {\bibfnamefont {W.}~\bibnamefont {Giarè}}, \bibinfo {author} {\bibfnamefont {O.}~\bibnamefont {Mena}}, \bibinfo {author} {\bibnamefont {Ruchika}}, \bibinfo {author} {\bibfnamefont {E.~D.}\ \bibnamefont {Valentino}}, \bibinfo {author} {\bibfnamefont {A.}~\bibnamefont {Melchiorri}},\ and\ \bibinfo {author} {\bibfnamefont {R.~C.}\ \bibnamefont {Nunes}},\ }\href {https://doi.org/10.1088/1475-7516/2024/05/097} {\bibfield  {journal} {\bibinfo  {journal} {Journal of Cosmology and Astroparticle Physics}\ }\textbf {\bibinfo {volume} {2024}}\bibinfo  {number} { (05)},\ \bibinfo {pages} {097}}\BibitemShut {NoStop}%
\bibitem [{\citenamefont {Mbewe}\ \emph {et~al.}(2024)\citenamefont {Mbewe}, \citenamefont {Mekuria}, \citenamefont {Sahlu},\ and\ \citenamefont {Abebe}}]{mbewe2024viscouscosmologicalfluidslargescale}%
  \BibitemOpen
\bibfield  {number} {  }\bibfield  {author} {\bibinfo {author} {\bibfnamefont {B.}~\bibnamefont {Mbewe}}, \bibinfo {author} {\bibfnamefont {R.}~\bibnamefont {Mekuria}}, \bibinfo {author} {\bibfnamefont {S.}~\bibnamefont {Sahlu}},\ and\ \bibinfo {author} {\bibfnamefont {A.}~\bibnamefont {Abebe}},\ }\href {https://arxiv.org/abs/2412.02276} {\bibinfo {title} {Viscous cosmological fluids and large-scale structure}} (\bibinfo {year} {2024}),\ \Eprint {https://arxiv.org/abs/2412.02276} {arXiv:2412.02276 [gr-qc]} \BibitemShut {NoStop}%
\bibitem [{\citenamefont {Arevalo}\ \emph {et~al.}(2012)\citenamefont {Arevalo}, \citenamefont {Bacalhau},\ and\ \citenamefont {Zimdahl}}]{Arevalo:2011hh}%
  \BibitemOpen
  \bibfield  {author} {\bibinfo {author} {\bibfnamefont {F.}~\bibnamefont {Arevalo}}, \bibinfo {author} {\bibfnamefont {A.~P.~R.}\ \bibnamefont {Bacalhau}},\ and\ \bibinfo {author} {\bibfnamefont {W.}~\bibnamefont {Zimdahl}},\ }\href {https://doi.org/10.1088/0264-9381/29/23/235001} {\bibfield  {journal} {\bibinfo  {journal} {Class. Quant. Grav.}\ }\textbf {\bibinfo {volume} {29}},\ \bibinfo {pages} {235001} (\bibinfo {year} {2012})},\ \Eprint {https://arxiv.org/abs/1112.5095} {arXiv:1112.5095 [astro-ph.CO]} \BibitemShut {NoStop}%
\bibitem [{\citenamefont {Li}\ \emph {et~al.}(2014)\citenamefont {Li}, \citenamefont {Zhang},\ and\ \citenamefont {Zhang}}]{Li_2014}%
  \BibitemOpen
  \bibfield  {author} {\bibinfo {author} {\bibfnamefont {Y.-H.}\ \bibnamefont {Li}}, \bibinfo {author} {\bibfnamefont {J.-F.}\ \bibnamefont {Zhang}},\ and\ \bibinfo {author} {\bibfnamefont {X.}~\bibnamefont {Zhang}},\ }\bibfield  {journal} {\bibinfo  {journal} {Physical Review D}\ }\textbf {\bibinfo {volume} {90}},\ \href {https://doi.org/10.1103/physrevd.90.063005} {10.1103/physrevd.90.063005} (\bibinfo {year} {2014})\BibitemShut {NoStop}%
\bibitem [{\citenamefont {Bolotin}\ \emph {et~al.}(2015)\citenamefont {Bolotin}, \citenamefont {Kostenko}, \citenamefont {Lemets},\ and\ \citenamefont {Yerokhin}}]{Bolotin_2015}%
  \BibitemOpen
  \bibfield  {author} {\bibinfo {author} {\bibfnamefont {Y.~L.}\ \bibnamefont {Bolotin}}, \bibinfo {author} {\bibfnamefont {A.}~\bibnamefont {Kostenko}}, \bibinfo {author} {\bibfnamefont {O.~A.}\ \bibnamefont {Lemets}},\ and\ \bibinfo {author} {\bibfnamefont {D.~A.}\ \bibnamefont {Yerokhin}},\ }\href {https://doi.org/10.1142/s0218271815300074} {\bibfield  {journal} {\bibinfo  {journal} {International Journal of Modern Physics D}\ }\textbf {\bibinfo {volume} {24}},\ \bibinfo {pages} {1530007} (\bibinfo {year} {2015})}\BibitemShut {NoStop}%
\bibitem [{\citenamefont {Li}\ and\ \citenamefont {Zhang}(2023)}]{Li_2023}%
  \BibitemOpen
  \bibfield  {author} {\bibinfo {author} {\bibfnamefont {Y.-H.}\ \bibnamefont {Li}}\ and\ \bibinfo {author} {\bibfnamefont {X.}~\bibnamefont {Zhang}},\ }\href {https://doi.org/10.1088/1475-7516/2023/09/046} {\bibfield  {journal} {\bibinfo  {journal} {Journal of Cosmology and Astroparticle Physics}\ }\textbf {\bibinfo {volume} {2023}}\bibinfo  {number} { (09)},\ \bibinfo {pages} {046}}\BibitemShut {NoStop}%
\bibitem [{\citenamefont {Lip}(2011)}]{Lip:2010dr}%
  \BibitemOpen
\bibfield  {number} {  }\bibfield  {author} {\bibinfo {author} {\bibfnamefont {S.~Z.~W.}\ \bibnamefont {Lip}},\ }\href {https://doi.org/10.1103/PhysRevD.83.023528} {\bibfield  {journal} {\bibinfo  {journal} {Phys. Rev. D}\ }\textbf {\bibinfo {volume} {83}},\ \bibinfo {pages} {023528} (\bibinfo {year} {2011})},\ \Eprint {https://arxiv.org/abs/1009.4942} {arXiv:1009.4942 [gr-qc]} \BibitemShut {NoStop}%
\bibitem [{\citenamefont {Paliathanasis}\ \emph {et~al.}(2024)\citenamefont {Paliathanasis}, \citenamefont {Duffy}, \citenamefont {Halder},\ and\ \citenamefont {Abebe}}]{paliathanasis2024compartmentalizationcoexistencedarksector}%
  \BibitemOpen
  \bibfield  {author} {\bibinfo {author} {\bibfnamefont {A.}~\bibnamefont {Paliathanasis}}, \bibinfo {author} {\bibfnamefont {K.}~\bibnamefont {Duffy}}, \bibinfo {author} {\bibfnamefont {A.}~\bibnamefont {Halder}},\ and\ \bibinfo {author} {\bibfnamefont {A.}~\bibnamefont {Abebe}},\ }\href {https://arxiv.org/abs/2409.05348} {\bibinfo {title} {Compartmentalization and coexistence in the dark sector of the universe}} (\bibinfo {year} {2024}),\ \Eprint {https://arxiv.org/abs/2409.05348} {arXiv:2409.05348 [gr-qc]} \BibitemShut {NoStop}%
\bibitem [{\citenamefont {van~der Westhuizen}\ \emph {et~al.}(2025{\natexlab{c}})\citenamefont {van~der Westhuizen}, \citenamefont {Figueruelo}, \citenamefont {Thubisi}, \citenamefont {Sahlu}, \citenamefont {Abebe},\ and\ \citenamefont {Paliathanasis}}]{vanderwesthuizen2025compartmentalizationdarksectoruniverse}%
  \BibitemOpen
  \bibfield  {author} {\bibinfo {author} {\bibfnamefont {M.}~\bibnamefont {van~der Westhuizen}}, \bibinfo {author} {\bibfnamefont {D.}~\bibnamefont {Figueruelo}}, \bibinfo {author} {\bibfnamefont {R.}~\bibnamefont {Thubisi}}, \bibinfo {author} {\bibfnamefont {S.}~\bibnamefont {Sahlu}}, \bibinfo {author} {\bibfnamefont {A.}~\bibnamefont {Abebe}},\ and\ \bibinfo {author} {\bibfnamefont {A.}~\bibnamefont {Paliathanasis}},\ }\href {https://arxiv.org/abs/2505.23306} {\bibinfo {title} {Compartmentalization in the dark sector of the universe after desi dr2 bao data}} (\bibinfo {year} {2025}{\natexlab{c}}),\ \Eprint {https://arxiv.org/abs/2505.23306} {arXiv:2505.23306 [astro-ph.CO]} \BibitemShut {NoStop}%
\bibitem [{\citenamefont {Zhang}\ \emph {et~al.}(2006)\citenamefont {Zhang}, \citenamefont {Wu},\ and\ \citenamefont {Zhang}}]{Zhang_2006}%
  \BibitemOpen
  \bibfield  {author} {\bibinfo {author} {\bibfnamefont {X.}~\bibnamefont {Zhang}}, \bibinfo {author} {\bibfnamefont {F.-Q.}\ \bibnamefont {Wu}},\ and\ \bibinfo {author} {\bibfnamefont {J.}~\bibnamefont {Zhang}},\ }\href {https://doi.org/10.1088/1475-7516/2006/01/003} {\bibfield  {journal} {\bibinfo  {journal} {Journal of Cosmology and Astroparticle Physics}\ }\textbf {\bibinfo {volume} {2006}}\bibinfo  {number} { (01)},\ \bibinfo {pages} {003–003}}\BibitemShut {NoStop}%
\bibitem [{\citenamefont {Sebastianutti}\ \emph {et~al.}(2024)\citenamefont {Sebastianutti}, \citenamefont {Hogg},\ and\ \citenamefont {Bruni}}]{Sebastianutti_2024}%
  \BibitemOpen
\bibfield  {number} {  }\bibfield  {author} {\bibinfo {author} {\bibfnamefont {M.}~\bibnamefont {Sebastianutti}}, \bibinfo {author} {\bibfnamefont {N.~B.}\ \bibnamefont {Hogg}},\ and\ \bibinfo {author} {\bibfnamefont {M.}~\bibnamefont {Bruni}},\ }\href {https://doi.org/10.1016/j.dark.2024.101546} {\bibfield  {journal} {\bibinfo  {journal} {Physics of the Dark Universe}\ }\textbf {\bibinfo {volume} {46}},\ \bibinfo {pages} {101546} (\bibinfo {year} {2024})}\BibitemShut {NoStop}%
\bibitem [{\citenamefont {Ma}\ \emph {et~al.}(2010)\citenamefont {Ma}, \citenamefont {Gong},\ and\ \citenamefont {Chen}}]{Ma_2010}%
  \BibitemOpen
  \bibfield  {author} {\bibinfo {author} {\bibfnamefont {Y.-Z.}\ \bibnamefont {Ma}}, \bibinfo {author} {\bibfnamefont {Y.}~\bibnamefont {Gong}},\ and\ \bibinfo {author} {\bibfnamefont {X.}~\bibnamefont {Chen}},\ }\href {https://doi.org/10.1140/epjc/s10052-010-1408-1} {\bibfield  {journal} {\bibinfo  {journal} {The European Physical Journal C}\ }\textbf {\bibinfo {volume} {69}},\ \bibinfo {pages} {509–519} (\bibinfo {year} {2010})}\BibitemShut {NoStop}%
\bibitem [{\citenamefont {Mazumder}\ \emph {et~al.}(2011)\citenamefont {Mazumder}, \citenamefont {Biswas},\ and\ \citenamefont {Chakraborty}}]{mazumder2011interactingholographicdarkenergy}%
  \BibitemOpen
  \bibfield  {author} {\bibinfo {author} {\bibfnamefont {N.}~\bibnamefont {Mazumder}}, \bibinfo {author} {\bibfnamefont {R.}~\bibnamefont {Biswas}},\ and\ \bibinfo {author} {\bibfnamefont {S.}~\bibnamefont {Chakraborty}},\ }\href {https://arxiv.org/abs/1106.4627} {\bibinfo {title} {Interacting holographic dark energy at the ricci scale and dynamical system}} (\bibinfo {year} {2011}),\ \Eprint {https://arxiv.org/abs/1106.4627} {arXiv:1106.4627 [gr-qc]} \BibitemShut {NoStop}%
\bibitem [{\citenamefont {Wang}\ \emph {et~al.}(2013)\citenamefont {Wang}, \citenamefont {Wands}, \citenamefont {Xu}, \citenamefont {De-Santiago},\ and\ \citenamefont {Hojjati}}]{Wang_2013}%
  \BibitemOpen
  \bibfield  {author} {\bibinfo {author} {\bibfnamefont {Y.}~\bibnamefont {Wang}}, \bibinfo {author} {\bibfnamefont {D.}~\bibnamefont {Wands}}, \bibinfo {author} {\bibfnamefont {L.}~\bibnamefont {Xu}}, \bibinfo {author} {\bibfnamefont {J.}~\bibnamefont {De-Santiago}},\ and\ \bibinfo {author} {\bibfnamefont {A.}~\bibnamefont {Hojjati}},\ }\bibfield  {journal} {\bibinfo  {journal} {Physical Review D}\ }\textbf {\bibinfo {volume} {87}},\ \href {https://doi.org/10.1103/physrevd.87.083503} {10.1103/physrevd.87.083503} (\bibinfo {year} {2013})\BibitemShut {NoStop}%
\bibitem [{\citenamefont {Carroll}(2021)}]{carroll2021quantumfieldtheoryeveryday}%
  \BibitemOpen
  \bibfield  {author} {\bibinfo {author} {\bibfnamefont {S.~M.}\ \bibnamefont {Carroll}},\ }\href {https://arxiv.org/abs/2101.07884} {\bibinfo {title} {The quantum field theory on which the everyday world supervenes}} (\bibinfo {year} {2021}),\ \Eprint {https://arxiv.org/abs/2101.07884} {arXiv:2101.07884 [physics.hist-ph]} \BibitemShut {NoStop}%
\bibitem [{\citenamefont {Valentino}\ \emph {et~al.}(2020)\citenamefont {Valentino}, \citenamefont {Gariazzo}, \citenamefont {Mena},\ and\ \citenamefont {Vagnozzi}}]{Valentino_2020_DE}%
  \BibitemOpen
  \bibfield  {author} {\bibinfo {author} {\bibfnamefont {E.~D.}\ \bibnamefont {Valentino}}, \bibinfo {author} {\bibfnamefont {S.}~\bibnamefont {Gariazzo}}, \bibinfo {author} {\bibfnamefont {O.}~\bibnamefont {Mena}},\ and\ \bibinfo {author} {\bibfnamefont {S.}~\bibnamefont {Vagnozzi}},\ }\href {https://doi.org/10.1088/1475-7516/2020/07/045} {\bibfield  {journal} {\bibinfo  {journal} {Journal of Cosmology and Astroparticle Physics}\ }\textbf {\bibinfo {volume} {2020}}\bibinfo  {number} { (07)},\ \bibinfo {pages} {045–045}}\BibitemShut {NoStop}%
\bibitem [{\citenamefont {Pavón}\ and\ \citenamefont {Wang}(2008)}]{Pav_n_2008}%
  \BibitemOpen
\bibfield  {number} {  }\bibfield  {author} {\bibinfo {author} {\bibfnamefont {D.}~\bibnamefont {Pavón}}\ and\ \bibinfo {author} {\bibfnamefont {B.}~\bibnamefont {Wang}},\ }\href {https://doi.org/10.1007/s10714-008-0656-y} {\bibfield  {journal} {\bibinfo  {journal} {General Relativity and Gravitation}\ }\textbf {\bibinfo {volume} {41}},\ \bibinfo {pages} {1–5} (\bibinfo {year} {2008})}\BibitemShut {NoStop}%
\bibitem [{\citenamefont {Poulin}\ \emph {et~al.}(2018)\citenamefont {Poulin}, \citenamefont {Boddy}, \citenamefont {Bird},\ and\ \citenamefont {Kamionkowski}}]{Poulin:2018zxs}%
  \BibitemOpen
  \bibfield  {author} {\bibinfo {author} {\bibfnamefont {V.}~\bibnamefont {Poulin}}, \bibinfo {author} {\bibfnamefont {K.~K.}\ \bibnamefont {Boddy}}, \bibinfo {author} {\bibfnamefont {S.}~\bibnamefont {Bird}},\ and\ \bibinfo {author} {\bibfnamefont {M.}~\bibnamefont {Kamionkowski}},\ }\href {https://doi.org/10.1103/PhysRevD.97.123504} {\bibfield  {journal} {\bibinfo  {journal} {Phys. Rev. D}\ }\textbf {\bibinfo {volume} {97}},\ \bibinfo {pages} {123504} (\bibinfo {year} {2018})},\ \Eprint {https://arxiv.org/abs/1803.02474} {arXiv:1803.02474 [astro-ph.CO]} \BibitemShut {NoStop}%
\bibitem [{\citenamefont {Wang}\ \emph {et~al.}(2018)\citenamefont {Wang}, \citenamefont {Pogosian}, \citenamefont {Zhao},\ and\ \citenamefont {Zucca}}]{Wang:2018fng}%
  \BibitemOpen
  \bibfield  {author} {\bibinfo {author} {\bibfnamefont {Y.}~\bibnamefont {Wang}}, \bibinfo {author} {\bibfnamefont {L.}~\bibnamefont {Pogosian}}, \bibinfo {author} {\bibfnamefont {G.-B.}\ \bibnamefont {Zhao}},\ and\ \bibinfo {author} {\bibfnamefont {A.}~\bibnamefont {Zucca}},\ }\href {https://doi.org/10.3847/2041-8213/aaf238} {\bibfield  {journal} {\bibinfo  {journal} {Astrophys. J. Lett.}\ }\textbf {\bibinfo {volume} {869}},\ \bibinfo {pages} {L8} (\bibinfo {year} {2018})},\ \Eprint {https://arxiv.org/abs/1807.03772} {arXiv:1807.03772 [astro-ph.CO]} \BibitemShut {NoStop}%
\bibitem [{\citenamefont {Visinelli}\ \emph {et~al.}(2019)\citenamefont {Visinelli}, \citenamefont {Vagnozzi},\ and\ \citenamefont {Danielsson}}]{Visinelli:2019qqu}%
  \BibitemOpen
  \bibfield  {author} {\bibinfo {author} {\bibfnamefont {L.}~\bibnamefont {Visinelli}}, \bibinfo {author} {\bibfnamefont {S.}~\bibnamefont {Vagnozzi}},\ and\ \bibinfo {author} {\bibfnamefont {U.}~\bibnamefont {Danielsson}},\ }\href {https://doi.org/10.3390/sym11081035} {\bibfield  {journal} {\bibinfo  {journal} {Symmetry}\ }\textbf {\bibinfo {volume} {11}},\ \bibinfo {pages} {1035} (\bibinfo {year} {2019})},\ \Eprint {https://arxiv.org/abs/1907.07953} {arXiv:1907.07953 [astro-ph.CO]} \BibitemShut {NoStop}%
\bibitem [{\citenamefont {Calder{\'o}n}\ \emph {et~al.}(2021)\citenamefont {Calder{\'o}n}, \citenamefont {Gannouji}, \citenamefont {L'Huillier},\ and\ \citenamefont {Polarski}}]{Calderon:2020hoc}%
  \BibitemOpen
  \bibfield  {author} {\bibinfo {author} {\bibfnamefont {R.}~\bibnamefont {Calder{\'o}n}}, \bibinfo {author} {\bibfnamefont {R.}~\bibnamefont {Gannouji}}, \bibinfo {author} {\bibfnamefont {B.}~\bibnamefont {L'Huillier}},\ and\ \bibinfo {author} {\bibfnamefont {D.}~\bibnamefont {Polarski}},\ }\href {https://doi.org/10.1103/PhysRevD.103.023526} {\bibfield  {journal} {\bibinfo  {journal} {Phys. Rev. D}\ }\textbf {\bibinfo {volume} {103}},\ \bibinfo {pages} {023526} (\bibinfo {year} {2021})},\ \Eprint {https://arxiv.org/abs/2008.10237} {arXiv:2008.10237 [astro-ph.CO]} \BibitemShut {NoStop}%
\bibitem [{\citenamefont {Guedezounme}\ \emph {et~al.}(2025)\citenamefont {Guedezounme}, \citenamefont {Dinda},\ and\ \citenamefont {Maartens}}]{guedezounme2025phantomcrossingdarkinteraction}%
  \BibitemOpen
  \bibfield  {author} {\bibinfo {author} {\bibfnamefont {S.~L.}\ \bibnamefont {Guedezounme}}, \bibinfo {author} {\bibfnamefont {B.~R.}\ \bibnamefont {Dinda}},\ and\ \bibinfo {author} {\bibfnamefont {R.}~\bibnamefont {Maartens}},\ }\href@noop {} {\bibinfo {title} {{Phantom crossing or dark interaction?}}} (\bibinfo {year} {2025}),\ \Eprint {https://arxiv.org/abs/2507.18274} {arXiv:2507.18274 [astro-ph.CO]} \BibitemShut {NoStop}%
\end{thebibliography}%

\end{document}